\pdfoutput=1
\documentclass[11pt]{article}

\usepackage{jheppub}
\usepackage{amsmath}
\usepackage{amssymb}
\usepackage{array}
\usepackage{physics}
\usepackage{dsfont}
\usepackage{silence}
\usepackage{caption}
\usepackage{mathrsfs}
\usepackage{mathtools}
\usepackage{pgfplots}
\pgfplotsset{compat=1.18}
\usepackage{tikz}
\usetikzlibrary{shapes,decorations.text}
\usepackage{pifont}
\usepackage{empheq}
\usepackage{orcidlink}
\usepackage{epiolmec}

\definecolor{darkgreen}{HTML}{006622}
\newcommand{\cmark}{\textcolor{darkgreen}{\text{\ding{51}}}}
\newcommand{\xmark}{\textcolor{red}{\text{\ding{55}}}}
\newcommand{\e}{\epsilon}
\newcommand{\be}[1]{\begin{equation}\label{#1} }
\newcommand{\ee}{\end{equation}}
\newcommand{\bes}[1]{\begin{subequations}\label{#1} }
\newcommand{\ees}{\end{subequations}}
\newcommand{\bea}[1]{\begin{eqnarray}\label{#1} }
\newcommand{\eea}{\end{eqnarray}}
\newcommand{\refb}[1]{(\ref{#1})}

\renewcommand{\L}{{\mathcal{L}}}

\renewcommand{\>}{\rangle}

\newcommand{\D}{\Delta}

\renewcommand{\(}{\left(}
\renewcommand{\)}{\right)}

\renewcommand{\t}{\tau}
\newcommand{\s}{\sigma}

\newcommand{\mt}[1]{\textrm{\tiny #1}}

\newcommand{\bcar}{\beta_{\mathfrak{car}}}
\newcommand{\ocar}{\Omega_{\mathfrak{car}}}
\newcommand{\Zcar}{{Z}_{\mathfrak{ccar}}}

\DeclareMathOperator{\extdm}{d}
\newcommand{\extd}{\extdm \!}

\newcommand{\eq}[2]{\begin{equation} #1 \label{#2} \end{equation}}

\newcommand{\killing}{\zeta} 

\newcommand{\mytitle}{Universal sectors of two-dimensional Carrollian CFTs}

\title{\mytitle}

\author[a,b]{\orcidlink{0000-0003-2093-2377}Ankit Aggarwal,} \author[b,c]{\orcidlink{0000-0002-6481-5933}Arjun Bagchi,}  \author[b,d]{\orcidlink{0000-0001-7529-7978}Stephane Detournay,} \author[a,b,e]{\orcidlink{0000-0001-7980-5394}Daniel Grumiller,} \author[a,f]{\orcidlink{0000-0002-0131-957X}Max Riegler,} \author[g]{and \orcidlink{0000-0002-7672-9389
}Joan Sim\'on} 

\affiliation[a]{Institute for Theoretical Physics, TU Wien, Wiedner Hauptstr.~8-10, A-1040 Vienna, Austria} 

\affiliation[b]{Erwin-Schr\"odinger Internation Institute for Mathematics and Physics, Boltzmanngasse 9, A-1090 Vienna, Austria}

\affiliation[c]{Indian Institute of Technology Kanpur, Kanpur 208016, India} 

\affiliation[d]{Physique Math\'ematique des Interactions Fondamentales, Universit\'e Libre de Bruxelles and International Solvay Institutes, Campus Plaine - CP 231, 1050 Bruxelles, Belgium}

\affiliation[e]{Theoretical Sciences Visiting Program (TSVP), Okinawa Institute of Science and Technology Graduate University, Onna, 904-0495, Japan} 

\affiliation[f]{Quantum
Technology Laboratories GmbH,
Clemens-Holzmeister-Str.~6/6, A-1100 Vienna, Austria}

\affiliation[g]{School of Mathematics and Maxwell Institute for Mathematical Sciences,\\
University of Edinburgh, Edinburgh EH9 3FD, UK}

\emailAdd{aggarwal@hep.itp.tuwien.ac.at}
\emailAdd{abagchi@iitk.ac.in}
\emailAdd{sdetourn@ulb.ac.be}
\emailAdd{grumil@hep.itp.tuwien.ac.at}
\emailAdd{rieglerm@hep.itp.tuwien.ac.at}
\emailAdd{J.Simon@ed.ac.uk}

\preprint{TUW--24--01}
\abstract{%
We revisit modular invariance in two-dimensional Carrollian conformal field theories from a geometric perspective. Focusing on the characters of the induced and highest-weight representations of the theory, we show that there are regions of parameter space where the vacuum character dominates in the dual channel. We use this property to zoom into different subsectors of the Carrollian theory. One of them is reminiscent of the Schwarzian sector of a relativistic CFT$_2$ and has a flat space holographic interpretation as $O$-plane orbifold. It exists only for the highest-weight representation. We prove that for all sectors with vacuum dominance in the dual channel, the specific heat is negative, concurrent with the holographic interpretation of the negative specific heat of asymptotically flat spacetimes with horizons.}

\begin{document}


\maketitle
\listoftables
\noindent\hrulefill
\bigskip
\listoffigures
\noindent\hrulefill
\bigskip


\renewcommand{\thefootnote}{\arabic{footnote}}
\setcounter{footnote}{0}


\section{Introduction}

\subsection*{Carrollian symmetries}

Carrollian symmetries arise from Poincar\'e symmetries in the vanishing speed of light limit \cite{Levy1965,Gupta1966}. After starting life as a mathematical curiosity, recent developments have shown that Carroll symmetries are ubiquitous. This is particularly true in a holographic context since they are associated with null hypersurfaces. Thus, (conformal) Carroll symmetries are relevant for flat space holography \cite{Bagchi:2010zz,Barnich:2010eb,Barnich:2012aw,Barnich:2012rz,Bagchi:2012yk,Bagchi:2012xr,Barnich:2012xq,Bagchi:2013lma,Duval:2014uva,Bagchi:2014iea,Bagchi:2015wna,Hartong:2015usd,Bagchi:2016bcd,Ciambelli:2018wre,Donnay:2022aba,Bagchi:2022emh,Donnay:2022wvx,Bagchi:2023fbj,Saha:2023hsl,Salzer:2023jqv,Saha:2023abr,Mason:2023mti,Chen:2023naw,Nguyen:2023vfz,Alday:2024yyj,Kraus:2024gso,Bagchi:2024efs,Bagchi:2024gnn,Ruzziconi:2024kzo,Chakrabortty:2024bvm,Fiorucci:2025twa}, near horizon holography \cite{Hawking:2016msc,Donnay:2015abr,Afshar:2016wfy,Grumiller:2019fmp} and causal diamond holography \cite{Adami:2023fbm,Freidel:2024emv}. 

Since they render time relative but space absolute, Carroll symmetries have found a variety of applications apart from holography, e.g.~in black holes \cite{Penna:2018gfx,Donnay:2019jiz,Marsot:2022qkx,Gray:2022svz,Redondo-Yuste:2022czg,Freidel:2022bai,Freidel:2022vjq,Bicak:2023rsz} and cosmology \cite{deBoer:2021jej,deBoer:2023fnj,Oling:2024vmq}, condensed matter physics in the context of fractons \cite{Kasikci:2023tvs,Figueroa-OFarrill:2023vbj,Figueroa-OFarrill:2023qty} and flat bands \cite{Bagchi:2022eui}, hydrodynamics of highly boosted fluids \cite{Petkou:2022bmz,Bagchi:2023ysc,Armas:2023dcz,Bagchi:2023rwd} as well as shallow water waves \cite{Bagchi:2024ikw}, and string theory (tensionless strings \cite{Bagchi:2013bga,Bagchi:2015nca,Bagchi:2021ban}, models for black hole microstates \cite{Bagchi:2022iqb} and strings near black holes as well as other Carrollian string theories \cite{Cardona:2016ytk,Bagchi:2023cfp}). By construction, Carroll symmetries are at the heart of Carroll gravity theories \cite{Hartong:2015xda,Bergshoeff:2017btm,Ciambelli:2018ojf,Gomis:2019nih,Ciambelli:2019lap,Grumiller:2020elf,Gomis:2020wxp,Hansen:2021fxi,Campoleoni:2022ebj,Ecker:2023uwm,Grumiller:2024dql} and Carroll (conformal) field theories \cite{Bagchi:2019xfx,Chen:2021xkw,Henneaux:2021yzg,Baiguera:2022lsw,Ecker:2024czx,Bagchi:2024qsb}. For a more detailed introduction to Carrollian symmetries and their applications, we refer the reader to the recent review \cite{Bagchi:2025vri}. 

In this work, we focus on the field theory side and address aspects of Carrollian conformal field theories in two dimensions (CCFT$_2$). We are interested in modular aspects of these theories (previously addressed in \cite{Bagchi:2012xr, Bagchi:2013qva, Bagchi:2019unf}) and the formulation of the Carrollian partition function in terms of characters of the Carroll conformal algebra. 

Our main goal is to exhibit regions in the space of Carroll chemical potentials where the partition function takes a universal form independent of the microscopic details of the underlying theory. Pursuing this goal is motivated by insights due to Ghosh, Maxfield, and Turiaci \cite{Ghosh:2019rcj}, who identified a universal Schwarzian sector in fairly generic 2d (Lorentzian) CFTs. Such a sector has also been found in non-unitary Warped Conformal Field Theories (WCFTs) \cite{Aggarwal:2022xfd,Aggarwal:2023peg}. It is well-known that both 2d CFTs and WCFTs contain a universal Cardy sector \cite{Cardy:1986ie, Detournay:2012pc}, where the partition function is dominated by the vacuum character in the dual channel. A Cardy sector also exists in CCFT$_2$ \cite{Bagchi:2012xr}, and in this paper, we shall address whether another, Schwarzian-like, universal sector also exists for these theories. 

\subsection*{Quantum field theories and universal sectors}

Let us make our line of thought more precise and more specific. Consider a 2d QFT with genus one partition function defined on a 2d surface:  
\begin{equation}\label{PartitionFunction}
 Z(\beta, \theta) = \mbox{Tr} \; e^{-\beta H + i \theta J}\,.
\end{equation}
$H$ and $J$ are the Hamiltonian and (angular) momentum of the theory associated with time- ($t$) and spatial ($\phi$) translations, respectively, while $\beta$ and $\theta$ are the inverse temperature and angular potential. 

Before specializing to CCFT$_2$ [see \eqref{CC-partitionfunction} below], we start with some general observations. The partition function plays a crucial role in capturing the thermodynamic properties of the theory, such as energy levels and their degeneracies, correlation functions, and the structure of excited states. Each term in this trace (taken over the full Hilbert space of the theory) corresponds to specific eigenvalues $(\xi,\Delta)$ of $H$ and $J$, respectively, and counts the number of states at a given energy and angular momentum, resulting in the degeneracy $\mathbb D(\xi,\Delta)$.  
This degeneracy is related to the entropy $S(\beta,\theta) = (1 - \beta \partial_\beta -\theta \partial_\theta) \ln Z$ after a Legendre transform trading chemical potentials for charges, $S(\xi,\Delta) = \ln \mathbb D(\xi,\Delta)$. 

In interacting QFTs, universal sectors for the density of states often emerge in specific limits of interest, such as high-temperature or large-charge limits, where the microscopic details of the theory become irrelevant and the system's behavior is dictated by symmetry principles and scaling laws. Examples include integrable field theories (such as the sine-Gordon model or massive Thirring model \cite{Zamolodchikov:1978xm, Yang:1968rm, Coleman:1974bu, Luther:1975wr}), large $N$ gauge theories \cite{'tHooft:1974bx, Witten:1998qj}, or supersymmetric field theories \cite{Witten:1986bf}.

Another class of examples, the one on which we shall focus in this work, involves 2d field theories with partition function of the form \eqref{PartitionFunction}, supplemented with some version of modular invariance or covariance. The best-known example is the Cardy formula for CFT$_2$ \cite{Cardy:1986ie}, yielding the asymptotic growth of states at high temperature ($\beta \rightarrow 0$), modulo certain assumptions such as unitarity, central charge $c>1$, energy spectrum bounded from below, sufficiently sparse spectrum, and modular invariance. The latter condition, connecting high-energy states to low-energy behavior, is essential in the derivation, allowing to project the whole partition function trace onto the vacuum state. In the context of AdS$_3$/CFT$_2$, the Cardy formula has been instrumental in allowing the matching of the Bekenstein--Hawking entropy of BTZ black holes \cite{Banados:1992gq} with that of a hot gas of strongly interacting particles \cite{Carlip:1994gy,Strominger}. 

Interestingly, similar universal results exist for different classes of 2d field theories endowed with a counterpart of modular invariance such as WCFTs \cite{Detournay:2012pc} and CCFT$_2$ \cite{Barnich:2012xq, Bagchi:2012xr}. There, the regime of validity is slow rotation, $\theta \rightarrow 0$, instead of a high-temperature one, but Cardy-like formulas for the asymptotic density can be obtained\footnote{%
The validity of the Cardy formula for CFT$_2$ was shown to be extendable, modulo a sparseness condition on the spectrum of primaries, in the so-called holographic regime $c \rightarrow \infty$, to $\beta = O(1)$ or energies $\xi \sim c$ \cite{Hartman:2014oaa}. We are not aware of a similar discussion for WCFTs or CCFTs.} 
under conditions similar to their CFT counterpart (in particular, boundedness of the spectrum) and matched to their gravitational counterparts. Results for theories with Lifshitz scaling and more general Carroll/Galilean theories --- of which WCFTs and CCFTs can be seen as special cases --- have been discussed in \cite{Gonzalez:2011nz,Shaghoulian:2015dwa,Perez:2016vqo,Afshar:2016kjj,Grumiller:2017jft,Melnikov:2018fhb, Chen:2019hbj}.

\subsection*{Schwarzian sector}

In the context of CFT$_2$, it was emphasized that yet another sector could be singled out where a universal behavior of the partition function emerges \cite{Ghosh:2019rcj}. This \textit{Schwarzian sector} (also known as near-extremal regime) is reached when 
\eq{
\textrm{Schwarzian\;sector\;in\;CFT}_2:\qquad \beta_L\sim c\gg 1\qquad\qquad 	\beta_R\sim c^{-\alpha} \ll 1
}{next_CFT}
with $\alpha>0$ \cite{Aggarwal:2022xfd} and left/right-moving chemical potentials defined through 
\eq{
 \beta = \frac{1}{2}(\beta_R + \beta_L) \qquad\qquad \theta = \frac{1}{2i}(\beta_R - \beta_L)\,.   
}{eq:lrpot}
The terminology originates from the fact that this sector is capturing a low temperature sector of the CFT$_2$ at large central charge, i.e., $T \sim 1/c \ll 1$ --- and also the behavior of near-extremal BTZ black holes in the semi-classical regime. In the range \eqref{next_CFT}, it can be shown that the partition function \eqref{PartitionFunction} is universal, in particular, dominated by the vacuum character with exponentially suppressed corrections as long as the spectrum exhibits a twist gap.\footnote{%
In QFT, the twist is the scaling dimension minus the spin. In CFT$_2$, it is given by the (smaller of the) left- or right-moving Virasoro zero mode eigenvalues.} 
It takes the form of a Schwarzian action in the left-moving inverse temperature $\beta_L$, resulting in a grand canonical entropy given by a Cardy entropy supplemented with a (large) logarithmic correction \cite{Ghosh:2019rcj}. 

Moving away from CFT$_2$, (non-unitary) WCFTs were in turn shown to display a similar universal near-extremal sector \cite{Aggarwal:2022xfd} (shown to describe the behavior of near-extremal Warped AdS$_3$ black holes in \cite{Aggarwal:2023peg}), 
\eq{
\textrm{Schwarzian\;sector\;in\;WCFT}:\qquad - i \theta \sim c \gg 1\qquad\qquad 	\beta\sim c^{\alpha} \gg 1\,.
}{next_WCFT}
In that case, the partition function \eqref{PartitionFunction} possesses a so-called warped Schwarzian sector \cite{Chaturvedi:2020jyy, Afshar:2019tvp}. Notice that this is only true in non-unitary WCFTs with an imaginary vacuum value for the Hamiltonian; unitary WCFTs do not have a near-extremal limit (even though they have a Cardy limit) where the vacuum
character dominates \cite{Aggarwal:2022xfd}.

\subsection*{Overture}

The goal of this paper is to investigate whether a generic CCFT$_2$ exhibits a similar universal near-extremal sector, different from the known Cardy-like sector of \cite{Barnich:2012xq, Bagchi:2012xr}. As we shall demonstrate, the answer is affirmative: There are six sectors (naturally grouped into three pairs, one with a Cardy sector, one with a Schwarzian sector, and one with a Boltzmann sector) where the partition function \eqref{PartitionFunction} in the $S$-dual channel is dominated by the Carroll vacuum character. As a byproduct, we shall prove the negativity of the specific heat for all these sectors, recover Cardy-like formulas from our partition functions, provide a holographic interpretation of our results, in particular, in terms of $O$-plane orbifolds for the Schwarzian sector, and highlight some key differences between induced and highest-weight representations. In particular, we argue that the Schwarzian sector only exists for highest-weight representations. 

The first six paragraphs of our conclusions, Section \ref{sec:6}, contain a more detailed version of this overture, with references to key formulas, Tables, and Figures. Readers who want to skip most of the details and get a summary of the bullet points are encouraged to start with this reprise.

\subsection*{Outline of the rest of the paper}

The remainder of this paper is organized as follows. In Section \ref{sec:2}, we summarize key aspects of CCFTs and Carroll modular invariance. In Section \ref{sec:3}, we discuss Carroll characters, their behavior under modular transformations, and their appearance in Carrollian partition functions. In Section \ref{sec:4}, we demonstrate in which regions of the parameter space the vacuum character dominates in the $S$-dual channel, permitting us to zoom into different subsectors of CCFTs. Key lessons from this Section are the existence of six such sectors and the negativity of specific heat for all of them. In Section \ref{sec:NEXT}, we elaborate in detail on the Schwarzian sector of CCFT$_2$ and prove that its microcanonical partition function does not yield log corrections to the leading order entropy. In Section \ref{sec:holo}, we provide a holographic interpretation of the Schwarzian sector as $O$-plane orbifolds. In Section \ref{sec:7}, we put our results in the context of existing literature. In Section \ref{sec:6}, we summarize our results and conclude. Appendix \ref{app:A} contains details on the saddle point approximation used in some of the ensembles other than the grand canonical. Appendix \ref{app:B} explicates the lowest-weight characters of 2d CFTs.


\section{Carrollian CFTs and Carroll modular invariance}\label{sec:2}

In this Section, we review symmetry aspects of CCFTs, with a particular focus on 2d. In Subsection \ref{sec:2.1}, we review key aspects of conformal Carroll symmetries and their spacetime interpretation. In Section \ref{sec:2.2}, we define the Carroll partition function. In Section \ref{sec:2.3}, we recover CCFT$_2$ results as a limit from Lorentzian CFT$_2$ results. Finally, in Section \ref{sec:2.4} we discuss Carroll modular transformations.

\subsection{Carroll symmetries and conformal Carroll symmetries}\label{sec:2.1}

Carroll symmetry arises geometrically on a $d$-dimensional Carroll manifold which is defined by the pair ($\tau^\mu, h_{\mu\nu}$) where $h_{\mu\nu}$ is a degenerate symmetric tensor and $\tau^\mu$ is a vector that generates the 1-dimensional kernel of $h_{\mu\nu}$, see e.g.~\cite{Bergshoeff:2022eog}. The Carroll algebra is generated by the isometries of this (weak) Carroll structure, $\L_\killing \t^{\mu}=\L_\killing h_{\mu\nu}=0$. When restricting to flat Carrollian spacetimes, $\t=\partial_t$ and $h_{\mu\nu}\,\extd x^\mu\extd x^\nu = \extd x_1^2 + \dots + \extd x_{d-1}^2$, the conformal isometry of this structure
\eq{
\L_\killing \t^{\mu}=-\lambda\,\t^{\mu}\qquad\qquad\L_\killing h_{\mu\nu}=2\lambda\,h_{\mu\nu}
}{eq:ccft1}
generates the $d$-dimensional conformal Carroll algebra, $\mathfrak{ccar}_d$, which is isomorphic to the $(d+1)$-dimensional Bondi--van~der~Burgh--Metzner--Sachs (BMS) algebra, $\mathfrak{bms}_{d+1}$. This isomorphism is the basis of the Carrollian approach to flat space holography \cite{Duval:2014uva}. From now on we focus on $d=2$, so that the algebra of relevance is $\mathfrak{bms}_3$ or, equivalently, $\mathfrak{ccar}_2$, and we have only one spatial coordinate, $x_1=x$.

The conformal Carroll Killing equations \eqref{eq:ccft1} are solved by the conformal Carroll vector fields
\eq{
\killing\big({\cal M},\,{\cal L}\big) = \big({\cal M}(x) + t\,\partial_x{\cal L}(x)\big)\,\partial_t+{\cal L}(x)\,\partial_x
}{eq:ccft2}
that depend on two arbitrary functions, $\cal M$ and $\cal L$, of the spatial coordinate $x$. Their Lie-bracket algebra
\eq{
    \big[\killing\big({\cal M}_1,\,{\cal L}_1\big),\,\killing\big({\cal M}_2,\,{\cal L}_2\big)\big]_{\textrm{\tiny Lie}} = \killing\big({\cal M}_{1\circ2},\,{\cal L}_{1\circ2}\big)
}{eq:ccft3}
with ${\cal M}_{1\circ2}={\cal M}_1 \partial_x{\cal L}_2-{\cal M}_2\partial_x{\cal L}_1+{\cal L}_1 \partial_x{\cal M}_2-{\cal L}_2\partial_x{\cal M}_1$ and ${\cal L}_{1\circ2}={\cal L}_1 \partial_x{\cal L}_2-{\cal L}_2\partial_x{\cal L}_1$, consists of a semi-direct sum of the Witt algebra (generated by $\cal L$) with an abelian algebra (generated by $\cal M$). Expanding in modes [for instance, Fourier modes $M_n=\killing(e^{inx},0)$, $L_n=\killing(0,e^{inx})$] and allowing for central extensions yields the centrally extended $\mathfrak{bms}_3$ (or, equivalently, centrally extended $\mathfrak{ccar}_2$) commutator algebra:
\bes{bms}
\begin{empheq}[box=\fbox]{align}\phantom{\bigg(}
[L_n,\,L_m] &= (n-m)\,L_{n+m} + \frac{c_{\mt L}}{12}\,\big(n^3-n\big)\,\delta_{n+m,\,0} \\ 
[L_n,\,M_m] &= (n-m)\,M_{n+m} + \frac{c_{\mt M}}{12}\,\big(n^3-n\big)\,\delta_{n+m,\,0} 
\phantom{\bigg)}\\
\phantom{\bigg)}[M_n,\,M_m] &=0
\end{empheq} 
\ees
In a gravity context, the algebra \eqref{bms} is the asymptotic symmetry algebra of 3d asymptotically flat spacetimes with Barnich--Comp\`ere boundary conditions \cite{Barnich:2006av}. In this case, $x=\phi$ parametrizes the celestial $S^1$ and $t=u$ ($t=v$) is the retarded (advanced) null direction along $\mathscr{I}^+$ ($\mathscr{I}^-$). The $L_n$ generate the diffeomorphisms of the celestial circle and are called superrotations. The $M_n$ are angle-dependent translations of the null direction and are called supertranslations. For Einstein gravity, the Virasoro central charge turns out to vanish, $c_{\mt L}=0$, while the mixed central charge is non-zero, $c_{\mt M}\neq 0$. The global subalgebra insensitive to the central terms is $\mathfrak{isl}(2,\mathbb{R})$ generated by $L_{\pm 1}, L_0, M_{\pm 1}, M_0$. In a holographic context, these six generators correspond to the Killing vectors of 3d Minkowski spacetime, the geometric dual to the CCFT$_2$ vacuum on the cylinder.

Since they are the prime candidates for the holographic dual of asymptotically flat gravity in 3d, we are interested in 2d quantum field theories that exhibit the symmetry \refb{bms}, i.e., CCFT$_2$. Now we go on to describe such field theories. 

\subsection{Carroll partition functions}\label{sec:2.2}

Given a CCFT$_2$ governed by the symmetries \eqref{bms}, it is useful to employ a vector field representation of these symmetries, like the one given in \eqref{eq:ccft2}, which in terms of the Fourier modes introduced above yields
\[ \label{carcy}
L_n = i e^{in\phi} (\partial_\phi + i n t \partial_t) \qquad\qquad M_n = ie^{in\phi} \partial_t\,.
\]
These are the generators on the null cylinder $\mathscr{I}^\pm$ with $t$ playing the role of retarded/advanced time. The zero-mode vector fields
\[
L_0 = i\partial_\phi=J \qquad\qquad M_0 = i\partial_t=H
\]
generate angular rotations and (null) time-translations of the dual field theory, respectively. Therefore, $M_0$ is to be identified with the Hamiltonian, while $L_0$ with the angular momentum generator. 

We define the partition function of the CCFT$_2$ as
\begin{equation}\label{CC-partitionfunction}
Z_\mathfrak{ccar}(\rho, \sigma) = \text{Tr} \, e^{2\pi i(\rho M_0+ \sigma L_0)} = \text{Tr} \, e^{-\beta_\mathfrak{car} H + i \theta_\mathfrak{car} J}\,.
\end{equation}
In the last equality, we introduced the inverse of the Carroll temperature $\beta_\mathfrak{car}$ and the angular potential $\theta_\mathfrak{car}$. This also gives us the mapping between chemical potentials
\be{eq:4}
\beta_\mathfrak{car} = - 2 \pi i \rho \qquad\qquad \theta_\mathfrak{car} = 2 \pi \sigma\,.
\ee 

\subsection[Carroll limit from 2d \texorpdfstring{CFT}{CFT2}]{Carroll limit from 2d \texorpdfstring{CFT}{CFT2}}\label{sec:2.3}

A CCFT$_2$ can be obtained by sending the speed of light to zero in a 2d relativistic CFT. Here, we recall how this works. We start with the generators
\be{eq:5}\L_n = ie^{in\omega} \partial_\omega \qquad\qquad {\bar{\L}}_n = ie^{in\bar\omega} \partial_{\bar\omega} \qquad\qquad \omega,\, \bar\omega = t \pm \phi
\ee
of a CFT$_2$ on a cylinder, which generate two copies of the Witt algebra. In the Euclidean version of the theory, we additionally compactify time, so that our CFT$_2$ is defined on a torus.

For the Carroll limit, we send the speed of light to zero, implying the scaling
\eq{
t \to \e\, t \qquad\qquad \phi \to \phi \qquad\qquad \e\to 0\,. 
}{eq:6}
In this limit, we need to scale the linear combination of generators
\eq{
L_n = \L_n - {\bar{\L}}_{-n} \qquad\qquad M_n = \e(\L_n + {\bar{\L}}_{-n})
}{eq:7}
to keep them finite. Starting with the expressions \refb{eq:5}, the above equation \refb{eq:7} results in \refb{carcy}. The original time direction is thus infinitely boosted to a null time. Replacing the Witt generators \eqref{eq:5} by Virasoro generators and defining the BMS generators through \eqref{eq:7} recovers the BMS algebra \refb{bms} with central charges 
\eq{
c_{\mt L} = c-\bar c\qquad\qquad c_{\mt M} = \lim_{\e\to 0} \e\,\big(c+\bar c\big)
}{eq:ccft4}
where $c$ and $\bar c$ are the original Virasoro central charges. Thus, if the original CFT$_2$ had no gravitational anomaly, $c=\bar c$, the CCFT$_2$ emerging from its contraction has no Virasoro central charge, $c_{\mt L}=0$. This explains why in the context of Einstein gravity we typically only care about the case $c_{\mt L}=0$. Another noteworthy aspect is that the mixed central charge $c_{\mt M}$ is non-zero and finite only if the original Virasoro central charges diverge linearly in $1/\e$. For Einstein gravity, this is indeed the case since the contraction parameter is the inverse AdS radius \cite{Barnich:2006av}, and the Brown--Henneaux central charges diverge linearly as the AdS radius tends to infinity \cite{Brown:1986nw}.

We now perform the same scaling at the level of Euclidean partition functions. The CFT partition function 
\be{eq:8}
Z_{\mt{CFT}} = \text{Tr} \, e^{2\pi i(\t \L_0- \bar{\t} \bar{\L}_0)} = \text{Tr} \, e^{-\beta_{\mt{CFT}} H_{\mt{CFT}} + i \theta_{\mt{CFT}} J_{\mt{CFT}}}
\end{equation}
entails the identifications
\be{eq:9}
\t = \frac{\theta_\text{\tiny CFT}}{2\pi} + i\, \frac{\beta_\text{\tiny CFT}}{2\pi} \qquad\qquad \bar{\t} = \frac{\theta_\text{\tiny CFT}}{2\pi} - i\, \frac{\beta_\text{\tiny CFT}}{2\pi}\,.
\end{equation}
While for real $\beta_\text{\tiny CFT},\theta_\text{\tiny CFT}$ the quantities $\tau, \bar\tau$ are complex conjugates of each other, this is no longer true if we complexify the chemical potentials, which is what we are going to do in later Sections.

We demand that the relativistic CFT partition function smoothly interpolates to the CCFT partition function as we send the speed of light to zero, 
\be{eq:10}
Z_\text{\tiny CFT} \longrightarrow Z_\mathfrak{ccar} \quad \text{as} \quad  \e \to 0
\end{equation}
implying
\be{eq:11}
Z_\text{\tiny CFT} = \text{Tr} \, e^{2\pi i(\t \L_0- \bar{\t} \bar{\L}_0)} \to \text{Tr} \, e^{\pi i[\t (L_0+ \frac{1}{\e}M_0) -\bar{\t}(-L_0+\frac{1}{\e} M_0)]} = \text{Tr} \, e^{2\pi i(\sigma L_0 +\rho M_0)} = Z_\mathfrak{ccar}\,.
\end{equation}
Thus, the modular parameters $\tau$, $\bar\tau$ decompose into an order unity and an infinitesimal part,
\be{mod}
2\s = \t + \bar\t\qquad\qquad 2\rho = \frac{1}{\e} (\t -\bar{\t}) \qquad\Rightarrow\qquad \t, \bar\t = \s \pm \e\,\rho\,. 
\end{equation}

Matching the angular potential and the inverse temperatures between relativistic and CCFTs yields
\begin{equation}
\theta_\text{\tiny CFT} = \theta_\mathfrak{car} = 2\pi\sigma\qquad\qquad \beta_\text{\tiny CFT} = \e\,\beta_\mathfrak{car}=-2\pi i\,\e\,\rho\,.
\label{eq:12}
\end{equation}
Since $\theta_\text{\tiny CFT}=i\beta_{\mt{CFT}}\Omega_{\mt{CFT}}$, we infer the Carrollian angular velocity is related to the CFT$_2$ angular velocity as
\begin{equation}
   \Omega_{\mathfrak{car}} = \epsilon\,\Omega_{\mt{CFT}}\,.
\label{eq:ocar}
\end{equation}

The formulas above allow different interpretations, depending on the specific application. If one keeps fixed the CFT temperature and angular velocity, the corresponding Carrollian quantities vanish. This can be understood as the signature Carrollian property that ``nothing can move (within the Carrollian lightcone)'' since the speed of light tends to zero. This interpretation is not particularly useful for CCFT$_2$ applications. Alternatively, if one keeps fixed the Carrollian temperature and angular velocity, the corresponding relativistic quantities tend to infinity. So, Carrollian rotation corresponds to faster-than-light movement, which is possible in a Carrollian context without necessarily violating energy bounds \cite{Ecker:2024czx}. In our work, we adopt the viewpoint that the Carrollian quantities are finite by default, though, as we shall see, the condition of vacuum dominance in the dual channel will drive us towards large or small values for at least one of the chemical potentials. Before we can discuss this, we need to introduce Carroll modular transformations, since they will provide the main tool to study vacuum dominance.

\subsection{Carroll modular transformations}\label{sec:2.4}

Starting with the CFT$_2$ modular transformations PSL$(2,\mathbb{Z})$,
\be{eq:13}
\t \to \frac{a \t + b}{c \t + d} \qquad\qquad ad-bc=1\quad\textrm{with}\quad a,b,c,d\in\mathbb{Z}
\end{equation} 
and exploiting the relation \eqref{mod} between original and Carrollian modular parameters yields the expansion
\eq{
\tau=\s+\e\,\rho\to\frac{a \s + b}{c \s + d} + \e\,\rho\,\frac{ad-bc}{(c\sigma+d)^2} +{\cal O}(\e^2)\,.
}{eq:ccft6}
Keeping only terms up to linear order in $\e$ establishes the Carroll modular transformation
\eq{\boxed{\phantom{\bigg(}
\s \to \frac{a \s + b}{c \s + d}\qquad\qquad \rho \to \frac{\rho}{(c\s +d)^2}\,.
\phantom{\bigg)}}}{eq:angelinajolie}
So we still have PSL$(2,\mathbb{Z})$ and a modular parameter $\s$ but the second parameter, $\rho$, transforms differently. Below, we explore the meaning of this transformation behavior.

The geometric picture associated with the small $\e$-limit is as follows. Assuming Im$(\tau)>0$, the original modular transformation acts on the complex upper half-plane $\mathcal{H}$. The Carroll limit zooms into the region slightly above the real axis, so that to leading order (LO) we see only the (standard) modular transformation of $\s$, while the first subleading term (given by $\rho$) transforms like an element in the tangent space.

A formal way to arrive at the same results as in the limiting construction and to explicitly implement the geometric picture advocated above is to extend the action of any modular transformation $g\in\textrm{PSL}(2,\mathbb{Z})$ at any given point $z\in\mathcal{H}$ to the tangent space of the base manifold $\text{T}_z\mathcal{H}$. One can then define the derivative action of $g$, $Dg:\text{T}_z\mathcal{H}\to\text{T}_{g(z)}\mathcal{H}$ as
\begin{equation}
        Dg(z,v) = (g(z),g'(z)v)\quad\text{where}\quad g'(z) = \frac{1}{(cz + d)^2}
\end{equation}
with $v\in\text{T}_z\mathcal{H}$. This is the same transformation behavior as in \eqref{eq:angelinajolie}. Hence, a sensible way to think about Carroll modular transformations is in terms of a simultaneous action of modular transformations on a given point of the base manifold and the resulting derivative action on the tangent space. So, similarly to a general CFT, it can be useful to think about CCFTs at finite temperature to be defined on the complex upper half-plane and the corresponding tangent space with some additional reality/imaginary conditions on the Carroll modular parameters $\s$ and $\rho$ (analogous to $\bar{z}=z^*$ in the Euclidean CFT case).

The Carroll modular group is generated by composing $S$- and $T$-transformations given by
\eq{
S:\; \s\to-\frac{1}{\s}\qquad\rho\to\frac{\rho}{\s^2}\qquad\qquad T:\;\s\to\s+1\qquad\rho\to\rho\,.
}{eq:ccft7}
The $S$-transformation acts on both modular parameters, while the $T$-transformation only acts on $\s$. In terms of the physical chemical potentials in \eqref{eq:12} and \eqref{eq:ocar}, the $S$-transformation acts as
\begin{equation}
S:\; \frac{(\bcar\ocar)^\prime}{2\pi} = \frac{2\pi}{\bcar\ocar}\qquad \qquad \ocar^\prime = -\ocar\,. 
\label{eq:pot-transf}
\end{equation}
The $T$-transformation leaves $\bcar$ invariant while shifting the Carrollian angular velocity
\begin{equation}
  T:\; \bcar^\prime = \bcar \qquad \qquad \ocar^\prime = \ocar - i\,\frac{2\pi}{\bcar}\,.
\label{eq:T-for}
\end{equation}
These modular transformations satisfy the usual identity relations
\eq{
S^2=\mathds{1}\qquad\qquad(ST)^3=\mathds{1}\,.
}{eq:ccft8}
The first identity is evident by inspection. The second identity follows from an explicit computation of consecutive actions of $ST$,
\eq{
ST:\;\s\to-\frac{1}{\s+1}\qquad\rho\to\frac{\rho}{(\s+1)^2}\qquad\qquad(ST)^2:\;\s\to-\frac{\s+1}{\s}\qquad\rho\to\frac{\rho}{\s^2}\,.
}{eq:ccft9}
Defining $\sigma_1=T^{-1}$ and $\sigma_2=TST$, the generators $\sigma_i$ obey the usual braiding relation $\sigma_1\sigma_2\sigma_1=\sigma_2\sigma_1\sigma_2$.

These modular transformations, especially the $S$-transformation defined in \eqref{eq:ccft7},
were used extensively to derive a BMS--Cardy formula \cite{Bagchi:2012xr} (see also \cite{Barnich:2012xq}), an asymptotic formula for structure constants \cite{Bagchi:2020rwb}, and two-point functions of probes in a thermal background \cite{Bagchi:2023uqm}. 


\section{Carrollian partition functions as sums over Carrollian characters}
\label{sec:3}

In this Section, we summarize aspects of Carrollian partition functions and how to represent them in terms of Carrollian characters. We start with a review of states and representations --- highest-weight and induced --- in Subsection \ref{sec:3.1} and then present Carrollian characters and partition functions in Subsection \ref{sec:3.2} using both of these representations. Subsection \ref{sec:3.3} displays the expectation values of the Carrollian charges, their variances, and the covariance extracted from the Carrollian partition function in the grand canonical ensemble. Subsection \ref{sec:3.4} collects some properties of a key ingredient in the Carrollian partition functions, the Dedekind $\eta$-function, that will prove to be useful for later Sections. Subsection \ref{sec:3.optional} derives the CCFT vacuum characters as limits from CFT vacuum characters; more specifically, we shall see that the standard CFT vacuum character yields the induced CCFT vacuum character and the so-called flipped vacuum character (defined in Subsection \ref{sec:3.1}) yields the highest-weight CCFT vacuum character.

\subsection{States and representations}\label{sec:3.1}

The states in a CCFT$_2$ are labelled with the eigenvalues of $L_0$ and $M_0$:
\eq{
L_0 |\xi,\Delta\> = \D |\xi,\Delta\> \qquad\qquad M_0 |\xi,\Delta\> = \xi |\xi,\Delta\>\,.
}{st}
We can construct two very different types of representations from these states. The first are the representations familiar from CFT$_2$, the highest-weight representations. Here we have primary states $|\xi,\Delta\>_p$ that are annihilated by positive modes of the generators:
\eq{
L_n |\xi,\Delta\>_p = M_n |\xi,\Delta\>_p =0 \qquad \forall n>0}
{hw}
This means that the weight $\D$ is bounded from below as the action of the positive modes is to lower the value of $\D$. Generic states in this representation are built by acting with negative modes of the generators $L, M$ on these primary states:
\eq{|\Psi\> = L_{-n_1}L_{-n_2}\ldots L_{-n_q} M_{-m_1}M_{-m_2}\ldots M_{-m_r} |\xi,\Delta\>_p \qquad n_i, m_j>0
}{eq:nolabel}

The second type is the induced representation, where all the supertranslations (except for $M_0$) annihilate the states:
\eq{
M_n|\xi,\Delta\>_I=0 \qquad \forall n \neq 0\,.
}{ind}
Generic states in this representation are obtained by acting with arbitrary combinations of $L_n$ generators (not necessarily $n>0$) on $|\xi,\Delta\>_I$:
\eq{
|\Phi\> = L_{n_1}L_{n_2}\ldots L_{n_m} |\xi,\Delta\>_I\,.
}{eq:whatever}

These induced representations naturally arise from the highest-weight representation of a relativistic CFT$_2$ via the map discussed in \refb{eq:7}. This is so because imposing the highest-weight conditions in the parent CFT$_2$,
\eq{
\mathcal{L}_n|\psi\rangle=0=\bar{\mathcal{L}}_n|\psi\rangle \qquad\forall n>0
}{eq:indhw1}
upon using \eqref{eq:7} induces the conditions
\begin{align}
     \L_n |\psi\> = \big(L_n + \frac{1}{\e} M_n \big) |\psi\> = 0  \qquad\qquad \bar{\L}_n |\psi\> &= \big(-L_{-n} + \frac{1}{\e} M_{-n} \big) |\psi\> = 0 \qquad\qquad \forall n>0 \nonumber \\
    \xRightarrow[\textrm{limit}]{\textrm{Carroll}}\quad M_n | \Psi\> &=0 \qquad\qquad \forall n\neq 0\,.
    \label{{eq:indhw0}}
\end{align}
The CCFT central charges associated with the limiting procedure yielding induced representations are given by \eqref{eq:ccft4} \cite{Campoleoni:2016vsh}.

By contrast, to get the highest-weight representations in the CCFT from a limit, we need to start with flipped representations in the CFT$_2$ \cite{Bagchi:2020fpr},
\eq{
\mathcal{L}_n|\psi\rangle=0=\bar{\mathcal{L}}_{-n}|\psi\rangle \qquad\forall n>0\quad
     \xRightarrow[\textrm{limit}]{\textrm{Carroll}}\quad L_n|\Psi\rangle=0=M_n|\Psi\rangle \qquad \forall n>0\,.
}{eq:indhw2}
If the state $|\psi\rangle=|0\rangle$ is the vacuum state and obeys the left conditions \eqref{eq:indhw2} then we refer to it as the ``flipped vacuum''. It is convenient to exploit the Virasoro algebra automorphism \cite{Bagchi:2019unf}
\eq{
    \bar{\L}_n \leftrightarrow - \bar{\L}_{-n}\qquad\qquad \bar c \leftrightarrow -\bar c 
}{eq:indhw3}
so that the flipped representation can be reinterpreted as highest-weight representation of the flipped Virasoro algebra. As a consequence of these sign flips, the CCFT central charges are given by
\eq{
c_{\mt L} = c+\bar c\qquad\qquad c_{\mt M}=\lim_{\epsilon\to 0} \epsilon\,(c-\bar c)\,.
}{eq:indhw6}
 We elaborate on the flipped representations in Subsections \ref{sec:3.optional}, \ref{sec:7.2}, \ref{sec:flip}, Table \ref{tab:2}, and Appendix \ref{app:B}.

\subsection{Characters and partition functions}\label{sec:3.2}

The two representations above have almost identical characters, up to small but significant differences involving the modulus \cite{Oblak:2015sea, Bagchi:2019unf}. For the highest-weight representations, the character for eigenvalues $\xi,\Delta$ in CCFT$_2$ with central charges $c_{\mt M}, c_{\mt L}$ is given by \cite{Bagchi:2019unf} 
\eq{\chi_{(c_{\mt M},c_{\mt L},\xi,\Delta)}(\rho, \sigma) = \frac{e^{-2\pi i(\sigma \frac{c_{\mt L}-2}{24}+\rho \frac{c_{\mt M}}{24})}e^{2\pi i(\sigma\Delta+\xi\rho)}}{\eta(\sigma)^2}
}{BMS_character3}
where $\eta(\s)$ is the Dedekind $\eta$-function, see Subsection \ref{sec:3.3} below for its definition and properties. The above formula is for non-vacuum states. For the vacuum $(\Delta=0,\xi=0)$, we have 
\eq{
\chi_{\mathds 1}(\rho, \sigma):=\chi_{(c_{\mt M},c_{\mt L},0,0)}(\rho, \sigma) =  \frac{e^{-2\pi i(\sigma \frac{c_{\mt L}-2}{24}+\rho \frac{c_{\mt M}}{24})}}{\eta(\sigma)^2}\,\big(1-e^{2\pi i \sigma}\big)^2\,.
}{eq:charactervacuum}
For induced representations, the characters above are slightly modified and get absolute values, see \cite{Oblak:2015sea}, Eqs.~(35) and (43), for details and also Eq.~\eqref{eq:lamourdemavie} below.

Notice that Carroll characters, \eqref{BMS_character3} and \eqref{eq:charactervacuum}, can be factored into functions $\chi^{\mt L}(\sigma)$, $\chi^{\mt M}(\rho)$ so that a generic Carroll character can be written as\footnote{%
This is analogous to the familiar split of CFT$_2$ characters into left and right characters.}
\begin{equation}
\chi_{(c_{\mt M},c_{\mt L},\xi,\Delta)}(\rho, \sigma) = \chi^{\mt L}_{(c_{\mt L}, \Delta)}(\sigma)\,\chi^{\mt M}_{(c_{\mt M}, \xi)}(\rho)~.
\end{equation}
Here,
\eq{
\chi^{\mt L}_{(c_{\mt L}, \Delta)}(\sigma)=\frac{e^{2\pi i\sigma(\Delta-\frac{c_{\mt L}-2}{24})}}{\eta(\sigma)^2}\,\big(1-\delta_{\textrm{\tiny vac}}e^{2\pi i \sigma}\big)^2 \qquad\qquad
\chi^{\mt M}_{(c_{\mt M}, \xi)}(\rho)=e^{2\pi i \rho(\xi-\frac{c_{\mt M}}{24})}
}{eq:characters LM}
and 
\be{eq:yetanother}
\delta_{\textrm{\tiny vac}}=\begin{cases}1, & \Delta=\xi=0\\
0, & \textrm{otherwise}  \end{cases}\,.
\end{equation}
In particular, the vacuum character \eqref{eq:charactervacuum} factorizes into $\chi_{\mathds 1}(\rho, \sigma)=\chi^{\mt L}_{\mathds 1}(\sigma)\,\chi^{\mt M}_{\mathds 1}(\rho)$ with
\eq{
\chi^{\mt L}_{\mathds 1}(\sigma):=  \frac{e^{-2\pi i\,\sigma \frac{c_{\mt L}-2}{24}}}{\eta(\sigma)^2}\,\big(1-e^{2\pi i \sigma}\big)^2 \qquad\qquad 
\chi^{\mt M}_{\mathds 1}(\rho):=  e^{-2\pi i\,\rho \frac{c_{\mt M}}{24}}\,.
}{eq:vacchar}

We focus on building the CCFT out of these characters. Our first observation is that the CCFT$_2$ partition function \eqref{CC-partitionfunction} can be constructed out of the characters described above.
\begin{empheq}[box=\fbox]{multline}
\label{partition_function}
\textrm{highest-weight:}\quad\sigma\in\mathcal{H}\qquad  Z_\mathfrak{ccar}(\rho, \sigma) = \sum_{\xi, \Delta}  D(\xi,\Delta) \, \chi_{(c_{\mt M},c_{\mt L},\xi,\Delta)}(\rho, \sigma) \\
= \frac{e^{-2\pi i(\sigma \frac{c_{\mt L}-2}{24}+\rho \frac{c_{\mt M}}{24})}}{\eta(\sigma)^2} \bigg(D(0,0)\,(1-e^{2\pi i \sigma})^2+\sum_{\xi, \Delta \neq 0}  D(\xi,\Delta) \,e^{2\pi i(\sigma\Delta+\rho\xi)}\bigg) 
\end{empheq}
Here $D(\xi,\Delta)$ is the multiplicity or density of primaries with weight $(\xi,\Delta)$, and we assume that the vacuum is non-degenerate, $D(0,0)= 1$. The partition function \eqref{partition_function} requires $\sigma$ to lie in the complex upper half-plane $\mathcal{H}$.

Finally, we express the CCFT$_2$ partition function in terms of induced characters,
\begin{empheq}[box=\fbox]{multline}
\textrm{induced:}\quad\rho\in i\cdot\mathbb{R}^+,\;\sigma\in\mathbb{R}\qquad  Z_\mathfrak{ccar}(\rho, \sigma) = \sum_{\xi, \Delta}  D(\xi,\Delta) \, \chi^{\textrm{\tiny ind}}_{(c_{\mt M},c_{\mt L},\xi,\Delta)}(\rho, \sigma) \\
 = \frac{e^{-2\pi i(\sigma \frac{c_{\mt L}}{24}+\rho \frac{c_{\mt M}}{24})}}{|\eta(\sigma+i\epsilon)|^2} \bigg(D(0,0)\,\big|1-e^{2\pi i\sigma-2\pi\epsilon}\big|^2 +\sum_{\xi, \Delta \neq 0}  D(\xi,\Delta)\, e^{2\pi i(\sigma\Delta+\rho\xi)}\bigg)
\label{eq:lamourdemavie}
\end{empheq}
with the same meanings of $D(0,0)=1$ and $D(\xi,\Delta)$ as before. The $\epsilon$-terms (with $\epsilon\ll 1$) are kept to regularize this expression. 

We shall demonstrate in Subsection \ref{sec:3.optional} that the restrictions of the ranges of the Carrollian chemical potentials imposed in \eqref{partition_function} and \eqref{eq:lamourdemavie} arise naturally in limits from CFT$_2$ vacuum characters together with properties of the Dedekind $\eta$-function discussed in Subsection \ref{sec:3.4}.

\subsection{Expectation values, variances, and covariance}\label{sec:3.3}

Here, we collect a few simple formulas that will be useful in applications in the next Section, such as our discussion of vacuum dominance, or for thermodynamical observables, such as the specific heat.

From the grand canonical partition function \eqref{CC-partitionfunction} and the eigenvalues \eqref{st}, the angular momentum and energy expectation values are determined by 
\eq{
    \langle\Delta\rangle\equiv\langle L_0\rangle=\frac 1 {2\pi i}\,\partial_\sigma \ln \Zcar \qquad\qquad \langle \xi\rangle\equiv\langle M_0\rangle=\frac 1 {2\pi i}\,\partial_\rho \ln \Zcar~.
}{eq:expectation_val}
Similarly, the variances are given by
\eq{
\langle\Delta^2\rangle-\langle\Delta\rangle^2 = \frac{1}{(2\pi i)^2}\,\partial_\sigma^2 \ln \Zcar\qquad\qquad \langle\xi^2\rangle-\langle\xi\rangle^2 = \frac 1 {(2\pi i)^2}\,\partial_\rho^2 \ln \Zcar
}{eq:variances}
and the covariance is given by
\eq{
\langle\Delta\xi\rangle-\langle\Delta\rangle\langle\xi\rangle = \frac 1 {(2\pi i)^2}\,\partial_\sigma\partial_\rho \ln \Zcar\,.
}{eq:covariance}

If the partition function \eqref{partition_function} --- possibly after some modular transformation \eqref{eq:angelinajolie} --- is dominated by a single primary state with multiplicity $D(\xi_0,\Delta_0)$ --- for instance, the vacuum state $D(0,0)$ --- then the logarithm of the partition function is approximately linear in the chemical potential $\rho$. This is so because the original partition function \eqref{partition_function} 
\eq{
Z_\mathfrak{ccar}(\rho, \sigma) = D(\xi_0,\Delta_0)\,e^{2\pi i\,\rho \big(\xi_0-\frac{c_{\mt M}}{24}\big)}\times g(\sigma) + \textrm{other\;primaries}\qquad\qquad g(\sigma) =\textrm{known}
}{eq:statedominance}
has an exponent linear in $\rho$, and modular transformations \eqref{eq:angelinajolie} preserve linearity in $\rho$. Thus, the quantity $\ln Z_\mathfrak{ccar}(\rho, \sigma)$ is approximately linear in $\rho$ whenever it is justified to neglect the contributions from the other primaries in \eqref{eq:statedominance}. This implies that the energy variance, the second equality \eqref{eq:variances}, is approximately zero. 

Therefore, if there is any modular frame where the Carrollian partition function \eqref{partition_function} is dominated by a single primary (which may or may not be the vacuum) the energy variance is approximately zero. This observation will play a key role in the next Section, in particular, in our derivation of the negativity of specific heat.

\subsection[Modular aspects of the Dedekind \texorpdfstring{$\eta$}{eta}-function]{Modular aspects of the Dedekind \texorpdfstring{$\boldsymbol{\eta}$}{eta}-function}\label{sec:3.4}

We list here key properties of the Dedekind $\eta$-function that we shall use in the remaining Sections of this paper. For more details and derivations of the formulas below see \cite{Dedekind}.
\begin{itemize}
\item \textbf{Definition.} Assuming $\s$ lies in the complex upper half-plane the Dedekind $\eta$-function is defined as
\eq{
\eta(\s)=e^{\frac{i\pi\s}{12}}\prod_{n=1}^\infty(1-e^{2\pi i n\s})\qquad\qquad\sigma\in\mathcal{H}\,.
}{eq:eta1}
\item \textbf{Cusp at infinity.} To investigate vacuum dominance, we are often interested in the limit where $\s$ has a very large (positive) imaginary part,
\eq{
\eta(\s)=e^{\frac{i\pi\s}{12}}\,\Big(1+\mathcal{O}\big(e^{2\pi i\s}\big)\Big)\qquad\textrm{if}\;\s\to i\infty\,.
}{eq:eta2}
\item $\boldsymbol{T}$\textbf{-transformation.} Under shifts by $1$ we get
\eq{
\eta(\s+1)=e^{\frac{i\pi}{12}}\,\eta(\s)\,.
}{eq:eta3}
\item $\boldsymbol{S}$\textbf{-transformation.} Inversion yields
\eq{
\eta(-1/\s)=\sqrt{-i\s}\,\eta(\s)\,.
}{eq:eta4}
\item \textbf{General modular transformation.} Combining arbitrary sequences of $T$- and $S$-transformations yields the behavior under general PSL$(2,\mathbb{Z})$ transformations
\eq{
\eta\bigg(\frac{a\s+b}{c\s+d}\bigg) = \tilde A(a,b,c,d)\,\sqrt{c\s+d}\,\eta(\s)
}{eq:eta5}
where $\tilde A(a,b,c,d)$ is a (known but for us irrelevant) phase, $ad-bc=1$ and $a,b,c,d\in\mathbb{Z}$.
\item \textbf{Approaching the origin.} For $\s=i\epsilon$ with $\epsilon\ll1$ after an $S$-transformation \eqref{eq:eta4} we use the cusp formula \eqref{eq:eta2} to obtain
\eq{
\eta(i\epsilon) \approx \epsilon^{-1/2}\,e^{-\frac{\pi}{12\epsilon}}\,.
}{eq:eta6}
\item \textbf{Approaching rational numbers.} 
For $\s=\frac{p+i\epsilon}{q}$ with coprime positive integers $p<q$ and $\epsilon\ll 1$ we apply a general modular transformation \eqref{eq:eta5} with $c=-q$ and $d=p$. For $b$ we choose the smallest positive integer so that $a=(1-bq)/p$ is also an integer (e.g., if $p=1$ or $p=2$ we choose $b=1$); the existence of such a $b$ is guaranteed by B\'ezout's identity. We get
\eq{
\eta\Big(\frac{p+i\epsilon}{q}\Big) \approx A(p,q)\,\epsilon^{-1/2}\,e^{-\frac{\pi}{12q\epsilon}}
}{eq:eta7}
where $A(p,q)$ is a (known but for us irrelevant) phase. Using $T$-transformations \eqref{eq:eta3} we see that the result \eqref{eq:eta7} holds for arbitrary rational numbers.
\item \textbf{Approaching irrational numbers.} The behavior for small imaginary part, $\epsilon\ll 1$, differs significantly from \eqref{eq:eta7} if an irrational number $\sigma\notin\mathbb{Q}$ is approached,
\eq{
\eta(\s+i\epsilon)\approx B(\s)\,\epsilon^{-1/4}
}{eq:eta8}
where $B(\s)$ is a complex $\mathcal{O}(1)$ coefficient.
\item \textbf{Logarithmic derivative.} The quantity $\eta^\prime/\eta$ can be expressed as
\eq{
\frac{1}{2\pi i}\,\partial_\sigma \ln\eta(\sigma) = \frac{1}{24}\,E_2(\sigma)
}{eq:eta9}
where $E_2$ is the second Eisenstein series defined as
\eq{
E_2(\sigma) = 1-24\,\sum_{n=1}^\infty \sigma_1(n)\,e^{2\pi i n\sigma}
}{eq:eta10}
with the divisor function $\sigma_1(n)=\sum_{d|n}d$, where $d|n$ means that $d$ is a divisor of $n$. We have the limiting behavior 
\eq{
E_2(\sigma) = 1 + \mathcal{O}\big(e^{2\pi i\sigma}\big)\qquad\textrm{if}\;\sigma\to +i \infty
}{eq:eta11}
and the quasi-modular property 
\eq{
E_2\Big(\frac{a\sigma+b}{c\sigma+d}\Big) = (c\sigma+d)^2\,E_2(\sigma) - \frac{6ic}{\pi}\,(c\sigma+d)\,.
}{eq:eta12}
For rational $\sigma=(p+i\epsilon)/q$, with coprime positive integers $p<q$, and $\epsilon\ll 1$, we get by the same arguments that led to \eqref{eq:eta7} the expansion
\eq{
E_2\Big(\frac{p+i\epsilon}{q}\Big) = -\frac{1}{\epsilon^2}\,\big(1+\mathcal{O}(\epsilon)\big)\,.
}{eq:eta13}
\end{itemize}

\subsection{Carrollian vacuum characters as limits from CFT vacuum characters}\label{sec:3.optional}

Starting from a CFT$_2$ with central charges $c,\bar c$ and assuming highest-weight representations in both chiral sectors, the total vacuum character $\chi^{\textrm{\tiny CFT}}_{c,0;\bar c, 0}(\tau;\bar\tau)$  is given by
\eq{
\chi_{c,0;\bar c, 0}^{\textrm{\tiny CFT}}(\tau;\bar\tau)=\chi_{c,0}^{\textrm{\tiny HW}}(\tau)\,\bar\chi_{\bar c,0}^{\textrm{\tiny HW}}(\bar\tau)
}{eq:lim1}
with the chiral vacuum characters
\eq{
\chi_{c,0}^{\textrm{\tiny HW}}(\tau)=\frac{1-e^{2\pi i\tau}}{\eta(\tau)}\,e^{-2\pi i\,\frac{c-1}{24}\,\tau}\qquad\qquad  \bar\chi_{\bar c,0}^{\textrm{\tiny HW}}(\bar\tau)=\frac{1-e^{-2\pi i\bar\tau}}{\eta(-\bar\tau)}\,e^{2\pi i\,\frac{\bar c-1}{24}\,\bar\tau}\,.
}{eq:lim2}
The expressions above are well-defined if both $\tau$ and $-\bar\tau$ lie in the complex upper half-plane. This happens, for instance, if $\tau$ lies in the complex upper half-plane and $\bar\tau$ is the complex conjugate of $\tau$ but it is not necessary that $\tau$ and $\bar\tau$ are complex conjugates of each other.

Inserting the Carroll expansions \eqref{eq:ccft4}, \eqref{mod} and dropping subleading terms in $\epsilon$, the vacuum character \eqref{eq:lim1} is given by
\eq{
\textrm{ill-defined:}\quad \chi_\mathds{1}^{\textrm{\tiny  ind}}(\rho,\sigma) = \frac{\big|1-e^{2\pi i\sigma}\big|^2}{\eta(\sigma)\,\eta(-\sigma)}\,e^{-2\pi i\,\frac{c_{\mt L}\sigma+c_{\mt M}\rho}{24}}\,.
}{eq:lim3}
This expression is not well-defined, because for generic $\sigma$ one of the $\eta$-functions has an argument in the wrong half-plane. The only exception is for real $\sigma$; however, then the $\eta$-function is still ill-defined unless we keep the $\epsilon$-contribution in the argument and assume $\rho$ lies in the complex upper half-plane. In this case, the CCFT vacuum character \eqref{eq:lim3},
\eq{
\textrm{induced:}\quad \chi_{\mathds 1}^{\textrm{\tiny  ind}}\big(\rho\in i\cdot\mathbb{R}^+,\,\sigma\in\mathbb{R}\big)  = \frac{\big|1-e^{2\pi i\sigma-2\pi\epsilon}\big|^2}{|\eta(\sigma+i\epsilon)|^2}\,e^{-2\pi i\,\frac{c_{\mt L}\sigma+c_{\mt M}\rho}{24}}
}{eq:lim4}
coincides with the induced CCFT character \cite{Oblak:2015sea}. 

So, the Carroll limit of the usual CFT vacuum character leads to the induced CCFT vacuum character. To obtain the highest-weight CCFT vacuum character we therefore have to commence from a different starting point, namely a CFT$_2$ with flipped representations, i.e., highest-weight in one chiral sector and lowest-weight in the other chiral sector. Details of lowest-weight representations are collected in Appendix \ref{app:B}.

Starting from a CFT$_2$ with central charges $c,\bar c$, the flipped vacuum character $\tilde\chi^{\textrm{\tiny CFT}}_{c,0;\bar c, 0}(\tau;\bar\tau)$ is given by
\eq{
\tilde\chi^{\textrm{\tiny CFT}}_{c,0;\bar c, 0}(\tau;\bar\tau)=\chi_{c,0}^{\textrm{\tiny HW}}(\tau)\,\bar\chi_{\bar c,0}^{\textrm{\tiny LW}}(\bar\tau)\,.
}{eq:lim5}
Mutatis mutandis, the Carroll limit analogous to above yields the CCFT vacuum character
\eq{
\textrm{highest-weight:}\quad \chi_{\mathds 1}(\rho,\sigma\in\mathcal{H}) = \frac{\big(1-e^{2\pi i\sigma}\big)^2}{\eta(\sigma)^2}\,e^{-2\pi i\,\frac{(c_{\mt L}-2)\sigma+c_{\mt M}\rho}{24}}
}{eq:lim6}
which is identical to \eqref{eq:charactervacuum}. This vacuum character is well-defined for $\sigma\in\mathcal{H}$. Similar results hold for massive characters.

In conclusion, we deduce that standard CFT characters yield induced CCFT characters with the restriction of real $\sigma$ and (positively) imaginary $\rho$, whereas flipped CFT characters yield highest-weight CCFT characters with the restriction of $\sigma$ to the complex upper half-plane. This conclusion will be important in the following Sections, where we investigate different regions of the complex $\sigma$- and $\rho$-planes to uncover sectors where the Carrollian partition function is dominated by one of the vacuum characters, \eqref{eq:lim4} or \eqref{eq:lim6}, in the $S$-dual channel. 


\section{Carrollian vacuum dominance in the dual channel}
\label{sec:4}

In this Section, we present six sectors where the Carroll partition function is dominated by the vacuum character in the ($S$-) dual channel. In this and the following Sections, we will deal mostly with CCFTs. Therefore, we drop the subscripts $\mathfrak{car}$ and $\mathfrak{ccar}$ to lighten the notation. All the variables without any subscript refer to the variables of the Carrollian theory.

Modular invariance of 2d Carrollian partition functions guarantees that under $S$-trans\-for\-ma\-tions \eqref{eq:ccft7} we get the partition function in the dual channel, e.g., for highest-weight representations
\begin{equation} \label{eq:Zmod}
        \!Z(\rho, \sigma)=Z \Big(\frac{\rho}{\sigma^2} ,-\frac{1}{\sigma}\Big) 
 = \frac{e^{2\pi i(\frac{c_\mt{L}-2}{24\sigma}-\rho \frac{c_\mt{M}}{24\sigma^2})}}{\eta(-1/\sigma)^2} \bigg((1-e^{-\frac{2\pi i}{\sigma}})^2+\!\sum_{\xi, \Delta \neq 0}\!  D(\xi,\Delta)\, e^{2\pi i\big(-\frac{\Delta}{\sigma}+\frac{\rho}{\sigma^2}\xi\big)}\bigg) \,. 
\end{equation}
It is sometimes convenient to write the $(\rho, \sigma)$-dependence in terms of the physical chemical potentials $\beta$ and $\Omega$. Using \eqref{eq:12} and \eqref{eq:ocar},
\begin{equation}
    \sigma= i\,\frac{\beta\Omega}{2\pi}\qquad \qquad \rho = i\,\frac{\beta}{2\pi}
\label{eq:dict}
\end{equation}
it follows that \textit{vacuum dominance in the dual channel} requires as a necessary condition 
\begin{equation}
\boxed{\phantom{\Bigg(}
\textrm{vacuum\,dominance, highest-weight:}\quad    
D(\xi,\Delta)\,\frac{\exp\left[-\frac{4\pi^2}{\beta\Omega}\Delta + \frac{4\pi^2}{\beta\Omega^2}\xi\right]}{\left(1-\exp\left[-\frac{4\pi^2}{\beta\Omega}\right]\right)^2} \to 0
    \phantom{\Bigg)}} 
\label{eq:vac-dom}
\end{equation}
for all non-vacuum primaries in the spectrum with CCFT weights given by $\Delta$ and $\xi$. The condition above holds for highest-weight representations. For induced representations, the denominator is replaced by the square of an absolute value \cite{Oblak:2015sea}.

When vacuum dominance \eqref{eq:vac-dom} holds, the partition function \eqref{eq:Zmod} is approximated by
\begin{equation}
  Z (\rho, \sigma) \approx \,\chi^{\mt{L}}_{\mathds 1}(-1/\sigma)\,\chi^{\mt M}_{\mathds 1}(\rho/\sigma^2)\,=e^{2\pi i(\frac{c_\mt{L}-2}{24\sigma}-\rho \frac{c_\mt{M}}{24\sigma^2})}\frac{ (1-e^{-\frac{2\pi i}{\sigma}})^2}{{\eta(-1/\sigma)^2}}\,.
\label{eq:tildeZ-mid}
\end{equation}
In this approximation, the energy expectation value
\begin{equation}
   \langle \xi \rangle \approx -\frac{c_{\mt{M}}}{24}\,\frac{1}{\sigma^2}= \frac{c_{\mt{M}}}{24} \Big(\frac{2\pi}{\beta\Omega}\Big)^2 
\label{eq:energy-ave}
\end{equation}
is entirely fixed by $\beta\Omega$. By contrast, the angular momentum expectation value  
\begin{align}
     \langle \Delta \rangle &\approx -\frac{c_{\mt{L}}-2}{24\sigma^2} + c_{\mt{M}}\,\frac{\rho}{12\sigma^3}-\frac{2}{\sigma^2}\frac{1}{e^{2\pi i/\sigma}-1} -\frac{1}{12\sigma^2}\,E_2(-1/\sigma)
\label{eq:momentum-ave}     \\
  &= \frac{\pi^2}{6}\,\frac{c_{\mt{L}}-2}{(\beta\Omega)^2}   + \frac{\pi^2}{3} \frac{c_{\mt{M}}\,|\beta|}{(\beta\Omega)^3}+\frac{8\pi^2}{(\beta\Omega)^2}\frac{1}{e^{4\pi^2/(\beta\Omega)}-1}+\frac{\pi^2}{3\beta^2\Omega^2}\,E_2\bigg(\frac{2\pi i}{\beta\Omega}\bigg) \nonumber 
\end{align}
depends both on $\beta\Omega$ and $\beta\Omega^2$. The second Eisenstein series $E_2$ is defined in \eqref{eq:eta9}. 

\subsection{Six sectors with vacuum dominance in the dual channel} \label{sec:4.sectors}

We search for all solutions of vacuum dominance \eqref{eq:vac-dom} assuming that the spectra can be unbounded, i.e., there are not necessarily upper and lower bounds on $\xi$ and $\Delta$ (though, as we shall see, we always need to require semi-boundedness, e.g., $\xi>0$). Besides the trivial option, $D(\xi,\,\Delta)\to 0$, there are several possibilities to achieve vacuum dominance \eqref{eq:vac-dom}, depending on whether $\beta\Omega$ or $\beta\Omega^2$ tend to zero from either above or from below. 

\paragraph{\underline{$\beta\Omega \to 0^\pm$:}} For the first set of possibilities, it is convenient to define 
\eq{
x=\exp[-4\pi^2/(\beta\Omega)]\in(0,\infty) 
}{eq:bellyache} 
in terms of which \eqref{eq:vac-dom} reads
\eq{
D(\xi,\,\Delta)\,\frac{x^{\Delta-\frac{\xi}{\Omega}}}{(x-1)^2} \to 0 
}{eq:vd1}
which is achieved for either $x\to\infty$ and $\Delta<2+\xi/\Omega$ or $x\to 0^+$ and $\Delta>\xi/\Omega$. Notice this is related to the expectation value of the energy \eqref{eq:energy-ave} by
\begin{equation}
    \langle \xi \rangle \approx \frac{c_{\mt{M}}}{24}\,\frac{\ln^2x}{4\pi^2}\,.
\end{equation}

\paragraph{\underline{$\beta\Omega^2 \to 0^\pm$:}} For the remaining possibilities, we define instead 
\eq{
y=\exp[4\pi^2/(\beta\Omega^2)]\in(0,\infty) 
}{eq:iamnotyourfriend} 
in terms of which \eqref{eq:vac-dom} reads
\eq{
D(\xi,\,\Delta)\,\frac{y^{\xi+(2-\Delta)\Omega}}{(y^\Omega-1)^2} \to 0 
}{eq:vd2}
which is achieved for either $y\to\infty$ and $\Delta>\xi/\Omega$ (for negative $\Omega$ we have instead $\Delta<2+\xi/\Omega$) or $y\to0^+$ and $\Delta<2+\xi/\Omega$ (for negative $\Omega$ we have instead $\Delta>\xi/\Omega$). Both sets of possibilities overlap with each other; e.g., for finite $\beta$, the condition $x\to\infty$ implies $y\to 0^+$ and vice versa.

Table \ref{tab:1} summarizes all possibilities to achieve vacuum dominance in the dual channel. As stated above, we allow for arbitrary spectra, i.e., values of $\xi$ and $\Delta$, subject to some convexity constraints following from vacuum dominance. In each case, there is a convexity condition on $\xi$, and we always assume $\xi>0$; otherwise, the Table would be doubled with an equal number of $\xi<0$ cases and corresponding changes in the various columns. 


\begin{table}[htb]
\begin{center}
\begin{tabular}{|c|>{$}c<{$}|>{$}c<{$}|>{$}c<{$}|>{$}c<{$}|>{$}c<{$}|>{$}c<{$}|>{$}c<{$}|>{$}c<{$}|}\hline
{\bf{Sector}} & \beta & \Omega & \theta & \beta\Omega^2 & \Delta & \xi & x & y \\\hline
Cardy & 0^- & \textcolor{lightgray}{>0\;\textrm{or}}<0 & i\cdot 0^{\textcolor{lightgray}{\mp}} & \leq 0 & \textcolor{lightgray}{\leq 2\;\textrm{or}}\geq 0 & >0 & \textcolor{lightgray}{\infty\;\textrm{or}}\;0 & \geq 0 \\
hard & <0 & \textcolor{lightgray}{ 0^+\;\textrm{or}}\;0^- & i\cdot 0^{\textcolor{lightgray}{\mp}} & 0^- & \textcolor{lightgray}{\leq 2\;\textrm{or}}\geq 0 & >0 & \textcolor{lightgray}{\infty\;\textrm{or}}\;0 & 0 \\ \hline
cold & -\infty & \textcolor{lightgray}{ 0^+\;\textrm{or}}\;0^- & \in i\,\mathbb{R} & 0^- & \textcolor{lightgray}{\leq 2\;\textrm{or}}\geq 0 & >0 & \textcolor{lightgray}{>1\;\textrm{or}} <1 & 0 \\
\textbf{Schwarzian} & \boldsymbol{-\infty} & \textcolor{lightgray}{ 0^+\;\textrm{or}}\;\boldsymbol{0^-} & \textcolor{lightgray}{\pm}\boldsymbol{i\cdot\infty} & 0^- & \textcolor{lightgray}{\leq 2\;\textrm{or}}\boldsymbol{\geq 0} & \boldsymbol{>0} & \boldsymbol{1} & \boldsymbol{0} \\ \hline
Boltzmann & >0 &  i\cdot0^\pm & \in\mathbb{R} & 0^- & \textrm{arbitrary} & >0 & \textrm{complex} & 0 \\
hot &  0^+ & \in i\,\mathbb{R} & 0 & 0^- &\textrm{arbitrary} & >0 & \textrm{complex} & 0 \\
\hline
\end{tabular}
\caption[Six sectors of vacuum dominance]{Six sectors of vacuum dominance grouped in three similar pairs}
\label{tab:1}
\end{center}
\end{table}

The six sectors in Table \ref{tab:1} are characterized by the behavior of temperature ($1/\beta$) and energy (essentially $\ln^2 x$, see below). The first four sectors all require a negative temperature. We labeled the high temperature sector ($\beta\to 0^-$) as ``Cardy'', which also features infinite energy. The ``hard'' sector has arbitrary (negative) temperature and infinite energy. The ``cold'' sector has vanishing temperature and finite energy. The ``Schwarzian'' sector is like the cold one but has additionally vanishing energy; we shall also refer to this sector as ``soft'' or ``near-extremal'', and explain in Subsection \ref{sec:Schw} below why we chose the label ``Schwarzian''. Finally, the fifth and sixth sectors are the only ones with positive temperature (hence the label ``Boltzmann'') and no spectral condition on $\Delta$; the only difference between them is that temperature remains finite in the fifth sector and becomes infinite in the sixth sector (the ``hot'' one). In the first four sectors, we will hereafter always assume the spectral constraints $\xi>0$ and $\Delta>0$, but it is understood that our discussion below generalizes to other choices of convexity conditions on $\xi$ and $\Delta$. We have grouped the six sectors in three pairs where members of a pair have similar behavior of $x$ and hence similar behavior of energy.

The conditions on $\theta=2\pi\,\sigma$ in Table \ref{tab:1} imply that not all sectors exist for partition functions \eqref{eq:lamourdemavie} built from induced representations. We summarize the allowed sectors depending on the use of induced vs.~highest-weight representations in Table \ref{tab:3}. Check (cross) marks mean that the corresponding sector exists (does not exist) for the partition function built from the respective representation. Entries with both a cross and a check mark mean that the corresponding sector has some issues, the discussion of which we postpone to Subsection \ref{sec:positive}. In addition, we list in that Table if the partition function $Z$ is real. A check mark means it is indeed real (as long as the central charges are real), a check mark with a condition on the central charges means it is real if that condition holds, a cross mark means it is not real in general (and cannot be made real by any choice of the central charges), and ``n.a.'' stands for ``not applicable'', i.e., the partition function does not exist in the corresponding sector.


\begin{table}[htb]
\centering
\begin{tabular}{||l||c|c||c|c||}\hline\hline
& induced & real $Z$ & highest-weight & real $Z$ \\\hline
Cardy & $\xmark$ & n.a. & $\cmark$ & $\cmark$ \\
hard & $\xmark$ & n.a. & $\cmark$ & $\cmark$ \\
cold & $\xmark$ & n.a. & $\cmark$ & $\cmark$ \\
Schwarzian & $\xmark$ & n.a. & $\cmark$ & $\cmark$ \\
Boltzmann & $\xmark\cmark$ & $\cmark$ if $c_{\mt L}=0$ & $\xmark\cmark$ & $\xmark$ \\
hot & $\xmark\cmark$ & $\cmark$ if $c_{\mt L}=0$ & $\xmark\cmark$ & $\xmark$  \\\hline\hline
\end{tabular}
\caption{CCFT sectors vs.~CCFT representations}
    \label{tab:3}
\end{table}

Let us introduce some notation before moving forward. Since in some contexts (e.g., holographic applications)  $c_{\mt{M}}$ has a dimension, it is natural to work with dimensionless chemical potentials by measuring them in $c_{\mt{M}}$ units as
\begin{equation}
  \beta = \frac{\tilde{\beta} }{c_{\mt{M}}} \qquad \qquad \Omega = c_{\mt{M}}\,\tilde{\Omega} 
\end{equation}
making sure that not only $x$ is dimensionless but also $y$. For convenience, we also define the dimensionless energy expectation value
\eq{
\tilde\xi=\frac{24\langle\xi\rangle}{c_{\mt{M}}} = \frac{4\pi^2}{\beta^2\Omega^2}=\frac{\ln^2x}{4\pi^2}\,.
}{eq:whenthepartyisover}

In the rest of this Section, we study the specific expressions for the CCFT$_2$ partition functions, their charges, and thermodynamic entropies in the six sectors listed in Table \ref{tab:1}. 

\subsection{Cardy and hard sector}
\label{sec:cardy}

For these sectors, we assume a partition function \eqref{partition_function} built from highest-weight representations (see Table \ref{tab:3}). Both in the Cardy and the hard sector, the energy expectation value \eqref{eq:energy-ave} tends to infinity, so these sectors both correspond to high energies. The Cardy sector additionally has a large (negative) temperature. Since our results below do not depend essentially on whether the (absolute value of) temperature is finite or large, we discuss both sectors jointly. 

First, let us approximate the partition function by imposing the vacuum dominance conditions from the first two rows of Table \ref{tab:1} and inserting them into the partition function \eqref{eq:tildeZ-mid}
\eq{
Z_{\textrm{\tiny Cardy}}(\rho, \sigma)=\chi^{\mt L}_{\mathds 1}\Big(-\frac{1}{\sigma}\Big)\,\chi^{\mt M}_{\mathds 1}\Big(\frac{\rho}{\sigma^2}\Big)\Big(1 + \mathcal{O}\Big(e^{-\frac{-\xi_{\rm gap}}{|\beta\Omega^2|}}\Big)\Big) \approx \exp\Big[\frac{\pi^2}{6}\,\Big(\frac{c_{\mt{L}}}{|\beta\Omega|}+ \frac{c_{\mt{M}}}{|\beta\Omega^2|}\Big)\Big]
}{eq:Z1}
where we used \eqref{eq:eta2} and $\xi_{\rm gap}$ is the weight of the primary with the lowest eigenvalue of $M_0$ other than $0$.

From the partition function \eqref{eq:Z1} we compute the thermodynamic entropy 
\begin{equation}
    S_{\textrm{\tiny Cardy}} =\left(1-\rho\partial_\rho-\sigma\partial_\sigma \right)\ln Z_{\textrm{\tiny Cardy}}=\left(1-\beta\partial_\beta \right)\ln Z_{\textrm{\tiny Cardy}}(\beta,\Omega) =\frac{\pi^2}{3}\,\left(\frac{c_{\mt L}}{|\beta\Omega|}+ \frac{c_{\mt M}}{|
\beta\Omega^2|}\right) 
\label{eq:Cardy-ent}
\end{equation}
which essentially coincides with the Cardy-like formula derived in \cite{Bagchi:2012xr}; we shall make this relation more precise below. We stress that there are no log corrections to the entropy \eqref{eq:Cardy-ent} in the grand canonical ensemble for the Cardy sector, as expected on general grounds from the gravity side, see \cite{Carlip:1994gc,Carlip:2000nv,Sen:2012dw} for the case of BTZ.
 
From \eqref{eq:Z1}, we find that the angular momentum is the sum of two terms 
\begin{equation}
  \tilde{\Delta} \equiv \langle \Delta \rangle \approx \frac{\pi^2\,c_{\mt{L}}}{6(\beta\Omega)^2} + \frac{\pi^2\,c_{\mt{M}}}{3|\beta^2\Omega^3|} \equiv \Delta_{\mt{L}} + \Delta_{\mt{M}}
\end{equation}
with ratio
\begin{equation}
  \frac{\Delta_{\mt{L}}}{\Delta_{\mt{M}}} = \frac{c_{\mt{L}}}{2}\,|\tilde{\Omega}|
\end{equation}
controlled by the dimensionless chemical potential $|\tilde{\Omega}|$.

We assume finite $c_{\mt{L}}$, implying $\Delta_{\mt{L}} \sim \tilde{\xi} \sim (\beta\Omega)^{-2}$. This allows us to organize the discussion in terms of the behavior of $|\tilde{\Omega}|$:
\begin{itemize}
\item \textbf{Finite} $|\tilde{\Omega} |$. This is the Cardy sector where the temperature $|\tilde{\beta} ^{-1}|$ is large, and all relevant charges scale in the same way $\tilde{\xi} \sim \Delta_\mt{L} \sim \Delta_\mt{M}$ and $\xi \sim \Delta$. This leads to an entropy
\begin{equation}
  S_{\textrm{\tiny Cardy}}  \approx 2\pi\,\Bigg(\frac{c_{\mt{L}}}{24}\,\sqrt{\tilde{\xi}} + \frac{\tilde{\Delta}}{\sqrt{\tilde{\xi}}} \Bigg) \qquad \qquad \text{finite}\,\,|\tilde{\Omega} |
\label{eq:cardy-finite}
\end{equation}
where the $c_{\mt M}$ dependence is implicit and can be read off from \eqref{eq:whenthepartyisover}. The above formula is known as the BMS--Cardy formula \cite{Bagchi:2012xr}.
\item \textbf{Large} $|\tilde{\Omega}|$. The temperature remains large, meaning that we stay in the Cardy sector. The physical charges
\begin{equation}
  \tilde{\Delta} \approx \Delta_\mt{L} = \frac{c_{\mt{L}}}{24}\,\tilde{\xi} \gg \Delta_{\mt{M}}\,.
\end{equation}
reduce the dominant contribution to the entropy to
\begin{equation}
  S_{\textrm{\tiny Cardy}}  \approx \frac{\pi}{6}\, c_{\mt{L}}\,\sqrt{\tilde{\xi}} \qquad \qquad |\tilde{\Omega} | \gg 1. 
\label{eq:cardy-large}
\end{equation}
\item \textbf{Small} $|\tilde{\Omega}|$. The temperature can be large or finite. Thus, we could either be in the Cardy or the hard sector. This case is characterized by an angular momentum $\tilde{\Delta} \approx \Delta_\mt{M} \gg \tilde{\xi}$ larger than the energy, leading to 
\begin{equation}
  S_{\textrm{\tiny Cardy}} \approx 2\pi\,\frac{\tilde{\Delta}}{\sqrt{\tilde{\xi}}} \qquad \qquad   |\tilde{\Omega} | \ll 1.
\label{eq:cardy-small}
\end{equation}
\end{itemize}

\subsection{Cold and Schwarzian sector}
\label{sec:Schw}

Also for these sectors, we assume a partition function \eqref{partition_function} built from highest-weight representations (see Table \ref{tab:3}). The cold and Schwarzian sectors from the middle part of Table \ref{tab:1} have a temperature that vanishes from below. The cold sector has finite energy, whereas the Schwarzian sector has vanishing energy. 

In the cold sector, there is no further simplification of the partition function \eqref{eq:tildeZ-mid} besides vacuum dominance,
\begin{equation} \label{eq:Z2}
       Z_{\textrm{\tiny cold}}(\beta, \Omega) \approx\exp\bigg[\frac{\pi^2}{6}\bigg(\frac{c_{\mt{L}}-2}{|\beta\Omega|}+ \frac{c_{\mt{M}}}{|\beta\Omega^2|}\bigg)\bigg]\,\Bigg(\frac{1-e^{-\frac{4\pi^2}{|\beta\Omega| }}}{\eta(\frac{2\pi i}{\beta\Omega})}\Bigg)^2 \,.
\end{equation}
This gives rise to an entropy
\begin{equation}
    S_{\textrm{\tiny cold}} =\left(1-\beta\,\partial_\beta\right)\ln Z_{\textrm{\tiny cold}} \approx 4\pi^2\,\bigg(\frac{c_{\mt L}-2}{12|\beta\Omega|}+ \frac{c_{\mt M}}{12|
\beta\Omega^2|}\bigg)+f\Big(\frac{i\beta\Omega}{2\pi}\Big)
\label{eq:Schw-ent}
\end{equation}
where 
\begin{multline}
    f\Big(\frac{i\beta\Omega}{2\pi}\Big)=f(\sigma)=2(1-\sigma\partial_\sigma)\ln\frac{1-e^{-\frac{2\pi i}{\sigma}}}{\eta(-\frac{1}{\sigma})}= 2\ln(1-e^{-4\pi^2/(\beta\Omega)}) 
    - \ln \frac{\beta\Omega}{2\pi} \\ - 2\ln \eta\Big(i\frac{\beta\Omega}{2\pi}\Big) + 1 
     + \frac{8\pi^2}{\beta\Omega}\,\frac{e^{-4\pi^2/(\beta\Omega)}}{1-e^{-4\pi^2/(\beta\Omega)}} -\frac{\beta\Omega}{12}\,E_2\Big(i\frac{\beta\Omega}{2\pi}\Big)\,.
\label{eq:f-sigma}
\end{multline}  
The finite energy $\xi$ is given by
\begin{equation}
    \langle \xi \rangle \equiv E_0 = \frac{c_{\mt{M}}}{24} \left(\frac{2\pi}{\beta\Omega}\right)^2\,.
\end{equation}
Both angular momentum and entropy split into a large and dominant contribution with finite subleading contributions that can be written as a function of $E_0$. Concretely, the angular momentum 
\begin{equation}
    \langle \Delta \rangle  = J_0 + \Delta_{\mt{finite}}
\end{equation}
splits into a dominant contribution
\begin{equation}
   J_0 = \frac{\pi^2 c_{\mt{M}}\,|\beta|}{3(\beta\Omega)^3}= \frac{\pi^2}{3}\,\frac{1}{\tilde{\beta} \tilde{\Omega} }\frac{1}{|\tilde{\beta} \tilde{\Omega}^2 |}
\end{equation}
and subleading finite contributions
\eq{
    \Delta_{\mt{finite}} = \frac{\pi^2}{6}\,\frac{c_{\mt{L}}-2}{(\beta\Omega)^2}  + \frac{8\pi^2}{(\beta\Omega)^2}\frac{1}{e^{4\pi^2/(\beta\Omega)}-1} + \frac{\pi^2}{3\beta^2\Omega^2}\,E_2\Big(\frac{2\pi i}{\beta \Omega}\Big)\,.
}{eq:whynolabel}

Similarly, the entropy has an analogous decomposition
\begin{equation}
  S_{\textrm{\tiny cold}}  \approx 2\pi\,\sqrt{\frac{c_{\mt{M}}}{24E_0}}\,J_0 + S_{\mt{finite}} \qquad\qquad
  S_{\mt{finite}} = \frac{\pi^2}{3}\, \frac{c_{\mt{L}}-2}{|\beta\Omega|} +f\Big(\frac{i\beta\Omega}{2\pi}\Big)
\label{eq:Sfinite}
\end{equation}
where the function $f$ is defined in \eqref{eq:f-sigma}.

A simple cross-check of our results is that the first law of thermodynamics holds,
\begin{equation}
\extd E_0 = T\,\extd S_{\textrm{\tiny cold}} - \Omega\,\extd J_0\,.
\end{equation}

If we additionally assume softness, i.e., near-extremality, $\beta\Omega\to\infty$, then the results above, while still valid, simplify quite a bit because the function \eqref{eq:f-sigma} reduces to
\eq{
f\Big(\frac{i\beta\Omega}{2\pi}\Big)\Big|_{\beta\Omega\gg 1} \approx -3\ln|\beta\Omega|+5\ln(2\pi)+3\,.
}{eq:whatwasimadefor}
To derive this result we used \eqref{eq:eta2}. This limit requires 
\begin{equation}
\label{eq:SchwarzianScaling}
-\tilde \beta\gg 1\qquad\qquad \tilde \Omega \sim -|\tilde \beta| ^{-\delta}\qquad\qquad  \frac 1 2 <\delta<1 ~ . 
\end{equation}
In this limit, we label the partition function \eqref{eq:Z2} as ``Schwarzian'', $Z_{\textrm{\tiny Schwarzian}}(\beta, \Omega)=Z_{\textrm{\tiny cold}}(\beta, \Omega)|_{\beta\Omega\gg 1}$, because the partition function
\eq{
\boxed{\phantom{\Bigg(}
 Z_{\textrm{\tiny Schwarzian}}(\beta, \Omega) \approx
 \frac{(2\pi)^5}{(\beta\Omega)^3}\, \exp\bigg[\frac{\beta\Omega}{12}+\frac{\pi^2}{6}\bigg(\frac{c_{\mt{L}}-2}{\beta\Omega} + \frac{c_{\mt{M}}}{|\beta\Omega^2|}\bigg)\bigg]
\phantom{\Bigg)}}
}{eq:badguy}
acquires a Schwarzian-like form. The dominant contributions to the entropy
\eq{
S_{\textrm{\tiny Schwarzian}} =\big(1-\beta\,\partial_\beta\big)\,\ln Z_{\textrm{\tiny Schwarzian}} \approx \frac{\pi^2\,c_{\mt{M}}}{3|\beta\Omega^2|}-3\,\ln|\beta\Omega| + \mathcal{O}(1)
}{eq:birdsofafeather}
consist of a leading divergent term (since $\beta\Omega^2\to 0^-$) and a logarithmic correction (since $\beta\Omega\to\infty$). If they are related as $\beta\Omega^2=-\alpha/(\beta\Omega)$ with some positive constant $\alpha$, then the entropy\footnote{%
This scaling corresponds to $\delta=\frac 2 3$ in \eqref{eq:SchwarzianScaling}.
} \eqref{eq:birdsofafeather} 
\eq{
S_{\textrm{\tiny Schwarzian}} =S_0 -3\,\ln S_0 + \mathcal{O}(1)\qquad\qquad S_0 = \frac{\pi^2\,c_{\mt{M}}}{3|\beta\Omega^2|}
}{eq:everythingiwanted}
shows gravity-like behavior, in the sense that there is an infinite LO piece (the ``Bekenstein--Hawking'' term) and a subleading logarithmic piece with a precise numerical prefactor (divided by two) that accounts for the number of generators (viz., six) in the wedge subalgebra of the CCFT$_2$ algebra \eqref{bms}. 

The partition function \eqref{eq:badguy} in the Schwarzian sector factorizes analogously to \cite{Ghosh:2019rcj},
\eq{
Z_{\textrm{\tiny Schwarzian}}(\beta,\Omega) \approx Z_0(\beta,\Omega) \times Z^{\textrm{\tiny Carroll}}_{\textrm{\tiny Schwarzian}}(\beta\Omega)\qquad\qquad Z_0(\beta,\Omega)=e^{S_0-\beta\Omega\,{\left(J_0-\frac1{12}\right)}-\beta E_0}
}{eq:trueschwarzian}
where $S_0$ is the LO entropy (see \eqref{eq:s0} below) and the fluctuations above the LO contribution are governed by the Carroll--Schwarzian partition function
\eq{
\boxed{\phantom{\Bigg(}
Z^{\textrm{\tiny Carroll}}_{\textrm{\tiny Schwarzian}}(\beta\Omega) = \frac{(2\pi)^5}{(\beta\Omega)^3}\, \exp\bigg[\frac{\pi^2}{6}\,\frac{c_{\mt{L}}-2}{\beta\Omega}\bigg]
\phantom{\Bigg)}}}{eq:carrollschwarzian}
which depends on temperature solely through the combination $\beta\Omega$. The result \eqref{eq:carrollschwarzian} shows a striking resemblance to the Schwarzian partition function (1.4) in \cite{Stanford:2017thb} and justifies post hoc our label ``Schwarzian'' for the soft sector. 

The Schwarzian sector describes small energy fluctuations $\langle \xi \rangle\to 0$, as can be understood from \eqref{eq:energy-ave}. The departure from zero in the energy can be interpreted as near-extremal quadratic corrections in the relevant chemical potential in the dual channel [see \eqref{eq:pot-transf}]\footnote{%
When expressed as a function of $\beta$ and $\theta$, there is no $\beta$-dependence in $\langle\xi\rangle$, defying the Schwarzian expectation that energy is quadratic in temperature, see, e.g., \cite{Almheiri:2016fws}. The reason for this is that we are not working with a fixed charge ensemble. Similar behavior was found in CFTs and WCFTs in \cite{Ghosh:2019rcj, Aggarwal:2022xfd, Aggarwal:2023peg}.}
\begin{equation}
  \langle \xi \rangle  \approx \frac{c_{\mt{M}}}{24} \left(\frac{2\pi}{\beta\Omega}\right)^2 \equiv M_{\mt{gap}}^{-1}\,\left(\frac{2\pi}{\beta\Omega}\right)^2 = \frac{4\pi^2\,M_{\mt{gap}}^{-1}}{-\theta^2}
\label{eq:xi-gap}
\end{equation}
identifying the mass gap scale \cite{Preskill:1991tb, Maldacena:1998uz,Page:2000dk} with
\eq{
M_{\mt{gap}}^{-1}=\frac{c_{\mt{M}}}{24}\,.
}{eq:displayedeq}
However, the angular momentum can have different behavior. Indeed, from \eqref{eq:momentum-ave} we get
\begin{equation}
  \langle \Delta \rangle +\frac 1 {12}\approx \frac{\pi^2}{6}\,\frac{c_{\mt{L}}-2}{(\tilde{\beta} \tilde{\Omega})^2} + 
  \frac{3}{\tilde{\beta} \tilde{\Omega} } + \frac{\pi^2}{3(\tilde{\beta} \tilde{\Omega} )|\tilde{\beta} \tilde{\Omega}^2 |}\,.
\label{eq:momentum-ex}
\end{equation}
For CCFTs with $c_{\mt{L}}=2$, we have $\langle \Delta \rangle + \frac{1}{12}= J_0 + \frac{3}{\beta\Omega}$. The angular momentum $J_0$ is large for $\frac23<\delta<1$ in \eqref{eq:SchwarzianScaling} and small for $\frac12<\delta<\frac23$. For the preferred value $\delta=\frac23$, $J_0$ is of order unity, though in principle, it can still be a small or large quantity. The entropy can again be split into two pieces. 
\eq{
    S  = S_0 + \delta S\qquad\qquad S_0 = \big(1-\beta\partial_\beta\big)\,\ln Z_0=2\pi\,\sqrt{\frac{c_{\mt{M}}}{24E_0}}\,J_0 = \beta\Omega\,J_0
}{eq:s0}
The first term is always large in the Schwarzian sector, no matter what the behavior of $J_0$ (as long as it does not vanish). The difference with the finite energy case is that entropy corrections $\delta S$ can be large. Concretely, the large $\beta\Omega$ behavior of \eqref{eq:Sfinite} equals
\eq{
    \delta S \approx - 3\ln (\beta\Omega)+ 3 + 5\ln 2\pi + \frac{\pi^2}{3}\,\frac{c_{\mt L}-2}{|\beta\Omega|} + \dots
}{eq:lovely}

\subsection{Boltzmann and hot sector}\label{sec:positive}

In the Boltzmann sector (and its limiting case, the hot sector), we cover both highest-weight and induced representations. Initially, we focus on the induced representations and comment on the highest-weight case in the end. We also assume $c_{\mt L}=0$ to get a real partition function. 

In these sectors, the parameter $\sigma$ is small and real, and hence the Dedekind $\eta$-function in the partition function \eqref{eq:lamourdemavie} requires the $i\epsilon$ prescription therein. The small $\epsilon$ limit depends on whether $\sigma$ is rational or irrational, see \eqref{eq:eta7} and \eqref{eq:eta8}. Note that the limiting formula \eqref{eq:eta7} has $\epsilon$ rescaled by a $\sigma$-dependent factor as compared to the expression in the partition function \eqref{eq:lamourdemavie}, which is necessary to get a cutoff-independent entropy.

\begin{figure}[htb]
\begin{center}
\includegraphics[width=0.45\linewidth]{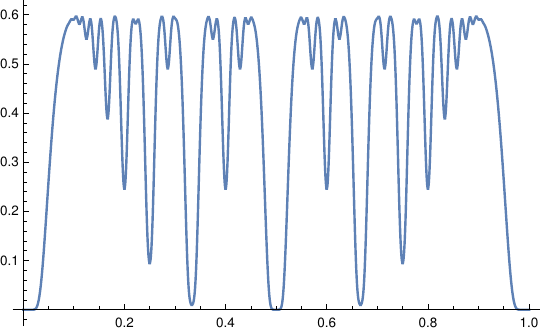}\qquad\includegraphics[width=0.45\linewidth]{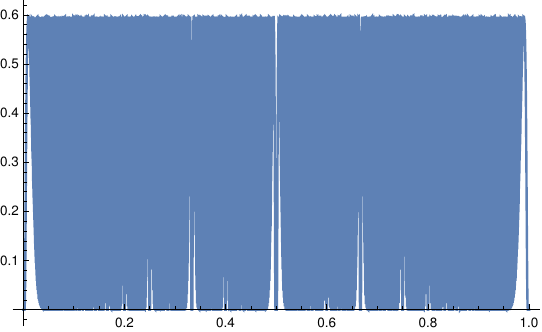}
\end{center}
\caption[Plots of $\sqrt{\epsilon}\,|\eta(x+i\epsilon)|^2$]{Plots of $\sqrt{\epsilon}\,|\eta(x+i\epsilon)|^2$ with $x\in[0,1]$. Left: $\epsilon=10^{-2}$. Right: $\epsilon=10^{-4}$.}
    \label{fig:Z}
\end{figure}

Applying this limit to the partition function \eqref{eq:tildeZ-mid} yields (we defined $\hat\Omega=i\Omega\in\mathbb{R}$)
\eq{
Z_{\textrm{\tiny Boltzmann}}(\beta,\hat\Omega)\big|_{\epsilon\ll 1} =\exp\bigg[\frac{\pi^2\,c_{\mt{M}}}{6\beta\hat\Omega^2}\bigg]\times Z_\epsilon\,.
}{eq:youshouldseemeinacrown}
with $Z_\epsilon=\mathcal{O}(1)\,\epsilon^3\,\exp\big[\tfrac{\pi}{6q\epsilon}\big]$ for rational $\sigma$ and $Z_\epsilon=\mathcal{O}(1)\,\sqrt{\epsilon}$ for irrational $\sigma$. To better visualize the behavior of the partition function we display in Fig.~\ref{fig:Z} the functions $\sqrt{10^{-2}}\,|\eta(x+i\cdot10^{-2})|^2$ and $\sqrt{10^{-4}}\,|\eta(x+i\cdot 10^{-4})|^2$ for $x\in[0,1]$. The left Fig.~shows the trend towards zero on rational numbers and the right Fig.~shows the plafond (slightly below 0.6) reached on irrational numbers.

 
The entropy is given by 
\eq{
S_{\textrm{\tiny Boltzmann}}=\big(1-\beta\,\partial_\beta\big)\,\ln Z_{\textrm{\tiny Boltzmann}}\approx\frac{\pi^2\,c_{\mt{M}}}{3\beta\hat\Omega^2}\,.
}{eq:happierthanever} 

A curious aspect of this sector is that the energy expectation value
\eq{
\langle\xi\rangle = -\frac{\pi^2\,c_{\mt{M}}}{6\beta^2\hat\Omega^2}
}{eq:mystrangeaddiction}
is a large negative number. This is surprising given that we assume a spectrum $\xi>0$. Our explanation for this is by analogy to the regularized sum over positive integers: even though each summand is positive, the ($\zeta$-)regularized sum evaluates to $-\frac{1}{12}$. Similarly, our partition function with real $\sigma$ is ill-defined and requires regularization, as explained above. This is why the energy expectation value turns out to be negative, despite the positivity of the energy spectrum. As expected from the imaginarity of $\Omega$, the angular momentum expectation value
\eq{
\langle\Delta\rangle=i\,\frac{\pi^2 \,c_{\mt{M}}}{3\beta^2\hat\Omega^3}
}{eq:nda}
is imaginary too. 

In the hot sector, additionally $\beta\to 0^+$ and thus the cutoff-independent term in the entropy \eqref{eq:happierthanever} tends to infinity, corresponding to some macroscopic limit. If we want the partition function based on induced representations to be real then we need to impose $c_{\mt L}=0$. There is no way of making the partition function based on highest-weight representations real since the denominator $1/|\eta(\sigma+i\epsilon)|^2$ is replaced by $1/(\eta(\sigma+i\epsilon))^2$. Thus, in the Boltzmann sector, it may be preferable to use induced representations.

\subsection{Carrollian vacuum dominance implies negative specific heat} \label{sec:neg-heat}

We conclude this Section with a result that is universal to all sectors discussed therein: vacuum dominance in the dual channel \eqref{eq:vac-dom} implies negative specific heat
\begin{equation}
  C_\Delta \equiv \left. \frac{\partial \langle\xi\rangle}{\partial T}\right|_\Delta < 0
\label{eq:Cxi}
\end{equation}
with fixed angular momentum, $\delta\langle\Delta\rangle=0$, and also negative specific heat
\eq{
C_\Omega \equiv \left. \frac{\partial \langle\xi\rangle}{\partial T}\right|_\Omega < 0
}{eq:lifeisfunny}
with fixed angular velocity, $\delta\Omega=0$. While negativity of specific heat is universal, the interpretation of why specific heat is negative depends on the sector. In the Cardy and Schwarzian sectors the culprit is temperature, which is negative, while the energy expectation value $\langle\xi\rangle$ is positive. By contrast, in the Boltzmann sector the culprit is the negative energy expectation value $\langle\xi\rangle$, while the temperature is positive.

Plugging the energy expectation value \eqref{eq:energy-ave} into the definition of specific heat at constant angular velocity \eqref{eq:lifeisfunny} yields
\eq{
C_\Omega \approx \frac{\pi^2\,c_{\textrm{\tiny M}}\,T}{3\Omega^2} < 0\,.
}{eq:illestofourtime}
In the Cardy and Schwarzian sectors, this expression is negative because $\Omega$ is real and $T$ is negative. In the Boltzmann sector, this expression is negative because $\Omega$ is imaginary and $T$ is positive. (We assume here positive central charge, $c_{\textrm{\tiny M}}>0$.)

To compute \eqref{eq:Cxi}, we split the derivative into two terms
\begin{equation}
  C_\Delta = \frac{\beta^2}{2\pi i}\left(\frac{\partial\langle\xi\rangle}{\partial\rho} + \left.\frac{\delta \sigma}{\delta\rho}\right|_\Delta\,\frac{\partial\langle\xi\rangle}{\partial\sigma}\right)
\end{equation}
where the ratio 
\eq{
\left.\frac{\delta \sigma}{\delta\rho}\right|_\Delta = - \frac{\langle\Delta\,\xi\rangle - \langle\Delta\rangle\langle\xi\rangle}{\langle\Delta^2\rangle - \langle\Delta\rangle^2}
}{eq:humble}
is obtained from requiring 
\eq{
  \delta\langle\Delta\rangle = 0 \qquad\Rightarrow\qquad \delta\sigma \left(\langle\Delta^2\rangle - \langle\Delta\rangle^2\right) + \delta\rho\left(\langle\Delta\,\xi\rangle - \langle\Delta\rangle\langle\xi\rangle\right)=0
}{eq:loco} 
by virtue of the variances \eqref{eq:variances} and covariance \eqref{eq:covariance}. Collecting all results yields the specific heat at constant angular momentum, 
\begin{equation}
  C_\Delta = \beta^2\,\bigg(\langle\xi^2\rangle - \langle\xi\rangle^2 - \frac{\left(\langle\Delta\,\xi\rangle - \langle\Delta\rangle\langle\xi\rangle\right)^2}{\langle\Delta^2\rangle - \langle\Delta\rangle^2}\bigg)\,.
\label{eq:Cxi2}
\end{equation}

Within our vacuum dominance approximation \eqref{eq:energy-ave}, fluctuations in the energy are suppressed,
\begin{equation}
    \langle\xi^2 \rangle - \langle\xi\rangle^2 = \frac{1}{2\pi i}\,\partial_\rho \langle\xi \rangle \approx 0
\end{equation}
as explained in Subsection \ref{sec:3.3}. Hence, the specific heat reduces to a manifestly negative expression 
\begin{equation}
\boxed{
\phantom{\Big(}
  C_\Delta \approx -\beta^2\,\frac{\left(\langle\Delta\,\xi\rangle - \langle\Delta\rangle\langle\xi\rangle\right)^2}{\langle\Delta^2\rangle - \langle\Delta\rangle^2} < 0\,.
\phantom{\Big)}
}
\label{eq:Cxi-vac}
\end{equation}
The above analysis is universal and holds in all sectors contained in Table \ref{tab:1}. It only relies on properties of CCFT$_2$ characters when computing the specific heat in the dual channel. 

In the Cardy and Schwarzian sectors, the negativity of \eqref{eq:Cxi-vac} is obvious since $\Delta$ and $\xi$ (as well as their expectation values, variances, and covariance) are real. In the Boltzmann sector, it is a bit less obvious (but still true) since $\langle\Delta\rangle$ is imaginary. In that sector, an alternative calculation of the specific heat at constant angular momentum starts directly with the expectation values \eqref{eq:mystrangeaddiction} and \eqref{eq:nda}, yielding $\frac{\delta\hat\Omega}{\delta\beta}|_{\Delta}\approx-\frac{2\hat\Omega}{3\beta}$ and 
\eq{
\textrm{Boltzmann\;sector:}\qquad C_\Delta \approx -\frac{\pi^2\,c_{\textrm{\tiny M}}\,T}{9\hat\Omega^2}
}{eq:ilomilo}
which is manifestly negative. Note that also the specific heat difference, $C_\Omega-C_\Delta=-\frac{2\pi^2c_{\textrm{\tiny M}}T}{9\hat\Omega^2}<0$, is negative.


\section{Schwarzian sector ensembles of 2d Carrollian theories}\label{sec:NEXT}

In this Section, we elaborate on the Schwarzian (or near-extremal) sector discussed in Section \ref{sec:Schw}. In particular, we consider three different ensembles, starting with the fixed $M$ ensemble in Section \ref{sec:M}, continuing with fixed $J$ ensembles in Appendix \ref{sec:J}, and finishing with microcanonical ensembles in Appendix \ref{app:micro} and Section \ref{sec:micro}. The reason for the plural will become clearer once we address these ensembles --- it has to do with different contours or saddle points that can change the characteristics of the corresponding ensemble. We also briefly discuss the microcanonical ensemble for other sectors in Section \ref{sec:micro}. Finally, in Section \ref{sec:5.4}, we derive the density of states for the Schwarzian sector using the Carroll modular S-matrix defined in this Section. 

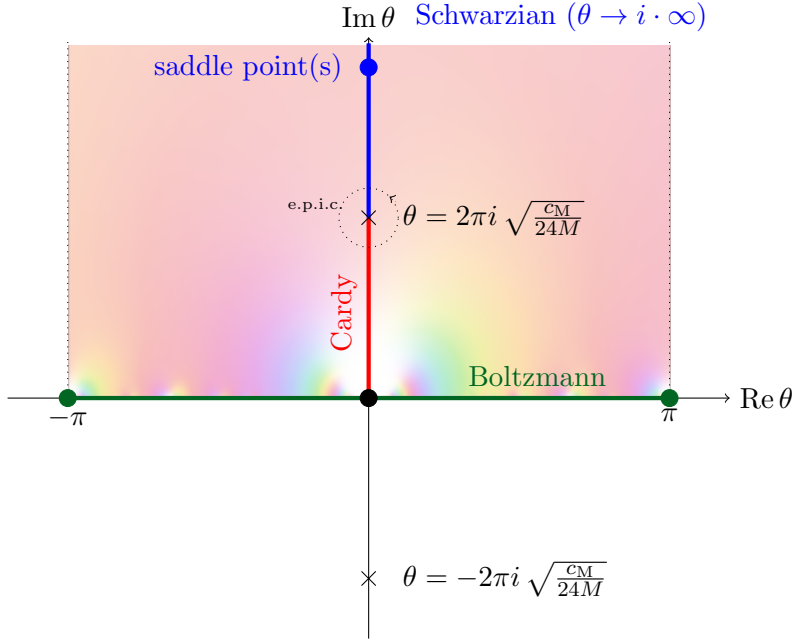
\begin{figure}[htb]
\begin{center}
\begin{tikzpicture}[scale=1.6]
  \node[anchor=south west, inner sep=0pt, opacity=0.4] at (-2.5, -0.02)
    {\includegraphics[width=8cm, height=4.715cm]{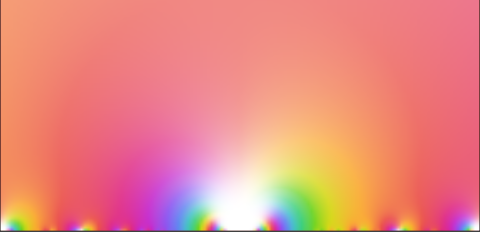}};
  \draw[->] (-3,0) -- (3,0) node[anchor=west] {$\mathrm{Re}\,\theta$};
  \draw[->] (0,-2) -- (0,3) node[anchor=south] {$\mathrm{Im}\,\theta$};
  \draw[ultra thick,darkgreen] (-2.5,0) -- (2.5,0) node[anchor=left,above,xshift=-50pt]{Boltzmann};
  \draw[ultra thick,red] (0,0) -- (0,1.5) node[anchor=left,above,rotate=90,xshift=-36pt]{Cardy};
  \draw[ultra thick,blue] (0,1.5) -- (0,2.95) node[anchor=left,above,xshift=73pt]{Schwarzian ($\theta\to i\cdot\infty$)};
  \node[draw=black, cross out, minimum size=5pt, inner sep=0pt] at (0,1.5) {};
  \node[anchor=west] at (0.2,1.5) {$\theta = 2\pi i\,\sqrt{\tfrac{c_{\mt{M}}}{24M}}$};
  \node[draw=black, cross out, minimum size=5pt, inner sep=0pt] at (0,-1.5) {};
  \node[anchor=west] at (0.2,-1.5) {$\theta = -2\pi i\,\sqrt{\tfrac{c_{\mt{M}}}{24M}}$};
  \node[blue] at (-1.0,2.75) {saddle point(s)};
  \filldraw[blue] (0,2.75) circle (2pt);
  \draw[thin,dotted] (-2.5,0) node[below]{$-\pi$} -- (-2.5,3.0);
  \filldraw[darkgreen] (-2.5,0) circle (2pt);
  \draw[thin,dotted] (2.5,0) node[below]{$\pi$} -- (2.5,3.0);
  \filldraw[darkgreen] (2.5,0) circle (2pt);
  \filldraw (0,0) circle (2pt);
  \draw[dotted] (0,1.5) circle (7pt) node[left,xshift=-5.5pt,yshift=4.85pt] {{\tiny e.p.i.c.}};
  \draw[<-,dotted] ([shift={(45:7pt)}]0,1.5) arc[start angle=60, end angle=45, radius=7pt];
\end{tikzpicture}
\end{center}
\caption[Analytic structure of partition function in complex $\theta$-plane]{Analytic structure of partition function in complex $\theta$-plane (see \protect\hyperlink{fig:description}{main text})}
\label{fig:theta}
\end{figure}

\hypertarget{fig:description}{As} a preparation for such a discussion, in Fig.~\ref{fig:theta} we show the analytic structure of the vacuum-dominated CCFT$_2$ partition function in the complex $\theta$-plane together with the location of the various sectors discussed in the previous Section. The background image is a complex plot of $1/\eta^2[\theta/(2\pi)]$ in a rectangle region with lower left corner $\theta=-\pi+2\pi i\cdot 10^{-2}$ and upper right corner $\theta=\pi+\pi i$. Single poles are marked by $\times$. In the microcanonical Subsection \ref{sec:micro} we shall \textbf{e}ncircle the \textbf{p}ole \textbf{i}n the \textbf{c}omplex upper half-plane, and the corresponding integration contour is depicted as a dotted circle with the arrow giving the sense of integration. The \textcolor{darkgreen}{\textbf{Boltzmann}} sector lies on/slightly above the real $\theta$ axis. The \textcolor{red}{\textbf{Cardy}} sector is on the imaginary $\theta$ axis with sufficiently small values of $|\theta|$. These two sectors meet at/slightly above the origin denoted as a \textbf{black dot}. The \textcolor{blue}{\textbf{Schwarzian}} sector is on the imaginary $\theta$ axis with sufficiently large values of $|\theta|$. The saddle points discussed in Appendix \ref{app:A} belong to the Schwarzian sector, and we depicted one of them schematically as a \textcolor{blue}{\textbf{blue dot}}. The \textcolor{darkgreen}{\textbf{green dots}} and vertical dotted lines are mapped to each other by a $T$-transformation.


\subsection[Fixed \texorpdfstring{$M$}{M} ensemble]{Fixed \texorpdfstring{$\boldsymbol{M}$}{M} ensemble}\label{sec:M}

To compare the field theory results to the bulk, one usually wants to take the near-extremal limit in the fixed charge sector (see, e.g., \cite{Almheiri:2016fws}). Therefore, we consider the mixed ensemble where the partition function depends upon $\langle \xi \rangle=M$ instead of $\beta$. This is achieved via a Legendre transform.

The vacuum dominant partition function \eqref{eq:badguy} written in terms of $\beta$ and $\hat\theta=-i\theta$ (i.e., $\hat\theta=\beta\Omega\in\mathbb{R}^+$)
\begin{equation}
    Z_{\textrm{\tiny Schwarzian}}(\beta,\,\hat\theta)=\frac{(2\pi)^5}{\hat\theta^3}\exp\bigg(\frac{\hat\theta}{12}+\frac{\pi^2}{6\hat\theta}\(c_{\mt{L}}-2\)-\frac{\pi^2}{6}\,\frac{\beta}{\hat\theta^2}\,c_{\mt{M}}\bigg)\,.
\label{eq:zbt}
\end{equation}
permits to define the fixed $M$ ensemble for $M>\frac{4\pi^2c_{\mt M}}{24\hat\theta^2}$ as
\eq{
Z_{\textrm{\tiny Schwarzian}}(M,\,\hat\theta)=\int\limits_{-\infty}^{0} {\rm d} \beta\, e^{\beta M}\, Z_{\textrm{\tiny Schwarzian}}(\beta,\,\hat\theta)
= \frac{(2\pi)^5\,e^{\frac{\hat\theta}{12}+\frac{\pi^2}{6\hat\theta}\(c_{\mt{L}}-2\)}}{(M\hat\theta^2-\frac{\pi^2}{6}\,c_{\mt M})\,\hat\theta}\, .
}{eq:lalapetz}
We have integrated only over negative $\beta$ because this is a defining property of the Schwarzian sector. The inequality $M>\frac{4\pi^2c_{\mt M}}{24\hat\theta^2}$ is needed for the convergence of the integral. For positive mass, $M>0$, it is obeyed automatically in the Schwarzian sector for generic values of $c_{\mt M}$ since $\hat\theta^2\gg1$, which further simplifies \eqref{eq:lalapetz} to 
\begin{equation}
Z_{\textrm{\tiny Schwarzian}}(M,\,\hat\theta)\approx \frac{(2\pi)^5}{M\hat\theta^3}\,\exp\bigg(\frac{\hat\theta}{12}+\frac{\pi^2}{6\hat\theta}\(c_{\mt{L}}-2\)\bigg)\, .
\end{equation}
The entropy in this ensemble is given by 
\begin{equation}
 S= (1-\hat \theta\partial_{\hat\theta})\ln   Z_{\textrm{\tiny Schwarzian}}(M,\,\hat\theta) =\frac{\pi^2}{3\hat\theta}\(c_{\mt{L}}-2\)-3\ln \hat\theta +3 -\ln M+5\ln(2\pi)
\end{equation}
The first term is the classical contribution to the entropy, assuming $c_{\mt{L}}\gg\hat\theta\gg 1$. The scale where log corrections to the entropy start competing with the classical is $\hat\theta\approx
c_{\mt L}/\ln{c_{\mt L}}$.  

\subsection{Microcanonical ensemble}\label{sec:micro}

The fixed $J$ ensemble leads to integrals that are ill-adapted to the saddle point approximation because of a wrong sign of the Gaussian; see Appendix \ref{app:A} for details. Similar remarks apply to the microcanonical ensemble. Physically, the reason for this sign is the negative specific heat.

In Appendix \ref{app:micro}, we tacitly assume some contour $C$ that is indifferent to the singularities at $\hat\theta=0,\infty$, has support at the saddle point, and avoids the two first-order poles in \eqref{eq:lalapetz} (see Fig.~\ref{fig:theta}). 

We drop now the last assumption and instead consider a different contour that encircles the pole\footnote{%
The pole is at the boundary of convergence of the fixed-$M$ partition function \eqref{eq:lalapetz}. Once we have the right-hand side of \eqref{eq:lalapetz}, we can analytically continue $\hat\theta$, and as long as we do not sit on the pole, the result is a meaningful expression. Thus, our contour around the pole is well-defined even though the partition function at the pole is ill-defined.}
at $\hat\theta=2\pi\,\sqrt{c_{\mt{M}}/(24M)}$ while avoiding all other singularities. For such contours, the microcanonical partition function evaluates to
\eq{\boxed{\phantom{\Big(}
Z_\odot(M,\,J) = \oint\extd\hat\theta\,\frac{(2\pi)^5\,e^{\hat\theta J+\frac{\hat\theta}{12}+\frac{\pi^2}{6\hat\theta}\(c_{\mt{L}}-2\)}}{2\pi i\,(M\hat\theta^2-\frac{\pi^2}{6}c_{\mt M})\,\hat\theta} = N\,e^{2\pi\Big(\(c_{\mt{L}}-2\)\,\sqrt{\frac{M}{24c_{\mt{M}}}} + \sqrt{\frac{c_{\mt{M}}}{24M}}\,\big(J+\tfrac{1}{12}\big)\Big)}
\phantom{\Big(}}}{eq:lostallfaith}
with $N=(2\pi)^3\frac{12}{c_{\mt M}}$. Assuming large $J$ and large $c_{\mt{L}}$ as well as inserting the definitions \eqref{eq:whenthepartyisover} with $\langle\xi\rangle = M$ and $\tilde\Delta = J$ recovers to LO precisely the Cardy entropy \eqref{eq:cardy-finite}
\eq{\boxed{\phantom{\big(}
S_\odot = \ln Z_\odot(M,\,J) \approx 2\pi\,\Bigg(\frac{c_{\mt{L}}}{24}\,\sqrt{\tilde\xi}+\frac{\tilde\Delta}{\sqrt{\tilde\xi}}\Bigg)\,.
\phantom{\big)}}}{eq:screechstale}
 
The fact that calculations using a saddle point approximation give only one part of the full microcanonical entropy \eqref{eq:screechstale} is unsurprising: in each saddle point calculation discussed in Appendix \ref{app:micro} we zoom into a region where either $c_{\mt M}$ or $c_{\mt L}$ dominates and thus end up with only the part of the entropy that features the dominant central charge associated with the corresponding saddle point. The only calculation where we never used a saddle point approximation was the one in the previous paragraph, which explains why, in this case, we get the full Cardy entropy containing both central charges.

It is curious that (for large Virasoro central charge and large angular momentum) the microcanonical Schwarzian partition function \eqref{eq:lostallfaith} is equivalent to the grand canonical Cardy partition function \eqref{eq:Z1}; in particular, neither of them produces log corrections to the LO entropy. Given that the grand canonical Schwarzian partition function \eqref{eq:badguy} does lead to such log corrections \eqref{eq:birdsofafeather}, it is natural to expect that the same would be true for the microcanonical Cardy partition function. We verify now that this is indeed the case by applying the analysis of this Section to more general partition functions with vacuum dominance in the dual channel \eqref{eq:tildeZ-mid}.

Since the Boltzmann sector $\beta>0$ is ill-defined mathematically, we confine ourselves to sectors with negative temperature. Similar to Section \ref{sec:M}, we first Laplace transform into a mixed ensemble where we integrate over the chemical potential $\rho$ and the partition function depends on the other chemical potential, $\sigma$, and on the extensive quantity $\xi$ associated with the chemical potential $2\pi i\rho$, see \eqref{eq:expectation_val}. Again, the integral is elementary since $\rho$ appears linearly in the exponent of the grand canonical partition function \eqref{eq:tildeZ-mid} and leads to two first-order poles in the complex $\sigma$-plane at the same loci as in Fig.~\ref{fig:theta}. Laplace transforming with a contour encircling the pole in the complex upper half-plane (e.p.i.c.) yields the result for the microcanonical partition function
\begin{align}
    Z_{\textrm{\tiny micro}}^{\textrm{\tiny vac.~dom.}}(\xi,\Delta) &= \tilde N\!\!\oint\limits_{\textrm{e.p.i.c.}}\!\! \extd\sigma\,e^{-2\pi i\sigma\,\Delta}\int\limits_{-i\infty}^0\!\extd\rho\,e^{-2\pi i\rho\,\xi}\,e^{2\pi i\big(\frac{c_\mt{L}-2}{24\sigma}-\rho \frac{c_\mt{M}}{24\sigma^2}\big)}\,\frac{ (1-e^{-\frac{2\pi i}{\sigma}})^2}{{\eta(-1/\sigma)^2}} \nonumber\\
    &=N\,\frac{\Big(1-e^{-2\pi\sqrt{\tilde\xi}}\Big)^2}{\tilde\xi^{3/2}\,\big(\eta(i\sqrt{\tilde\xi})\big)^2}\,\exp[2\pi\Bigg( \frac{c_{\mt L}-2}{24}\,\sqrt{\tilde\xi} + \frac{\Delta}{\sqrt{\tilde\xi}}\Bigg)]
\end{align}
where $N,\tilde N$ are state-independent normalization factors and the dimensionless mass $\tilde\xi$ is defined in \eqref{eq:whenthepartyisover}.

In the limit of small $\tilde\xi$, we use Eq.~\eqref{eq:eta6} and recover the Schwarzian result \eqref{eq:lostallfaith},
\eq{
 Z_{\textrm{\tiny micro}}^{\textrm{\tiny vac.~dom.}}(\xi\to 0,\Delta) \approx N\,\exp[2\pi\Bigg( \frac{c_{\mt L}-2}{24}\,\sqrt{\tilde\xi} + \frac{\Delta+\frac{1}{12}}{\sqrt{\tilde\xi}}\Bigg)]\,.
}{eq:wildflower}
In the limit of large $\tilde\xi$, we use the Eq.~\eqref{eq:eta2} and produce the Cardy result \cite{Bagchi:2013qva}
\eq{
 Z_{\textrm{\tiny micro}}^{\textrm{\tiny vac.~dom.}}(\xi\to\infty,\Delta) \approx N\,\tilde\xi^{-3/2}\,\exp[2\pi\Bigg( \frac{c_{\mt L}}{24}\,\sqrt{\tilde\xi} + \frac{\Delta}{\sqrt{\tilde\xi}}\Bigg)]\,.
}{eq:hotline}
As anticipated, in the Cardy sector, the microcanonical partition function \eqref{eq:hotline} leads to log corrections to the LO entropy. In particular, in the large $|\tilde\Omega|$ regime with the LO entropy $S_{\textrm{\tiny Cardy}}$ given by \eqref{eq:cardy-large}, the microcanonical Cardy entropy 
\eq{
S = \ln Z_{\textrm{\tiny micro}}^{\textrm{\tiny vac.~dom.}}(\xi\to\infty,\Delta)\big|_{\Delta\approx\tfrac{c_{\mt L}}{24}\,\tilde\xi} \approx S_{\textrm{\tiny Cardy}} - 3\,\ln S_{\textrm{\tiny Cardy}} + \dots
}{eq:blue}
has precisely the same numerical factor in front of the log correction as the grand canonical Schwarzian entropy \eqref{eq:birdsofafeather}.

\subsection{Density of states from modular S-matrix}\label{sec:5.4}

In this Subsection, we derive the density of states/partition function in the microcanonical ensemble for the Schwarzian sector using the Carroll modular S-matrix. This will also provide an alternative way of deriving the microcanonical results obtained in the previous Subsection.

Carroll modular S-matrices are defined as the objects relating the characters in the standard and the dual channels according to \cite{Aggarwal:2025hkb}
\begin{equation} \label{eq:Smatrixdef}
    \chi_{\xi',\Delta'}\left(\frac \rho{\sigma^2}, -\frac{1}\sigma\right)=\int \dd \Delta\,\frac{\dd\xi}{4\sqrt{\xi-\frac{c_{\mt M}}{24}}}\,\mathbb S(\xi',\Delta';\xi,\Delta) \chi_{\xi, \Delta}(\rho, \sigma)\,.
\end{equation}
When writing the full CCFT$_2$ partition function as an integral over $(\xi,\,\Delta)$ with some density of states $\mathbb D(\xi,\Delta)$, we can approximate it
\begin{equation} 
  Z (\rho, \sigma)=: \int \dd\Delta\,\dd\xi\,\mathbb D(\xi,\Delta)e^{2\pi i(\rho \xi+\sigma \Delta)} \approx \chi_\mathds{1}(\rho/\sigma^2, -1/\sigma)
\label{eq:conj}
\end{equation}
using vacuum dominance in the dual channel. Comparing \eqref{eq:Smatrixdef} and \eqref{eq:conj}, together with the expression of characters \eqref{BMS_character3}, we find for the highest-weight representation
\begin{equation} \label{eq:dos-smatrix}
    \mathbb D_{\rm Schwarzian}(\xi,\Delta) \approx \frac{\mathbb S(\mathds{1}; \xi+\frac{c_{\mt M}}{24}, \Delta+\frac{c_{\mt L}}{24})}{4\sqrt{\xi}} 
\end{equation}
where we used that $\eta(\sigma)^2\approx e^{\frac{i\pi\sigma}{6}}$ in the Schwarzian sector. Thus, knowledge of the vacuum modular S-matrix determines the density of states in the Schwarzian sector. The above result is not true in the Cardy sector because $\eta(\sigma)^2\approx \frac{i}{\sigma}~e^{-\frac{i\pi}{6\sigma}}$ in that sector. In other words, the contribution of descendants becomes important at high energies, and the logarithmic corrections cannot be trusted. This is also true for the Cardy sector of 2d CFTs \cite{Ghosh:2019rcj}.

CCFT$_2$ modular S-matrices have been recently computed in \cite{Aggarwal:2025hkb}.\footnote{%
Our conventions for the central charges here differ from \cite{Aggarwal:2025hkb} by factors of 12.} 
Using a Liouville-inspired parametrization,
\begin{equation}
    \xi-\frac{c_\mt{M}}{24}=:P_\mt{M}^2(\xi)\qquad\qquad \Delta-\frac{c_\mt{L}-2}{24}=:P_\mt{L}(\Delta)\qquad\qquad \frac{c_\mt{L}-2}{12}=:Q_\mt{L}
\label{eq:l-mom}
\end{equation}
non-vacuum modular S-matrices for \textit{highest-weight} representations equal
\eq{
    \mathbb S(P_{\mt M}',P_{\mt L}';P_{\mt M},P_{\mt L})=2\frac {|P_{\mt M}'|} {|P_{\mt M}|^2}\frac{P_{\mt M}}{P_{\mt M}'}\sin\left[2\pi \left(\frac{P_{\mt M}'}{P_{\mt M}}P_{\mt L}+\frac{P_{\mt M}}{P_{\mt M}'}P_{\mt L}'\right)\right]
}{eq:vmsm}
whereas the vacuum modular S-matrix is
\begin{equation} \label{eq:SP1}
    \mathbb S(\mathds{1};P_{\mt M},P_{\mt L})=
    \frac {8 P_{\mt M} } {|P_{\mt M}|^2}\sinh\left[2\pi \left(\frac{P_{\mt L}}{P_{\mt M}}\sqrt{\frac{c_{\mt M}}{24}}+\frac{P_{\mt M}}{\sqrt{c_{\mt 
    M}/24}} \frac{Q_{\mt L}-2}{2}\right)\right]\sinh^2\left(\frac{\pi P_{\mt M}}{\sqrt{ c_{\mt M}/24}}\right)~.
\end{equation}

Since $P_{\mt M}$ carries units, both the modular S-matrix and the density of states in the continuum formulation carry units. The relations
\eq{
  \frac{\langle P_\mt{M}(\xi+\frac{c_{\mt M}}{24})\rangle}{\sqrt{c_\mt{M}/24}} = \sqrt{\tilde{\xi}}\qquad \qquad \left\langle P_\mt{L}\left(\Delta+\frac{c_{\mt L}}{24}\right) \right\rangle = \tilde{\Delta}+\frac{1}{12}\,.
}{eq:observe}
will be useful below. For the highest-weight representation, when the vacuum dominates in the dual channel, the vacuum modular S-matrix \eqref{eq:vmsm} with \eqref{eq:observe} reads
\begin{equation}
    \mathbb S\left(\mathds{1}; , \xi+\frac{c_{\mt M}}{24}, \Delta+\frac{c_{\mt L}}{24}\right)\approx\frac{8\sinh^2\big(\pi \sqrt{\tilde \xi}\big)} {\sqrt{\tilde \xi\,c_{\mt M}/24}}\sinh\bigg[2\pi \bigg(\frac{c_{\mt L}-26}{24}\,\sqrt{\tilde \xi} +\frac{\tilde \Delta+\frac{1}{12}}{\sqrt{\tilde \xi}}\bigg)\bigg]~.
\end{equation}
In the Schwarzian sector, $\sqrt{\tilde \xi}\ll 1$ and $\tilde\Delta\gg 1$, the argument for the second $\sinh$ is large and can be approximated by an exponential 
   \begin{equation} \label{eq:smatrixapprox}
    \mathbb S\left(\mathds{1}; , \xi+\frac{c_{\mt M}}{24}, \Delta+\frac{c_{\mt L}}{24}\right)\approx\frac {8\sinh^2\big(\pi \sqrt{\tilde \xi}\big)} {\sqrt{\tilde \xi\,c_{\mt M}/24}}\exp\bigg[2\pi \bigg(\frac{c_{\mt L}-26}{24}\,\sqrt{\tilde \xi} +\frac{\tilde \Delta}{\sqrt{\tilde \xi}}\bigg)\bigg]~.
\end{equation}
The remaining $\sinh^2$ can be approximated for small $\sqrt{\tilde \xi}$,
\begin{equation}
    \mathbb S_{\rm Schwarzian}\left(\mathds{1}; , \xi+\frac{c_{\mt M}}{24}, \Delta+\frac{c_{\mt L}}{24}\right)\approx\frac {8\pi^2\sqrt{\tilde \xi } } {\sqrt{c_{\mt M}/24}}\exp\bigg[2\pi \bigg(~\frac{c_{\mt L}}{24}\,\sqrt{\tilde \xi}+\frac{\tilde \Delta}{\sqrt{\tilde \xi}}\bigg)\bigg]~.
\end{equation}
The first term in the exponent only survives when $c_{\mt L}\gg 1$. Finally, the Carroll--Schwarzian density of states becomes
\eq{\boxed{\phantom{\Bigg(}
    \mathbb D_{\rm Schwarzian}\left(\mathds{1}; \xi,\Delta\right)\approx\frac {48\pi^2} {c_{\mt M}}\exp\bigg[2\pi \bigg(~\frac{c_{\mt L}}{24}\,\sqrt{\tilde \xi}+\frac{\tilde \Delta}{\sqrt{\tilde \xi}}\bigg)\bigg]~.
\phantom{\Bigg)}}}{eq:thediner}
This matches the microcanonical Schwarzian partition function in \eqref{eq:lostallfaith} (up to the prefactor, which is just a normalization), since $\tilde \Delta$ is large in the Schwarzian sector. 


\section{Holographic interpretation}\label{sec:holo}

In this Section, we put our field theory results from the previous Sections into the context of flat space holography. The main purpose of this Section is to add a new entry in the Flat$_3$/CCFT$_2$ dictionary, namely a geometric interpretation of the Schwarzian sector of a (holographic) CCFT$_2$, which we shall do in Section \ref{sec:6.2}. Before that, we summarize in Section \ref{sec:6.1} salient aspects of flat space cosmologies. We postpone a more general discussion of 3d gravity partition functions and comparison with earlier literature to Section \ref{sec:3d-grav}.

\subsection{Summary of flat space cosmologies}\label{sec:6.1}

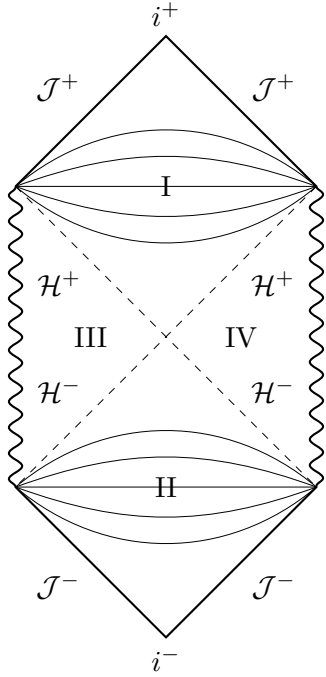
\begin{figure}[htb]
\centering
\begin{tikzpicture}
\node (I)   at ( 0, 2) {I};
\node (II)  at ( 0,-2) {II};
\node (III) at (-1, 0) {III};
\node (IV)  at ( 1, 0) {IV};

\path (I) +( 90:2) coordinate[label=90:$i^+$] (Itop)
          +(-90:2) coordinate (Ibot)
          +(  0:2) coordinate (Iright)
          +(180:2) coordinate (Ileft);

\path (II) +( 90:2) coordinate (IItop)
           +(-90:2) coordinate[label=-90:$i^{-}$] (IIbot)
           +(180:2) coordinate (IIleft)
           +(  0:2) coordinate (IIright);

\draw [thick,decoration={snake,segment length=0.343cm}] (Ileft)
  --(Itop)   node[midway,above left]  {$\cal{J}^+$}
  --(Iright) node[midway,above right] {$\cal{J}^{+}$}
  decorate { --(IIright) }
  --(IIbot)  node[midway,below right] {$\cal{J}^{-}$}
  --(IIleft) node[midway,below left]  {$\cal{J}^{-}$}
  decorate { --cycle };

\draw [very thin,dashed]
  (Ileft) --(IIright) node[near start,below left]  {$\cal{H}^+$}
                      node[near end,above right]   {$\cal{H}^-$}
  (Iright) --(IIleft) node[near start,below right] {$\cal{H}^+$}
                      node[near end,above left]    {$\cal{H}^-$};

\draw [very thin] (Ileft) --(Iright);
\draw [very thin] (Ileft) to [out=20,in=160] (Iright);
\draw [very thin] (Ileft) to [out=40,in=140] (Iright);
\draw [very thin] (Ileft) to [out=-20,in=-160] (Iright);
\draw [very thin] (Ileft) to [out=-40,in=-140] (Iright);

\draw [very thin] (IIleft) --(IIright);
\draw [very thin] (IIleft) to [out=20,in=160] (IIright);
\draw [very thin] (IIleft) to [out=40,in=140] (IIright);
\draw [very thin] (IIleft) to [out=-20,in=-160] (IIright);
\draw [very thin] (IIleft) to [out=-40,in=-140] (IIright);
\end{tikzpicture}
\caption{2d slice of Penrose diagram for flat space cosmologies}
\label{fig:Penrose}
\end{figure}

Thermal states in a CCFT$_2$ are dual to flat space cosmologies (FSC) \cite{Bagchi:2012yk,Bagchi:2012xr,Barnich:2012xq}, which are orbifolds of 3d Minkowski space \cite{Cornalba:2002fi}. A 2d slice through their Penrose diagram looks like the Schwarzschild Penrose diagram with time and radius exchanged, see Fig.~\ref{fig:Penrose}. The dashed line corresponds to the (cosmological) horizon, which bifurcates into future and past horizons $\mathcal{H}^\pm$. Region I (II) describes an expanding (contracting) universe that is locally flat. In regions III and IV, there are singularities in the causal structure, namely the onset of closed timelike curves, denoted by the wiggly lines.


We start from the BTZ metric with mass $M>0$ and angular momentum $J$ ($|J|<\ell\,M$),
\begin{equation}
\label{eq:btz}
  \extd s_{\textrm{\tiny BTZ}}^2=\bigg(8 G M-\frac{r^2}{\ell^2}\bigg)\extd t^2 + \frac{\extd r^2}{-8 G M + \frac{r^2}{  \ell^2} + \frac{16 G^2 J^2}{r^2}}
  - 8 G J \extd t \extd\varphi + r^2\extd\varphi^2\qquad \varphi\sim \varphi + 2\pi
\end{equation}
whose outer and inner horizons $r_\pm$ are given by 
\eq{
  M = \frac{r_+^2 + r_-^2}{8 G \ell^2} \qquad\qquad J =\frac{r_+ r_-}{4 G \ell}\,.
}{eq:MJ}
The flat space scaling limit 
\eq{
r_+\to\hat r_+\,\ell\qquad\qquad r_-\to r_0\qquad\qquad\textrm{with\;} \ell\to\infty \quad\textrm{keeping\;fixed\;} \hat r_+, r_0 
}{eq:FSClimit}
yields the Lorentzian FSC metric in the notation of \cite{Bagchi:2013lma},
\eq{
\extd s^2 = \hat r_+^2\,\Big(1-\frac{r_0^2}{r^2}\Big)\extd t^2 - \frac{\extd r^2}{\hat r_+^2\,\big(1-\frac{r_0^2}{r^2}\big)} + r^2\,\Big(\extd\varphi-\frac{\hat r_+ r_0}{r^2}\extd t\Big)^2\qquad\qquad\varphi\sim\varphi+2\pi\,.
}{eq:FSC1}
The outer horizon was scaled to infinity with the AdS radius, and the inner, cosmological horizon remains at a finite value. Note, however, that the time coordinate is denoted by $r$ and $t$ is the radial coordinate (indeed, the $r=\rm const.$ lines in regions I and II of Fig.~\ref{fig:Penrose} are Cauchy hypersurfaces at different values of time). Moreover, as opposed to BTZ spacetimes, there is (at most) one Killing horizon and thus, there appear to be no extremal FSCs (see, however, the next Subsection).

To compare with our CCFT$_2$ discussion, we first continue to Euclidean signature. There are three different ways to do this: 1.~We Wick rotate the radius, $t\to i\tau$, and analytically continue $\hat r_+\to -i r_+$; this was done in \cite{Bagchi:2013lma}. 2.~We Wick rotate the time, $r\to i\rho$, analytically continue $\hat r_+ \to i \rho_+, r_0\to i\rho_0$, keep $t=\tau$, and make an overall flip of the signature from $(-,-,-)$ to $(+,+,+)$. 3.~After changing to mostly minus signature, we Wick rotate both spatial coordinates, $t\to i\tau$, $\varphi\to i\phi$, without any analytic continuation of parameters but insisting that the new coordinate $\phi$ is now $2\pi$-periodic.

In all three cases, we end up with the same Euclidean metric (up to notations)
\eq{
\extd s^2 = r_+^2\,\Big(1-\frac{r_0^2}{r^2}\Big)\extd\tau^2 + \frac{\extd r^2}{r_+^2\,\big(1-\frac{r_0^2}{r^2}\big)} + r^2\,\Big(\extd\varphi-\frac{r_+ r_0}{r^2}\extd\tau\Big)^2\qquad\qquad\varphi\sim\varphi+2\pi
}{eq:FSC2}
which is regular at the locus $r=r_0$ provided we impose the periodicities
\eq{
(\tau,\varphi)\sim(\tau+\beta,\varphi+\theta)
}{eq:FSC3}
with real $\beta,\theta$ given by
\eq{
\beta=\frac{2\pi r_0}{r_+^2}\qquad\qquad \theta=\frac{2\pi}{r_+}\,.
}{eq:FSC4}

We stress that the geometries \eqref{eq:FSC2} for $r_+^2\neq 1$ have a conical defect asymptotically, in the sense that $\tau=\rm const.$ circles for large $r$ have an area of $r^2 \pi/r_+^2$. At the same time, these geometries are free from conical singularities at $r=r_0$: the Euclidean manifold caps off smoothly and globally yields a filled torus. One can understand the asymptotic conical defect as a flat space limit of the conical singularity obtained at the outer horizon of a Euclidean BTZ black hole patch, where the inner horizon is kept free from conical singularities.  

The chemical potential $\beta$ has an interpretation as Carrollian inverse temperature from a boundary perspective. However, from the bulk perspective, it is a velocity that is the Legendre-dual of the momentum associated with the radial coordinate $\tau$. The chemical potential $\theta$ is the twist angle of the torus, both in the boundary CCFT$_2$ and the bulk.

Let us now return to the construction in our paper inspired by the GMT analysis \cite{Ghosh:2019rcj}. In the BTZ case, the holographic interpretation of the extremal sector was related to zooming in towards extremal BTZ black holes. However, as reviewed above, there is no extremal FSC in the usual sense, so there is no evident bulk interpretation of the Schwarzian CCFT$_2$ sector discussed in the present work. 

In the next Subsection, we resolve this issue by taking a different flat space scaling limit of BTZ that rescales both horizon radii democratically and pushes them towards infinity, thereby allowing a flat space notion of extremality and recovering a Schwarzian sector of a CCFT$_2$.

\subsection{Schwarzian sector of flat space cosmologies}\label{sec:6.2}

The flat space limit \eqref{eq:FSClimit} cannot capture the notion of a near-extremal limit in the parent AdS$_3$/CFT$_2$ discussion since it scales the outer and the inner horizon differently. This suggests considering the more democratic limit
\begin{equation}
    r_\pm \to \sqrt{\ell}\,\hat{r}_\pm\,.
\label{eq:Sch-limit}
\end{equation}
The original BTZ charges become
\begin{equation}
    M \to 0 \qquad \qquad J=\frac{\hat{r}_+\hat{r}_-}{4G}
\end{equation}
whereas the resulting flat metric equals
\eq{\boxed{\phantom{\Big(}   
O\textrm{-folds:}\;\extd s^2=-8GJ\,\extd t\extd\varphi + r^2\extd\varphi^2 + \frac{r^2}{(4GJ)^2}\,\extd r^2 \qquad\qquad \varphi\sim\varphi+2\pi\,.
\phantom{\Big)}}}{eq:2flat-BTZ}
After the diffeomorphism
\begin{equation}
    \extd\varphi = \extd\psi + \frac{1}{4GJ}\,\extd r\qquad \qquad \extd t = \extd u + \frac{r^2}{(4GJ)^2}\,\extd r
\end{equation}
the flat metric becomes
\begin{equation}
    \extd s^2 = -8GJ\,\extd u\extd\psi - 2\extd u\extd r + r^2\extd\psi^2\qquad\qquad \psi\sim\psi+2\pi\,.
\end{equation}
This allows comparing with the general asymptotically flat metric satisfying Barnich--Comp\`ere boundary conditions \cite{Barnich:2006av}
\begin{equation}\label{bms3metric}
 \extd s^2 = \Theta(\psi)\extd u^2  -2\extd r\extd u + 2[\Xi(\psi) + \frac{u}{2}\Theta^\prime] \extd\psi \extd u + r^2 \extd\psi^2\qquad\qquad \psi\sim\psi+2\pi\, .
\end{equation}
Restricting to the zero mode sector, i.e. $\Theta = 8GM$ and $\Xi=4GJ$, we conclude the spacetime \eqref{eq:2flat-BTZ} indeed has zero mass and non-vanishing angular momentum. 

The geometries \eqref{eq:2flat-BTZ} were studied by Cornalba and Costa in \cite{Cornalba:2005je}, who identified them as $O$-plane orbifolds (abbreviated here as $O$-folds). They are locally and asymptotically flat (indeed, the Riemann tensor of \eqref{eq:2flat-BTZ} vanishes everywhere) and have naked closed timelike curves without any horizon. See Fig.~\ref{fig:schematic} for their Penrose diagram and their position in the Barnich--Comp\`ere state space. 


\begin{figure}[htb]
    \begin{center}
\def\L{3.2}
\begin{tikzpicture} 
\begin{scope}[shift={(-0.1*\L,0)}]
  \coordinate (A) at (-0.9*\L, 0.9*\L);
  \coordinate (B) at (0.9*\L, 0.9*\L);
  \coordinate (O) at (0,0);
  \fill[blue!20] (A) -- (O) -- (B) -- cycle;
\draw[black,thin,->] (-\L,0) -- (\L,0) node[right]{$J$};
\draw[black,thin,->] (0,-0.4*\L) -- (0,\L) node[above]{$M$};
\draw[red] (O) -- (B) node[below,rotate=45]{\hspace*{-2.2truecm}\footnotesize extremal BTZ};
\draw[red,thin] (O) -- (A);
\filldraw[darkgray,thick] (0,-0.3*\L) circle(0.01*\L) node[right]{\tiny global\,AdS$_3$}; 
\draw [decorate,decoration={brace},xshift=-2pt] (0,-0.3*\L) -- (0,0) node[midway,xshift=-3pt,left] {\tiny gap};
\draw[black,->] (0.1*\L,0.4*\L) -- (0.75*\L,0.85*\L) node[above,midway,sloped] {\!\!Schwarzian};
\end{scope}
\begin{scope}[scale=0.6,shift={(-1.75*\L,0.75*\L)}]
\draw[red,thin,dashed] (0,1.5) coordinate (ip) -- (1.5,0) coordinate (in);
\draw[red,thin,dashed] (0,-1.5) coordinate (im) -- (in);
\draw[red,thin,decoration={snake,segment length=0.343cm}] (im) decorate {-- (ip)};
\draw[red,thin,dashed] (im) -- (1.5,-3) coordinate (scri);
\draw[red,thin] (scri) -- (in);
\end{scope}
\draw[->] (\L,0.3*\L) -- (1.5*\L,0.3*\L) node[above]{\hspace*{-1.3truecm}$\ell\to\infty$};
\begin{scope}[shift={(2.5*\L,0)}]
  \coordinate (A) at (-0.9*\L,0);
  \coordinate (B) at (-0.9*\L, 0.9*\L);
  \coordinate (C) at (0.9*\L,0.9*\L);
  \coordinate (D) at (0.9*\L,0);
  \fill[blue!20] (A) -- (B) -- (C) -- (D) -- cycle;
  \draw[black,thin,->] (-\L,0) -- (\L,0) node[right]{$J$};
  \draw[black,thin,->] (0,-0.4*\L) -- (0,\L) node[above]{$M$};
  \draw[red] (A) -- (D) node[below,red] {\hspace*{-2.2truecm}\footnotesize $O$-fold};
  \filldraw[darkgray,thick] (0,-0.3*\L) circle(0.01*\L) node[right]{\tiny global\,Flat$_3$};
  \draw [decorate,decoration={brace},xshift=-2pt] (0,-0.3*\L) -- (0,0) node[midway,xshift=-3pt,left] {\tiny gap};
\draw[black,->] (0.5*\L,0.85*\L) -- (0.5*\L,0.05*\L) node[above,midway,sloped] {Schwarzian};
\end{scope}
\begin{scope}[scale=0.6,shift={(3.2*\L,0.75*\L)}]
\draw[red,thin] (0,1.5) coordinate (ip) -- (1.5,0) coordinate (in);
\draw[red,thin] (0,-1.5) coordinate (im) -- (in);
\draw[red,thin,decoration={snake,segment length=0.343cm}] (im) decorate {-- (ip)}; 
\end{scope}
\end{tikzpicture}
\caption[Gravity side of Schwarzian sectors in AdS$_3$/CFT$_2$ \& Flat$_3$/CCFT$_2$]{Gravity side of Schwarzian sectors in AdS$_3$/CFT$_2$ (left) and Flat$_3$/CCFT$_2$ (right)}
\label{fig:schematic}
\end{center}
\end{figure}

Below, we review some aspects of the $O$-fold. The metric \eqref{eq:2flat-BTZ} is a quotient of $\mathbb{R}^{1,2}$ 
\begin{equation}
    \extd s^2 = -2\extd x^+\,\extd x^- + \extd x^2
\end{equation}
by the finite action generated by the Killing vector field \cite{Cornalba:2005je}
\begin{equation}
    \xi = -\left(x^-\partial_x + x\partial_+\right) + \gamma\,\partial_- 
\end{equation}
that combines null translations and boosts. In adapted coordinates
\begin{equation}
       x^+ = \frac1\gamma\,y^+ - y y^- + \frac{\gamma}{6}\,(y^-)^3\qquad\qquad x^- = \gamma\,y^-\qquad\qquad x = y - \frac{\gamma}{2}\,(y^-)^2
\end{equation}
the Killing vector simplifies to a coordinate Killing vector, $\xi=\partial_{y^-}$, whereas the metric becomes
\eq{
    \extd s^2 = -2\extd y^+ \extd y^- + \extd y^2 + 2\gamma\, y\, (\extd y^-)^2\qquad\qquad y^-\sim y^-+1\,.
}{eq:bohemiangravity}
The latter is equivalent to \eqref{eq:2flat-BTZ} using the further map $\gamma\,y= 2\pi^2\,r^2$, $2\pi y^+ = \gamma\,t$ and $2\pi y^-=\varphi$ with $\frac{\gamma}{4\pi^2}=4G|J|$. The locus $r=0$ (or, equivalently, $y=0$) bounds the region of spacetime with closed timelike curves. 

Motivated by our Schwarzian sector discussion in Section \ref{sec:Schw}, it is natural to explore the behavior of the Lorentzian FSC \eqref{eq:FSC1} when $M\to 0$ at fixed $J$, so that we explore low energies, as in \eqref{eq:xi-gap}. Since
\begin{equation}
    \hat{r}_+ = \sqrt{8GM} \qquad \qquad \frac{r_0}{G} = \sqrt{\frac{2}{GM}}\,|J|
\end{equation}
this corresponds to 
\begin{equation}
    \hat{r}_+\to 0\quad\textrm{and}\quad\frac{r_0}{G} \to \infty \quad \text{with} \quad \hat{r}_+ r_0=4G|J|\quad\text{fixed}\,.
\label{eq:gsch-lim}
\end{equation}
Thus, the horizon size $r_0$ grows. In the strict limit, it is pushed to infinity and the Lorentzian FSC \eqref{eq:FSC1} 
reduces to \eqref{eq:2flat-BTZ}. More importantly, the Lorentzian FSC thermodynamical potentials
\begin{equation}
    \beta_{\text{FSC}}=2\pi \frac{r_0}{\hat{r}_+^2}\qquad \qquad \Omega_{\text{FSC}} = \frac{\hat{r}_+}{r_0}
\end{equation}
satisfy the scaling relations
\begin{equation}
    (\beta\Omega)_{\text{FSC}}\sim \frac{1}{\hat{r}_+} \to \infty\qquad \qquad (\beta\Omega^2)_{\text{FSC}} \sim \frac{G}{r_0}\to 0
\end{equation}
when $M\to 0$. The latter corresponds to the Schwarzian sector discussed earlier.


\section{Connections and comparisons}\label{sec:7}

Since our results overlap with earlier work, in this Section, we comment on some of the existing literature. In Subsection \ref{sec:7.2}, we summarize key differences between induced and highest-weight representations and why they matter for CCFTs. In Subsection \ref{sec:flip}, we contrast the CCFT Schwarzian sector with the CFT Schwarzian sector, uncovering many similarities, and obtain the former as a specific limit of a relativistic CFT. In Subsection \ref{sec:valid}, we compare with earlier results on CCFT partition functions. Finally, we briefly comment on the literature regarding 3d gravity partition functions in \ref{sec:3d-grav}, thereby extending our previous Section.

\subsection{Induced vs.~highest-weight representations}\label{sec:7.2}

Consider CCFTs as Carrollian limits of CFTs. If the parent CFT$_2$ has finite Virasoro central charges without gravitational anomaly, $0<c=\bar c<\infty$, i.e., essentially for any CFT$_2$ in the ``yellow book'' \cite{diFrancesco}, the two different limits discussed in Subsection \ref{sec:3.1} yield rather different CCFTs, summarized in Table \ref{tab:2}. The upper half provides results for the standard highest-weight (HW) representations in the CFT$_2$, yielding the induced representations in the CCFT$_2$. The lower half shows analogous results for the flipped representations with the flipped central charges given in \eqref{eq:indhw6}. The last entry in the Table assumes for illustration that the classical Brown--Henneaux central charges $c=-\bar c=3\ell/(2G)$ both receive quantum corrections $c\to c+a$ and $\bar c\to\bar c+a$,\footnote{%
These quantum corrections are obtained from quantization of diff$(S^1)/U(1)$ Virasoro coadjoined orbits, yielding $a=1$, and diff$(S^1)/PSL(2,\mathbb{R})$ Virasoro coadjoined orbits, yielding $a=13$, see Section 5 in \cite{Cotler:2018zff}.
} leading to a non-zero Virasoro central charge $c_{\mt L}=2a$ after taking the CCFT limit. Non-vanishing $c_{\mt L}$ in a CCFT$_2$ implies there must have been a gravitational anomaly in the parent CFT$_2$. This is indeed the case for quantization in the flipped representation since the latter discriminates between left- and right-moving sectors. Therefore, quantum Einstein gravity (highlighted in italics in Table \ref{tab:2}) has $c_{\mt L}=0$ based on a CCFT with induced representations (third entry in Table) and $c_{\mt L}=2a$ based on a CCFT with highest-weight representations (last entry in Table).


\begin{table}[htb]
\centering
\begin{tabular}{|l|l|l|l|}
\hline
CFT$_2$ rep. & Virasoro central charge & CCFT$_2$ rep. & BMS central charge \\\hline
\textbf{HW} & $\boldsymbol{0<\bar c=c<\infty}$ & \textbf{Induced} & $\boldsymbol{c_{\mt L}=0}$\textbf{,} $\boldsymbol{c_{\mt M}=0}$ \\
HW & $0<\bar c<c<\infty$ & Induced & $c_{\mt L}=c-\bar c$, $c_{\mt M}=0$ \\
\textit{HW} & $c=\bar c\to+\infty$ & \textit{Induced} & $c_{\mt L}=0$, $c_{\mt M}=\textit{finite}$\\
HW & $c=-\bar c\to+\infty$ & Induced & $c_{\mt L}\to+\infty$, $c_{\mt M}=0$  \\\hline
\textbf{Flipped} & $\boldsymbol{0<\bar c=c<\infty}$ & \textbf{HW} & $\boldsymbol{c_{\mt L}=2c}$\textbf{,} $\boldsymbol{c_{\mt M}=0}$ \\
Flipped & $0<\bar c<c<\infty$ & HW & $c_{\mt L}=c+\bar c$, $c_{\mt M}=0$ \\
Flipped & $c=\bar c\to+\infty$ & HW & $c_{\mt L}\to+\infty$, $c_{\mt M}=0$ \\
Flipped & $c=-\bar c\to+\infty$ & HW & $c_{\mt L}=0$, $c_{\mt M}=\textrm{finite}$  \\
\textit{Flipped} & $c-\bar c\to+\infty$, $c+\bar c=2a$ & \textit{HW} & $c_{\mt L}=2a$, $c_{\mt M}=\textit{finite}$  \\
\hline
\end{tabular}
\caption[Induced \& highest-weight representations in CCFT$_2$ as limit of CFT$_2$]{Induced and highest-weight representations in CCFT$_2$ as limit of CFT$_2$. \textbf{Bold:} vanilla CFT$_2$ and its CCFT$_2$ limit. \textit{Italics:} flat space Einstein gravity (with quantum corrected central charges)}
\label{tab:2}
\end{table}

The case concerning vanilla CFT$_2$ is highlighted in bold-faced letters in the Table. For such CFTs, we need to use the flipped vacuum \eqref{eq:indhw2} to get a CCFT with at least one non-vanishing central charge. Indeed, explicit examples of quantum CCFTs such as the free BMS boson \cite{Hao:2021urq} or the free BMS fermion (a.k.a.~BMS--Ising model) \cite{Yu:2022bcp,Hao:2022xhq,Banerjee:2022ocj,Bergshoeff:2023vfd} typically involve the flipped vacuum in the parent CFT and always lead to non-zero $c_{\mt L}$ twice the value of the parent Virasoro central charge.\footnote{
The best-known example of a CCFT with $c_{\mt M}\neq0$ where induced representations appear naturally is the classical BMS--Liouville model \cite{Barnich:2012rz}. Using instead highest-weight representations yields an additional quantum shift for $c_{\mt L}$ from $0$ to $2$ \cite{Merbis:2019wgk}, see the last entry in Table \ref{tab:2}.
}

As we shall demonstrate in the next Subsection, the existence of a Schwarzian sector in a CCFT also relies on the flipped representation in the parent CFT. Formulated intrinsically, the Schwarzian sector exists for CCFT highest-weight representations but not for CCFT induced representations. 

\subsection{Schwarzian sectors: From relativistic to Carroll CFTs}
\label{sec:flip}

Comparing our CCFT$_2$ Schwarzian partition function \eqref{eq:badguy}, reexpressed here in terms of $\beta$ and $\theta$ as 
\eq{
Z_{\textrm{\tiny Schwarzian}}^{\textrm{\tiny CCFT}}(\beta,\,\theta) \approx \frac{(2\pi)^5}{(-i\theta)^3}\, \exp\bigg[\frac{-i\theta}{12}+\frac{\pi^2}{6}\,\bigg(\frac{c_{\mt{L}}-2}{-i\theta} + \frac{c_{\mt{M}}\beta}{\theta^2}\bigg)\bigg]
}{eq:GMT0}
with the CFT$_2$ Schwarzian partition of \cite{Ghosh:2019rcj},
\eq{
Z_{\textrm{\tiny Schwarzian}}^{\textrm{\tiny CFT}}(\beta,\,\theta) \approx \frac{(2\pi)^{5/2}}{(\beta-i\theta)^{3/2}}\, \exp\bigg[\frac{\beta-i\theta}{24}+\frac{\pi^2}{6}\,\frac{c-1}{\beta-i\theta}\bigg]
}{eq:GMT1}
we see some striking resemblance. To show that this is no coincidence, we consider here a CFT$_2$ limit that directly obtains \eqref{eq:GMT0}. 

To obtain the CCFT answer in the limit, we need to start with the flipped representation (reviewed in the Appendix \ref{app:B}), since taking highest-weight characters in both Virasoro sectors does not yield a well-defined limit as explained in Subsection \ref{sec:3.optional}. 

We start with a CFT$_2$ partition function that is dominated by the Virasoro vacuum character in the dual channel [see Eq.~\ref{eq:appB8}],\footnote{%
Vacuum dominance is achieved here for $\theta\to+i\infty$ and arbitrary $\beta$. In the CCFT context, $\beta$ is negative. As a consequence of $\theta\to+i\infty$, both $\tau$ and $\bar\tau$ lie in the complex upper half-plane. 
} 
\eq{
\tilde Z^{\textrm{\tiny CFT}}(\tau,\,\bar\tau) \approx \chi_{c,0}^{\textrm{\tiny HW}}(-1/\tau)\,\bar\chi_{\bar c,0}^{\textrm{\tiny LW}}(-1/\bar\tau)
}{eq:GMT2}
with the modular parameters
\eq{
\tau = \frac{1}{2\pi}\,\big(\theta+i\epsilon\beta\big)\qquad\qquad \bar\tau = \frac{1}{2\pi}\,\big(\theta-i\epsilon\beta\big)
}{eq:GMT3}
where we anticipated already the Carrollian rescaling of inverse temperature by the contraction parameter $\epsilon$ that we intend to send to 0 from above. In addition to the standard highest-weight Virasoro vacuum character 
\eq{
\chi_{c,0}^{\textrm{\tiny HW}}(\tau)= \frac{e^{-2\pi i\tau\frac{(c-1)}{24}}}{\eta(\tau)}\,\big(1-e^{2\pi i\tau}\big)
}{eq:GMT4}
we use above the lowest-weight Virasoro vacuum character \eqref{eq:appB7} (obtained from \eqref{eq:GMT4} by replacing on the right-hand side $\tau\to\bar\tau$ and $c\to-\bar c$) 
\eq{
\bar\chi_{\bar c,0}^{\textrm{\tiny LW}}(\bar\tau)= \frac{e^{2\pi i\bar\tau\frac{(\bar c+1)}{24}}}{\eta(\bar\tau)}\,\big(1-e^{2\pi i\bar\tau}\big)
}{eq:GMT5} 
because we want to smoothly obtain the CCFT highest-weight representation \eqref{hw}, as explained in the Subsection \ref{sec:3.optional}. The Virasoro central charges can be recast in terms of the CCFT central charges as
\eq{
c=\frac{1}{2\epsilon}\,c_{\mt M}+\frac12\,c_{\mt L} \qquad\qquad 
\bar c=\frac{1}{2\epsilon}\,c_{\mt M}-\frac12\,c_{\mt L} \,.
}{eq:GMT6}
Our claim is
\eq{
Z_{\textrm{\tiny Schwarzian}}^{\textrm{\tiny CCFT}}(\beta,\,\hat\theta) =  \lim_{\epsilon\to 0} \lim_{\hat\theta \gg 1}
\tilde Z^{\textrm{\tiny CFT}}(\tau,\,\bar\tau) 
}{eq:GMT7}
where $\hat\theta=-i\theta\in\mathbb{R}^+\gg 1$. To show this, we insert the relation \eqref{eq:eta6} into the CFT partition function \eqref{eq:GMT2} exploiting $\hat\theta\gg 1$, with the characters \eqref{eq:GMT4}, \eqref{eq:GMT5}, yielding
\eq{
\tilde Z^{\textrm{\tiny CFT}}(\tau,\,\bar\tau) \approx \frac{2\pi i}{\tau}\,\frac{2\pi i}{\bar\tau}\,\frac{e^{-\frac{i\pi\tau}{12}}e^{-\frac{i\pi\bar\tau}{12}}}{\sqrt{-i\tau}\,\sqrt{-i\bar\tau}}\,\exp\Big(\frac{2\pi i}{\tau}\,\frac{c-1}{24}-\frac{2\pi i}{\bar\tau}\,\frac{\bar c+1}{24}\Big)\,.
}{eq:GMT9}
Plugging in the definitions of the modular parameters \eqref{eq:GMT3} and the central charges \eqref{eq:GMT6} and expanding in $\epsilon$ obtains
\eq{
\tilde Z^{\textrm{\tiny CFT}}(\beta,\,\hat\theta) \approx \frac{(2\pi)^5}{\hat\theta^3}\,\exp\Big(\frac{\hat\theta}{12}+\frac{\pi^2}{6\hat\theta}\,\big(-c_{\mt M}\frac{\beta}{\hat\theta}+c_{\mt L} -2 
+ {\cal O}(\epsilon)\big)\Big)
}{eq:GMT10}
where notably the $1/\epsilon$ terms in the exponent cancel precisely. Taking the limit $\epsilon\to 0$ of \eqref{eq:GMT10} and replacing $\hat\theta=-i\theta$ establishes our claim above.

Note that for the induced representations, a construction similar to the one above would not work since for imaginary $\theta$, one of the $\eta$-functions will always have an argument that is not in the complex upper half-plane. In other words, there is no Schwarzian sector in a CCFT that is built on induced representations. By contrast, induced representations naturally lead to the Boltzmann sector discussed in Section \ref{sec:positive}. We have discussed these aspects in Subsections \ref{sec:3.optional} and \ref{sec:4.sectors}, and summarized these statements, together with a corresponding one for the Cardy sector, in Table \ref{tab:3}.

\subsection{Validity of Carrollian partition functions}\label
{sec:valid}

A superficial look at the literature on Carrollian partition functions and the quantization of CCFTs may lead to puzzling conclusions: on the one hand, Carrollian partition functions and thermodynamics are considered to be ill-defined \cite{deBoer:2023fnj} and the quantum Carroll theories might be either trivial or non-existing \cite{Cotler:2024xhb}, on the other hand, the Cardy formulas and partition functions summarized in this work (and related earlier work) make sense both on the field theory side and on the gravity side, there is a version of thermodynamics that works (albeit with unusual signs in the first law and specific heat), and we know explicit examples of quantum CCFTs, like a free BMS invariant scalar field \cite{Hao:2021urq} or the BMS invariant version of the Ising model \cite{Yu:2022bcp,Hao:2022xhq}. For details, we refer to the first six Sections of the current paper and Refs.~therein.

Here, we attempt a more refined look at the facts, with the intention to clarify these apparent contradictions. One crucial aspect is whether or not the angular potential $\theta$ is assumed to be real. If it is real, we have to face the technical issue highlighted in Section \ref{sec:positive} on the Boltzmann sector: the argument of the Dedekind $\eta$-function becomes real and hence the partition function is ill-defined. This issue was first pointed out in Oblak's work on characters of the BMS group \cite{Oblak:2015sea}. The resolution proposed therein is essentially the resolution we used in Section \ref{sec:positive}, namely analytic continuation $\theta\to\theta+i\epsilon$ together with the limit $\epsilon\to0^+$.\footnote{%
This regularization reintroduces an effective AdS radius $\sim 1/\epsilon$, so secretly the partition function is associated with AdS$_3$ rather than with flat space \cite{Prohazka:pc}.
} 
Physically, the source of the infinities associated with the ill-defined Dedekind $\eta$-function can be attributed to the mundane IR divergences on the gravity side --- Minkowski space has an infinite volume.

Recently, these IR divergences were resolved differently, by explicitly isolating them through a volume term of the form $\delta(0)$ \cite{Cotler:2024cia}. Our conclusions of Section \ref{sec:positive} agree with their analysis. In particular, there is a dichotomy in the partition function depending on whether $\theta$ is rational or irrational. The finite part of the partition function has a meaningful thermodynamical interpretation summarized in Section \ref{sec:positive}.

In summary, for real $\theta$, the naive partition function is ill-defined, but it is physically clear why, namely because of the infinite volume of Minkowski space. The result for the partition function depends on whether $\theta$ is rational or irrational.

For the Schwarzian sector considered in the present work, we used instead imaginary $\theta$. In this case, none of the issues above arise since the Dedekind $\eta$-function is well-defined in the complex upper half-plane. It was independently suggested in \cite{Poulias:2025eck} to consider imaginary chemical potentials as a way to get a well-defined partition function. In the present work, we have seen that this is not a choice but forced upon us if we want to zoom into the Schwarzian sector or the Cardy sector. However, this only appears to work for the highest-weight representations and not for the induced representations. Another potentially important aspect in defining Carroll partition functions is that the Hamiltonian is central in the Carroll algebra \cite{deBoer:2023fnj}. Here, we emphasize that our focus is on the conformal Carroll algebra, where the representation theory differs significantly, and the Hamiltonian is no longer central.

\subsection{Comparison with 3d gravity}
\label{sec:3d-grav}

The conformal Carrollian partition function \eqref{partition_function} is determined by the conformal Carrollian symmetries and the choice of representations (induced vs.~highest-weight). Given the BMS$_3$/CCA$_2$ isomorphism, there can be a gravity interpretation of our field theory results, some of which we uncovered in Section \ref{sec:holo}. 

In this Subsection, we consider some aspects of the gravitational path integral --- divergences, their regularization, different 1-loop results that exist in the literature, and how these calculations can be reproduced within a CCFT$_2$ setting.

The 1-loop partition function of 3d gravity around Minkowski with vanishing angular velocity $(\Omega=0)$ was first computed in \cite{Barnich:2015mui} using heat kernel methods. The result
\eq{
Z_{\mt{1-loop}}(\beta) = e^{S_0} \lim_{\sigma\to i0^+}\frac{|e^{i\pi\sigma/12}|^2}{|\eta(\sigma)|^2}\,\big|1-e^{2\pi i\sigma}\big|^2
}{eq:Blaja1loop}
depends on the Euclidean action $I_{\mt{E}}$ through $S_0=-I_{\mt{E}}=\frac{\beta}{8G}$ and coincides precisely with the field theory partition function \eqref{eq:lamourdemavie} with $D(0,0)=1$, $D(\xi,\Delta)=0$ otherwise, $c_{\mt M}=3/G$, and $c_{\mt L}=0$. Note that the Dedekind $\eta$-function evaluated at zero vanishes [see Eq.~\eqref{eq:eta6}], 
\eq{
\lim_{\epsilon\to 0^+}\eta(i\epsilon)=\lim_{\epsilon\to 0^+} \frac{1}{\sqrt{\epsilon}}\,e^{-\frac{\pi}{12\epsilon}} = 0
}{eq:etaiepsilon}
and hence the partition function \eqref{eq:Blaja1loop} has a divergent (but state-independent) factor exactly like in the Boltzmann sector, see Section \ref{sec:positive}. The same work already anticipated that the ``partition function $Z[\beta,\theta]$ is most naturally viewed as a function of complex angular potential,'' which we exploited systematically in our discussion of Cardy- and Schwarzian sectors.

An alternative more recent calculation  \cite{Cotler:2024cia} yields the 1-loop partition function\footnote{%
The restriction to $n>1$ stems from the symmetries of Minkowski. In terms of modes, one must subtract the contribution from $n=0$ and $n=\pm 1$, with $n=\pm 1$ contributing in the same way, i.e., $(\beta/G)^{-1/2}$.}
\eq{
    Z_{\mt{1-loop}}(\beta) = e^{S_0}\,\delta(0)\,\prod_{n>1} \frac{n^2-1}{8\pi\beta/G}
}{eq:1loop}
with the same expression for $S_0$ as in \eqref{eq:Blaja1loop}. The divergence above is represented by $\delta(0)$, which may be interpreted as an IR divergence due to the non-compactness of Minkowski space. Additionally, the infinite product in \eqref{eq:1loop} requires regularization, see below. 

Analogous divergences were encountered in \cite{Merbis:2019wgk} using geometric action methods. References \cite{Barnich:2015mui,Merbis:2019wgk} used different regularizations, both yielding sensible results, but when matching the regularized actions with BMS characters, they yield different quantum shifts of the central charges. In particular, \cite{Merbis:2019wgk} obtains
\eq{
Z_{\mt{1-loop}}(\beta) = e^{S_0} \lim_{\sigma\to 0}\frac{e^{-2\pi i\sigma}}{\eta(\sigma)^2}\,\big(1-e^{2\pi i\sigma}\big)^2
}{eq:Wout1loop}
which differs from \eqref{eq:Blaja1loop} by the absence of absolute values and agrees with the highest-weight character \eqref{eq:charactervacuum} for $c_{\mt L}=26$. 

According to our Table \ref{tab:2}, we understand the differences between \eqref{eq:Blaja1loop} and \eqref{eq:Wout1loop} as coming from the different representations that were used in the CCFT description. While somewhat implicit in \cite{Barnich:2015mui} due to their efficient use of heat kernel methods and zeta-function regularization, their final result in their last equation clearly shows the appearance of a CCFT$_2$ character in the induced representation. By contrast, the calculations in \cite{Merbis:2019wgk} that led to a quantum shift of the central charge $c_{\mt L}$ by $26$ used the highest-weight representations. These results agree precisely with the respective italicized lines in Table \ref{tab:2}. Thus, our explanation for the different results of the 1-loop partition functions in \cite{Barnich:2015mui,Cotler:2024cia} and \cite{Merbis:2019wgk} is that the first set of Refs.~employed induced representations while the last Ref.~used highest-weight representations.

The geometric action methods developed in \cite{Merbis:2019wgk} were used in \cite{Cotler:2024cia,Simon:2024dwm} to prove that the 1-loop contribution to the Minkowski and flat space cosmologies partition function is exact. In addition, these works pointed out the existence of a non-trivial structure depending on whether $\beta\Omega$ was rational, irrational, or purely imaginary, a structure that we recovered in our discussion of Carrollian vacuum dominance in the dual channel in Section \ref{sec:4}. 

One may expect that stripping the divergent part in \eqref{eq:1loop} due to the non-compactness of Minkowski space yields physical answers. However, observables may still depend on the regularization of the infinite tower of modes. To explore this, let us compute the specific heat. The latter is negative when vacuum dominance in the dual channel holds, as shown in Subsection \ref{sec:neg-heat}. Let $Z_\infty$ stand for the divergent $\delta (0)$ and the $\beta$ independent contribution from the $n>1$ modes in \eqref{eq:1loop}. Since
\eq{
    \ln Z_{\mt{1-loop}} = S_0 - \Big(\ln \frac{\beta}{G}\Big)\,\sum_{n>1}^\infty 1 + \ln Z_\infty
}{eq:anotheronebitesthedust}
it follows the expectation value of the energy equals
\begin{equation}
    \langle \xi \rangle = -\partial_\beta \ln Z_{\mt{1-loop}} = -\frac{1}{8G} + \frac{1}{\beta}\,\sum_{n>1}^\infty 1\,.
\end{equation}
This has two contributions: a $\beta$-independent one from the Euclidean action and a $\beta$-dependent and divergent one from the $n>1$ modes. Zeta function regularization, $\sum_{n=1}^\infty = -\frac{1}{2}$, yields
\eq{
    \left.\langle \xi \rangle\right|_{\mt{reg}} = -\frac{1}{8G} - \frac{3}{2\beta} \qquad \Rightarrow \qquad  \left.\frac{\partial \langle \xi \rangle}{\partial T}\right|_{\mt{reg}} = -\frac{3}{2}\,. 
}{eq:yourpower}
Thus, after regularization, the specific heat is negative, compatible with our general analysis in Section \ref{sec:neg-heat}. Note, however, that we cannot directly compare the regularized result \eqref{eq:yourpower} with our general expression for specific heat \eqref{eq:illestofourtime} since the latter assumes $\Omega\neq 0$.


\section{Conclusions}
\label{sec:6}

Before addressing some of the open questions implied by our paper, we recap with a guided tour through its boxed equations, Tables, and Figures.

\subsection*{Reprise}

The Carroll conformal symmetries \eqref{bms} and the Carroll modular transformations \eqref{eq:angelinajolie} are the key technical ingredients to write Carroll partition functions as sums over highest-weight characters \eqref{partition_function} or induced characters \eqref{eq:lamourdemavie} and to establish the necessary condition for vacuum dominance in the dual channel \eqref{eq:vac-dom}, the analysis of which leads to six different sectors summarized in Table \ref{tab:1}. 

In particular, the Schwarzian sector ($\theta$ is the angular potential and $\beta$ inverse temperature), 
\eq{
\textrm{Schwarzian\;sector\;in\;CCFT}_2:\qquad -i\theta\gg 1 \qquad -\beta\gg 1 
}{eq:final}
produces the partition function \eqref{eq:badguy}, which splits into a LO term and a fluctuation term, the Carroll--Schwarzian partition function \eqref{eq:carrollschwarzian}. Remarkably, the Schwarzian sector requires the use of highest-weight representations rather than induced ones, see Tables \ref{tab:3} and \ref{tab:2}. 

Assuming a positive energy spectrum of primaries, the Boltzmann sector is the only such sector with positive temperature, but it comes with technical difficulties encoded in the Dedekind $\eta$-function evaluated at real values, see Fig.~\ref{fig:Z}. 

Physically, a striking aspect of all sectors is the negativity of specific heat \eqref{eq:Cxi-vac}, which has a natural holographic interpretation on the gravity side: asymptotically flat spacetimes with horizons typically have negative specific heat, including Kerr black holes and flat space cosmologies as prominent examples. 

Exploiting the analytic structure of the various partition functions in the complex $\theta$-plane (see Fig.~\ref{fig:theta}), we also discussed ensembles different from the grand canonical one (depending on inverse temperature $\beta$ and angular potential $\theta)$, especially the microcanonical ensemble, where the partition function \eqref{eq:lostallfaith} correctly reproduces the BMS-Cardy entropy \eqref{eq:screechstale}. The microcanonical partition function is recovered from the Carroll--Schwarzian density of states \eqref{eq:thediner} employing the Carroll modular S-matrix.

Holographically, the gravity side of the Schwarzian sector differs from the better-known flat space cosmologies (see Fig.~\ref{fig:Penrose}) and instead has an interpretation through $O$-folds \eqref{eq:2flat-BTZ}, see Fig.~\ref{fig:schematic}.

\subsection*{Comments on induced vs.~highest-weight representations}

The relevance of the two types of representations --- induced or highest-weight --- of the dual CCFT$_2$ for holography in 3d asymptotically flat spacetimes has been debated for a while in the literature. The most obvious argument is that since induced representations are explicitly unitary, while the highest-weight representations are not, it is the induced representation that should play more of a role in the construction of flat holography. Interestingly, though, most calculations for matching bulk and boundary observables had explicitly or implicitly used highest weights, e.g.~\cite{Bagchi:2012xr, Bagchi:2014iea}. 

Our present constructions leave little room for doubt that the highest-weight representations of the dual CCFT$_2$ are vitally important for 3d flat holography. These are the representations that describe the Schwarzian sector and hence the ``extremal'' FSCs, or, more precisely, the $O$-folds are holographically dual to CCFT$_2$ in these representations. The Cardy sector is also included in the highest-weight representations. It is thus natural to conjecture that the non-trivial orbifolds of 3d Minkowski spacetimes, the analogs of the BTZ black holes in the AdS case, fall into highest-weight representations of the dual CCFT. 

On the other hand, the induced representations appear to be more suited for scattering problems \cite{Hao:2025btl} and for the Boltzmann sector described in Section \ref{sec:positive}. Thus, it is possible that a complete holographic description of asymptotically flat spacetimes needs to employ both types of representations, the induced ones to describe scattering and the highest-weight ones to describe thermal states, e.g., in the Cardy and Schwarzian sectors.

Another pressing issue is the apparent non-unitarity of the highest-weight representations in the context of holography. It is worth noting that also for WCFTs, the Schwarzian sector of the theory is connected to non-unitary representations \cite{Aggarwal:2022xfd}. We comment more on WCFTs below. On the other hand, if all the non-trivial zero-mode solutions of 3d asymptotically flat spacetimes are, indeed, dual to highest-weight representations of the CCFT$_2$, it is worthwhile trying to investigate if one needs to modify the definition of unitarity for these theories. A possible clue lies in the definition of conjugation and in- and out-states in a CCFT$_2$. On the complex plane, the reality condition $\bar{z} = z^\ast$ is not preserved in the Carroll limit, and one needs to consider a different reality condition $\bar{z} = \frac{1}{z^\ast}$ \cite{Hao:2021urq}. This may provide hints at a different notion of unitarity in 2d CCFTs. See also \cite{Hao:2025btl} for a curious definition of conjugates, albeit for the induced representations. 

There is a related and equally, if not more, intriguing point that emerges from our discussion of highest-weight vs.~induced representations in flat holography. We have seen that the induced representations naturally follow from the highest-weight representations of two copies of the Virasoro algebra. The Carroll highest weights, however, descend from the ``flipped'' Virasoro representations, where one half is highest-weight while the other half is lowest-weight. While the usual AdS$_3$/CFT$_2$ correspondence is built on the well-known highest-weight representations of the Virasoro algebra, the immediate question is what role these flipped representations have to play within AdS$_3$ holography. 

The Cardy formula of the dual theory relates the Virasoro highest weights to the area of the outer horizon of the BTZ black hole. Interestingly, the Cardy analysis with the flipped representations relates them to the \textit{inner} horizon of the non-extremal BTZ. It is then natural that the flat space limit relates the Carroll highest-weight representations to the horizon area of the FSC, which itself comes as a limit of this inner horizon \cite{Riegler:2014bia,Fareghbal:2014oba}. This raises the tantalizing prospect that one can go beyond the outer horizon to inside the BTZ black hole and access the inner horizon by looking at the flipped representations of the dual CFT$_2$. This hints at the fact that the dual CFT$_2$ knows about the physics in the interior of black holes, accessible through an automorphism of the symmetry algebra. 

\subsection*{Relations to WCFTs}

It is interesting to compare and contrast the near-extremal limits considered in this paper with those of closely related 2d field theories already mentioned in the introduction. Carrollian CFTs encompass a broader class of field theories, for which the conformal isometry equations (\ref{eq:ccft1}) can be generalized to allow for an anisotropic scaling between the degenerate two-tensor field $h_{\mu\nu}$ and nowhere-vanishing kernel vector field $\tau^\mu$ \cite{Duval2014a}. The special case where the kernel vector does not scale corresponds to WCFTs (up to global subtleties \cite{Despontin:2025dog}). The main difference between these theories and the CCFTs considered in this paper is that the Virasoro generators $L_n$, instead of being supplemented with the spin-2 modes $M_n$, come along with spin-1 (current) generators $P_n$. Near-extremal limits of WCFTs were studied in \cite{Aggarwal:2022xfd}, and we are now in the position to highlight similarities and differences between these two setups. 

The Cardy sector in WCFTs\footnote{%
We are focusing here on WCFTs in the so-called \textit{canonical ensemble}, closer in structure to CCFTs, in opposition to the \textit{quadratic ensemble}. We refer the reader to \cite{Aggarwal:2022xfd} for more details. } corresponds to $\beta \Omega \rightarrow 0^+$. When $L_0$ is bounded from below, this allows to project the partition function onto the vacuum character. This is precisely (one instance of) the Cardy sector for CCFTs described in Table \ref{tab:1} (first line, second choice). 

New conditions for vacuum dominance, and the appearance of a (warped) Schwarzian sector for WCFTs were identified in \cite{Aggarwal:2022xfd} with a somewhat unexpected result: the latter could be achieved in the limit $\Omega \rightarrow 0^+$, but \textit{only in a non-unitary theory with imaginary vacuum value $P_0^{vac}$}. In essence, this allowed to project the partition function on the state bounding the imaginary $P_0$ charge. The further limit $\beta \Omega \rightarrow \infty$ then gives rise to the Schwarzian-like sector. This resonates with our findings for CCFTs: vacuum dominance, then emergence of a Schwarzian-like sector, require highest-weight representations, which are non-unitary in the context of holography, along with condition $\beta \Omega^2 \rightarrow 0^-$ followed by $\beta \Omega \rightarrow \infty$. It could be interesting to investigate both the relevance of induced representations for WCFTs and the possibility of having imaginary vacuum charges for CCFTs. Furthermore, near-extremal sectors could further exist for the more general anisotropic Conformal Carroll field theories appearing in \cite{FarahmandParsa:2018ojt, Chen:2019hbj, Despontin:2025dog}.

\subsection*{Loose ends}

We conclude with a couple of loose ends and further possible checks. It could be rewarding to analyze the $O$-folds in more detail on the gravity side. In particular, it would be interesting to calculate the (holographically renormalized) Euclidean on-shell action associated with $O$-fold saddle points and to verify under which conditions they dominate the Euclidean partition function in the semi-classical approximation. Moreover, the evaluation of this partition function must match our Schwarzian partition function \eqref{eq:badguy} if flat space holography works as expected.

A more technical loose end is to make formal sense of the saddle point calculations in Appendix \ref{app:A}. This might be possible by a suitable analytic continuation of the integration contour.

Finally, it seems worthwhile to investigate whether one can derive on the field theory side (or at least motivate from the gravity side) from first principles the property
\eq{
i\theta^3 \propto \beta 
}{eq:reallyfinal}
that led to the gravity-like log corrections to the entropy  \eqref{eq:everythingiwanted} in the Schwarzian sector of 2d Carrollian CFTs. 

\bigskip

\begin{center}
{\LARGE{\EOstar}}
\end{center}

\newpage
\enlargethispage{0.5truecm}


\section*{Acknowledgments} \addcontentsline{toc}{section}{Acknowledgments}

\paragraph{Opening remarks.}
This project has been \textit{long} and we have many people to thank! We are grateful to Harald Skarke for sharing his insights on modular transformations in discussions at TU Wien in 2023, which entered the first three paragraphs of Section \ref{sec:2.4}. 

We thank Aritra Banerjee, Nabamita Banerjee, Glenn Barnich, Rudranil Basu, Jordan Cotler, Matthew Headrick, Shruti Menon, Mark Mezei, Blagoje Oblak, Sabrina Pasterski, Stefan Prohazka, Romain Ruzziconi, Amartya Saha, Muktajyoti Saha, Jakob Salzer, Atul Sharma, Shahin Sheikh-Jabbari, Stefan Vandoren, Dima Vassilevich, Girish Vishwa, Gabriel Wong, and Akshay Yelleshpur Srikant for useful discussions on various aspects of CCFT$_2$ and 3d flat space. 

AA and DG additionally thank the participants of the ``Carroll Group Meetings'' at TU Wien, March--June 2025, namely Alois Altenburger, Florian Ecker, Logan Fisher, Lucas H\"orl, Matt\'eo Leturcq-Daligaux, Iva Lovrekovic, Saikat Mondal, Luciano Montecchio, Mohanna Shams-Nejati, and Kaiyu Zhang for numerous presentations and discussions. 

\paragraph{Funding information.} AA, DG, and MR were supported by the Austrian Science Fund (FWF) [Grants DOI: \href{https://www.fwf.ac.at/en/research-radar/10.55776/P32581}{10.55776/P32581},
DOI: \href{https://www.fwf.ac.at/en/research-radar/10.55776/P33789}{10.55776/P33789}, and DOI: \href{https://www.fwf.ac.at/en/research-radar/10.55776/P36619}{10.55776/P36619}]. AA was partially supported by the Fonds de la Recherche Scientifique F.R.S.-FNRS (Belgium), IISN – Belgium (convention 4.4503.15), and by the Delta ITP consortium, a program of the NWO that is funded by the Dutch Ministry of Education, Culture and Science (OCW). AA and DG were supported by the OeAD travel grant IN 04/2022 and the grant DST/IC/Austria/P-9/202 (G). This research was conducted while visiting the Okinawa Institute of Science and Technology (OIST) through the Theoretical Sciences Visiting Program (TSVP).  

AB was partially supported by a Swarnajayanti Fellowship from the Science and Engineering Research Board
(SERB) of India under grant SB/SJF/2019-20/08 and also by an ANRF grant CRG/2022/006165. 
AB also acknowledges the support of the Royal Society of London for two international exchange grants with the University of Edinburgh (2022-2024) and Durham University (2024-2026). AB and DG acknowledge support from the ESI Research in Teams project ``Chaos, Butterflies, and Entanglement in Flat Space'' in June/July 2022.

SD is a Senior Research Associate of the Fonds de la Recherche Scientifique F.R.S.-FNRS (Belgium).
He acknowledges support of the Fonds de la Recherche Scientifique F.R.S.-FNRS (Belgium) through the following projects: CDR project C 60/5 - CDR/OL ``Horizon holography : black holes and field theories'' (2020-2022), PDR/OL C62/5 ``Black hole horizons: away from conformality'' (2022-2025) and CDR n$^\circ$40028632 (2025-2026). This work is supported by the F.R.S.-FNRS (Belgium) through convention IISN 4.4514.08 and benefited from the support of the Solvay Family. SD is a member of BLU-ULB, the interfaculty research group focusing on space research at ULB.

JS was supported by the Science and Technology Facilities Council [grant numbers ST/T000600/1, ST/X000494/1].

\paragraph{Venue acknowledgments in preparation phase.} AB, SD, and DG thank Marc Henneaux and the hospitality of Coll\`ege de France, Paris, in May 2022, where discussions about the current project were initiated.

AB thanks CPHT, Ecole Polytechnique, Paris, and NORDITA, Stockholm, for visiting professorships, and NBI Copenhagen for support during his sabbatical year (2022-23). He is grateful to the theory groups and individuals in these institutes for wonderful discussions on various aspects of flat holography. AB also acknowledges the support of ULB Brussels and TU Wien during various visits in these couple of years.

AB and DG thank the organizers of the 2nd Carroll Workshop, Nicolas Boulanger, Andrea Campoleoni, Laura Donnay, Adrien Fiorucci, Yannick Herfray, and Romain Ruzziconi, for the hospitality in Mons in September 2022.

DG thanks the Simons Collaboration on Celestial Holography for travel support and the participants of the Kickoff Workshop at Harvard University in October 2023 for extensive and insightful remarks on flat space holography. 

AB, DG, and JS thank the ICMS and the organizers, Jos\'e Figueroa-O'Farrill and Jelle Hartong, for the hospitality during the workshop ``Beyond Lorentzian Geometry II'' in Edinburgh in February 2023. 

AB and DG thank Nordita and the organizers, Niels Obers, Gerben Oling, and Ziqi Yan, for the hospitality during the workshop ``Non-Relativistic Strings and Beyond'' in Stockholm in May 2023. 

DG acknowledges an OIST TSVP scholarship and thanks Yasha Neiman and his group for the hospitality and the students and postdocs for asking numerous questions in his lecture series on asymptotic symmetries in July/August 2023 in Okinawa. 

DG thanks the organizers of the 3rd Carroll Workshop, in particular, Adrien Fiorucci, Anastasious Petkou, and Konstantinos Siampos, for the excellent atmosphere and the stimulating discussions in Thessaloniki in October 2023.

DG thanks Luca Ciambelli and C\'eline Zwikel for the invitation to Perimeter Institute in October 2023, and additionally thanks Jacqueline Caminiti, Rob Myers, and Sabrina Pasterski for hours of discussions on various aspects of flat space holography, especially in three bulk dimensions.

AA, AB, SD, DG, and MR thank the Erwin--Schr\"odinger Institute (ESI) for the hospitality in April 2024 during the program ``Carrollian physics and holography'' and we are grateful to the participants of this program for numerous insightful discussions.

\paragraph{Venue acknowledgments in presentation phase.} AA and JS would like to thank the organizers of the Solvay workshop on ``Near-Extremal Black Holes and Holography'' for the invitation in September 2024 and the participants for many interesting discussions; AA presented a preliminary version of this work at this workshop. 

DG thanks Ahmed Almheiri and his team for organizing Strings in Abu Dhabi in January 2025 and for the generous invitation to present the Carrollian approach to flat space holography. 

AA thanks the String theory group at ICTS-TIFR for the invitation and hospitality, where a preliminary version of this work was presented in February 2025. He would especially like to thank Ramesh Ammanamanchi, Nava Gaddam, R. Loganayagam, and Ashoke Sen for interesting discussions during this visit.

AA, AB, SD, and DG acknowledge travel support by BITS Pilani, Goa, and thank the participants of the workshop ``Holography, strings and other fun things II'' for enjoyable discussions in February 2025. 

AA and DG thank the organizers, Eric Bergshoeff and Andrea Fontanella, and the participants of the workshop ``Non-Lorentzian Geometries and their Applications'' at the Hamilton Mathematics Institute, Trinity College Dublin, in April-May 2025 for the hospitable environment.

DG presented results of this work at the GGI conference ``From Asymptotic Symmetries to Flat Holography: Theoretical Aspects and Observable Consequences'' in June 2025 and acknowledges local support from GGI. He thanks the organizers, Andrea Campoleoni, Laura Donnay, Dario Francia, Sabrina Pasterski, Andrea Puhm, and Simone Speziale, and the participants for an inspiring ambience and extensive discussions.

DG thanks the organizers, Niklas Garner, Lionel Mason, Romain Ruzziconi, and Akshay Yelleshpur-Srikant, and the participants of the workshop ``From Good Cuts to Celestial Holography'' by the Oxford Geometry Group in July 2025 for the hospitality at St.~Antony's College, the punt, and The Perch.

JS would like to thank the organizers and participants of the workshops ``Gravity -- New quantum and string perspectives'' in Benasque (Spain) from July 7-18th, 2025 and ``Quantum Gravity, Holography, Strings \& Quantum Information,'' at the International Institute of Physics in Natal (Brazil) from July 21st to August 1st, 2025, for their hospitality and feedback during the final stages of this project.


\appendix

\section*{APPENDIX}


\section[\texorpdfstring{\!}{}saddle point approximation]{\texorpdfstring{\!\!\!\!}{}saddle point approximation}\label{app:A}

In this Appendix, we evaluate various integrals appearing in the Schwarzian sector in the saddle point approximation
\eq{
\int_C {\rm d}x\, e^{N K(x)}\approx e^{N K(x_{0})}\,\sqrt{\dfrac{2\pi}{N |K''(x_0)|}}
}{eq:troubles}
where $C$ is a suitable contour, $N$ is large, and $x_0$ is the saddle point, $K'(x_0)=0$. We apply this to the fixed $J$ ensemble in Section \ref{sec:J} and to the microcanonical ensemble in Section \ref{app:micro}.

The above equation is only valid if $K''(x_0)<0$, which is not the case here when $\hat\theta_0>0$. This is related to negativity of specific heat discussed in Section \ref{sec:neg-heat}. So, we cannot technically use the saddle point approximation since the Gaussian integral would diverge. Nevertheless, if we still use the approximation, we get reasonable results that we present below, suggesting there is a suitable analytic continuation of the integral contour $C$. In the main text, we present a way to avoid the saddle point approximation in the microcanonical ensemble by doing the integral exactly for a suitable contour in Section \ref{sec:micro}, where we first go to a fixed $M$ ensemble and then to a fixed $J$ ensemble. 

\subsection[Fixed \texorpdfstring{$J$}{J} ensemble]{Fixed \texorpdfstring{$\boldsymbol{J}$}{J} ensemble}\label{sec:J}

We consider here the mixed ensemble where the partition function depends upon $\langle \Delta \rangle=J$ instead of $\hat\theta$. Like in the main text, this is again achieved via a Legendre transform using some suitable\footnote{%
The natural-looking contour over the positive real axis does not work because the integrand in \eqref{eq:hiren} does not converge for $\hat\theta\to+\infty$. In Fig.~\ref{fig:theta}, such a contour corresponds to the positive imaginary axis.
} contour $C$ 
\eq{
Z_{\textrm{\tiny Schwarzian}}(\beta,\,J)=\int_C {\rm d} \hat\theta\, e^{\hat\theta J}\, Z_{\textrm{\tiny Schwarzian}}(\beta,\,\hat\theta) \,.
}{eq:hiren} 
Our saddle point equation 
\eq{
 \Big( J+ \frac{1}{12} \Big)-\frac{\pi^2}{6\hat\theta^2}\,(c_{\mt{L}}-2)+\frac{\pi^2\beta}{3\hat\theta^3}\,c_{\mt{M}} - \frac{3}{\hat\theta}=0
}{eq:sickboi}
in general has three solutions for $\hat\theta$. However, typically only one of them, $\hat\theta=\hat\theta_0(\beta,J)$, is compatible with reality and positivity of $\hat\theta$. The result for the partition function evaluated on this saddle point is given by
\eq{
Z_{\textrm{\tiny Schwarzian}}(\beta,\,J)\approx
e^ {\hat\theta_0 J+\frac{\hat\theta_0}{12}+\frac{\pi^2}{6\hat\theta_0}(c_{\mt{L}}-2)-\frac{\pi^2}{6}\,\frac{\beta}{\hat\theta_0^2}\,c_{\mt{M}}}\,\sqrt{\frac{(2\pi)^{11}}{\big|3\hat\theta_0^4+\frac{\pi^2}{3}\hat\theta_0^3(c_{\mt{L}}-2)-\pi^2\beta\hat\theta_0^2 c_{\mt{M}}\big|}}\,.
}{eq:violetstale}

There are six possibilities to achieve a large $N$ factor:
\begin{enumerate}
\item $-\beta c_{\mt{M}}\gg J$ and $c_{\mt{L}}\ll (-\beta c_{\mt{M}})^{2/3}$: In this case, the relevant saddle point given by
\eq{
\hat\theta_0= \Big(-\frac{4\pi^2\beta c_{\mt{M}}}{12J+1}\Big)^{1/3} + \mathcal{O}(1) \gg 1
}{eq:chalkoutlines}
yields the partition function
\eq{
\hat Z_{\textrm{\tiny Schwarzian}}(\beta,\,J)\approx N_1(J)\,\big(-\beta c_{\mt{M}}\big)^{-5/6}\,e^{\frac32\,\big(2\pi(J+\tfrac{1}{12})\big)^{2/3}\,(-\beta  \frac{c_{\mt{M}}}{12})^{1/3}}
}{eq:moneygame}
with $N_1(J)\propto(J+\tfrac{1}{12})^{1/3}$. 
\item $c_{\mt{L}}\gg J$ and $c_{\mt{L}}\gg(-\beta c_{\mt{M}})^{2/3}$: In this case, the relevant saddle point given by
\eq{
\hat\theta_0 = 2\pi\,\sqrt{\frac{c_{\mt{L}}}{24J+2}} + \mathcal{O}(1) \gg 1
}{eq:howtobeme}
yields the $\beta$-independent partition function
\eq{
Z_{\textrm{\tiny Schwarzian}}(\beta,\,J)=N_2(J)\,c_{\mt{L}}^{-5/4}\,\exp\(2\pi\sqrt{\frac{c_{\mt{L}}}{6} \big(J+\tfrac{1}{12}\big)}\)
}{eq:jennystale}
with $N_2(J)\propto(J+\tfrac{1}{12})^{3/4}$. 
\item $-\beta c_{\mt{M}}=-\tilde\beta\gg 1$, $c_{\mt{L}}\sim(-\tilde\beta)^{\delta}\gg 1$, $J\sim (-\tilde\beta)^{-2+3\delta}\gg 1$, and $\frac 23<\delta<1$: In this case, we introduce a parameter $\alpha=-\tilde\beta c_{\mt{L}}^{-3/2}$ and find the saddle point value
\eq{
\hat\theta_0 = 2\pi^{2/3}\,\sqrt{\frac{c_{\mt{L}}}{12}}\,\frac{2\pi^{2/3}(12J+1)^{1/3}+R^{2/3}}{(12J+1)^{2/3}R^{1/3}} + \mathcal{O}(1) \gg 1
}{eq:power}
with $R=3\alpha(12J+1)+\sqrt{(12J+1)(9\alpha^2(12J+1)-8\pi^2)}$. 

In the Schwarzian sector, we need $-\tilde \beta\gg 1$ and $\tilde \Omega \sim -|\tilde \beta| ^{ -\delta}$. 
Therefore, we have 
\begin{equation}
    \hat \theta\sim |\tilde \beta|^{1-\delta}\gg1~\qquad\qquad -\tilde{\beta}\tilde{\Omega}^2\sim |\tilde \beta^{1-2\delta}|\ll 1
\end{equation}
and
\begin{equation}
    c_{\mt L}\sim|\tilde \beta^{\delta}|\gg 1\qquad\qquad J\sim|\tilde \beta^{-2  +3\delta}|\gg 1 
\end{equation} 
such that the first three terms of the \eqref{eq:sickboi} are large and scale in the same way $\sim |\tilde \beta^{-2 +3\delta}|\gg1
$. With these scalings, the saddle point is located at
\begin{equation}
\hat\theta_0 = 2\pi^{2/3}\,\sqrt{\frac{c_{\mt{L}}}{12}}\,\bigg({\frac{2\pi^{2/3}}{(12RJ)^{1/3}}+\frac{R^{1/3}}{(12J)^{2/3}}} \bigg) \sim (-\tilde\beta)^{1-\delta} \gg 1
\end{equation}
with $R=36\alpha J+4\sqrt{3J(27\alpha^2J-2\pi^2)}$. We also assume that $27\alpha^2 J\geq2\pi^2$ for a real saddle point. The partition function is
\begin{equation}
    Z_{\textrm{\tiny Schwarzian}}(\beta,\,J)\approx
e^ {\hat\theta_0 J+\frac{\pi^2}{6\hat\theta_0}c_{\mt{L}}-\frac{\pi^2\beta}{6\hat\theta_0^2}\,c_{\mt{M}}}\,\sqrt{\frac{(2\pi)^{11}}{\big|\frac{\pi^2}{3}\,\hat\theta_0^3\,c_{\mt{L}}-\pi^2\beta\hat\theta_0^2 c_{\mt{M}}\big|}}\,.
\end{equation}
\item $-\beta c_{\mt{M}}=-\tilde\beta\gg 1$, $c_{\mt{L}}\sim(-\tilde\beta)^{\frac23}\gg 1$, $J\sim \mathcal{O}(1)$: this case is similar to the previous one with $\delta=\frac23$. The partition function is given by \eqref{eq:violetstale}. 
\item $-\beta c_{\mt{M}} \gg  c_{\mt{L}}$ and $ c_{\mt{L}}^3\gg J(-\beta c_{\mt{M}})^2$: In this case, the saddle point $\hat\theta_0=\beta c_{\mt{M}}/c_{\mt{L}}<0$ violates our assumption of positive $\hat\theta$ unless the central charge ratio is negative. We dismiss this case as unphysical.
\item $J\gg (-\beta c_{\mt{M}})^{1/4}$ and $J\gg c_{\mt{L}}^{1/3}$: This case leads to saddle points where $\hat\theta\propto 1/J$ is small and hence outside the range of validity of the Schwarzian sector.
\end{enumerate}

\subsection{Microcanonical ensemble}\label{app:micro}

Starting with the fixed $M$ ensemble we trade here also $\hat\theta$ for $J$ by defining
\eq{
Z(M,\,J) = \frac{(2\pi)^5}{M}\,\int_C\extd\hat\theta\,e^{\hat\theta(J+\tfrac{1}{12})+\frac{\pi^2}{6\hat\theta}\,(c_{\mt{L}}-2)-3\ln\hat\theta}
}{eq:micro1}
with some suitable contour $C$ and exploiting again the saddle point approximation \eqref{eq:troubles}, yielding
\eq{
Z(M,\,J) \approx \frac1M\,e^{\hat\theta_0(J+\tfrac{1}{12})+\frac{\pi^2}{6\hat\theta_0}(c_{\mt{L}}-2)}\,\sqrt{\frac{(2\pi)^{11}}{\big|\frac{\pi^2}{3}\hat\theta_0^3(c_{\mt{L}}-2)+3\hat\theta_0^4\big|}}\,.
}{eq:micro2}
The only possibility for a saddle point in the Schwarzian sector arises for $c_{\mt{L}}\gg 1$, yielding
\eq{
\hat\theta_0 = \sqrt{\frac{2\pi^2 c_{\mt{L}}}{12J+1}} + \mathcal{O}(1) \gg 1\,.
}{eq:micro3}
Inserting this result into the partition function \eqref{eq:micro2} establishes
\eq{
Z(M,\,J) \approx N_3(M,\,J)\,c_{\mt{L}}^{-5/4}\,e^{2\pi\sqrt{\frac{c_{\mt{L}}}{6}(J+\tfrac{1}{12})}}
}{eq:micro4}
with $N_3(M,\,J)\propto \frac1M\,(J+\tfrac{1}{12})^{3/4}$. 

Alternatively, the microcanonical partition function \eqref{eq:micro4} is obtained from the fixed $J$ partition function \eqref{eq:jennystale} by
\eq{
Z(M,\,J) = \int\limits_{-\infty}^0\extd\beta\,e^{\beta\,M}\,Z_{\textrm{\tiny Schwarzian}}(\beta,\,J)\approx N_3(M,\,J)\,c_{\mt{L}}^{-5/4}\,e^{2\pi\sqrt{\frac{c_{\mt{L}}}{6}(J+\tfrac{1}{12})}}
}{eq:micro42}
with the same expression for $N_3(M,\,J)$ as above.

For large $J$, the LO microcanonical entropy
\eq{
S=\ln Z(M,\,J) \approx 2\pi\sqrt{\frac{c_{\mt{L}}J}{6}}
}{eq:micro5}
is compatible with the cold result \eqref{eq:Schw-ent} if we use $c_{\mt{L}}\gg 1$ and apply the approximation\footnote{%
This term comes from the ``finite'' part defined in \eqref{eq:whynolabel}. However, if $c_{\mt{L}}$ tends to infinity, the first term in the ``finite'' part is infinite and dominates all other terms.
} 
\eq{
\langle\Delta\rangle \approx \frac{\pi^2}{6}\, \frac{c_{\mt{L}}}{\hat\theta^2}\,.
}{eq:micro6}

If instead we use the fixed $J$ partition function \eqref{eq:moneygame} as a starting point, we end up with a different microcanonical partition function
\eq{
\hat Z(M,\,J) = \int\limits_{-\infty}^0\extd\beta\,e^{\beta\,M}\,\hat Z_{\textrm{\tiny Schwarzian}}\approx N_4\,c_{\textrm{\tiny M}}^{-1}\,e^{2\pi\sqrt{\frac{c_{\textrm{\tiny M}}}{6M}}(J+\tfrac{1}{12})}
}{eq:micro41}
with $N_4=\rm const$. In the limit of large $J$, the associated LO entropy
\eq{
\hat S=\ln\hat Z(M,\,J)\approx 2\pi\,\frac{\tilde\Delta}{\sqrt{\tilde\xi}}\qquad\qquad \tilde \Delta = J\qquad \tilde\xi = \frac{6M}{c_{\textrm{\tiny M}}} 
}{eq:micro40}
coincides with the Cardy entropy \eqref{eq:cardy-finite} when the $c_{\textrm{\tiny L}}$ term therein is negligible.


\section{Lowest-weight characters of 2d CFTs}\label{app:B}

Consider a lowest-weight representation denoted by $V_{\textrm{\tiny LW}}(c,h)$. This is built upon a lowest-weight state $|h\rangle_{\textrm{\tiny LW}}$
\eq{
L_0 |h\rangle_{\textrm{\tiny LW}} = h|h\rangle_{\textrm{\tiny LW}}\qquad\qquad
L_n |h\rangle_{\textrm{\tiny LW}} = 0 \quad \text{for all } n < 0
}{eq:appB1}
annihilated by all lowering operators $L_n$ with $n<0$.

Descendants of lowest-weight states are created by applying raising operators $L_n$ with $n>0$. A state at level $N$ is of the form $L_{n_1} \dots L_{n_k} |h\rangle_{\textrm{\tiny LW}}$, where $n_1 + \dots + n_k = N$. Using the commutation relation $[L_0, L_n] = -nL_n$, the $L_0$ eigenvalue of such a state is $h-N$. Therefore, the spectrum of $L_0$ is bounded from above: $\{h, h-1, h-2, \dots\}$.

The character for this lowest-weight module is
\eq{
\chi_{c,h}^{\textrm{\tiny LW}}(\tau) = \text{Tr}_{V_{\textrm{\tiny LW}}(c,h)} \,q^{L_0 - c/24} \qquad\qquad q=e^{2\pi i \tau}\,.
}{eq:appB2}
The algebraic structure of the raising operators $\{L_{n>0}\}$ is identical to that of the lowering operators $\{L_{n<0}\}$. This means the number of independent states at level $N$, $d(N)$, is the same as in the highest-weight case. The only difference is the $L_0$ eigenvalue.
\eq{
\chi_{c,h}^{\textrm{\tiny LW}}(\tau) = \sum_{N=0}^{\infty} d(N) q^{(h-N) - c/24} = q^{h-c/24} \sum_{N=0}^{\infty} d(N) q^{-N}=q^{h-c/24} \prod_{k=1}^{\infty} \frac{1}{1- q^{-1}}
}{eq:appB3}
Thus, the lowest-weight character is related to the highest-weight character by flipping signs,
\eq{
    \chi_{c,h}^{\textrm{\tiny LW}}(\tau) =q^{h-(c+1)/24}\frac{1}{\eta(-\tau)}=\chi_{-c,-h}^{\textrm{\tiny HW}}(-\tau) \,.
}{eq:appB4}

We are interested in the lowest-weight anti-holomorphic sector 
\eq{
\bar \chi_{\bar c,\bar h}^{\textrm{\tiny LW}}(\bar \tau) =\text{Tr}_{V_{\textrm{\tiny LW}}(\bar c,\bar h)} \bar q^{\bar L_0 - \bar c/24}= \text{Tr}_{V_{\textrm{\tiny LW}}(\bar c,\bar h)} e^{-2\pi i \bar \tau(\bar L_0 - \bar c/24)}=\chi_{\bar c,\bar h}^{\textrm{\tiny LW}}(-\bar \tau)
}{eq:appB5}
and note that this is essentially the same computation as we did above for the holomorphic sector with $q\to \bar q$ and $\tau \to -\bar \tau$. We have
\eq{
    \bar\chi_{\bar c,\bar h}^{\textrm{\tiny LW}}(\bar \tau) =\bar q^{\bar h-(\bar c+1)/24}\frac{1}{\eta(\bar \tau)}=\chi_{\bar c,\bar h}^{\textrm{\tiny LW}}(-\bar \tau)
}{eq:appB6}
The vacuum character in the lowest-weight representation is 
\eq{
    \bar\chi_{\bar c,0}^{\textrm{\tiny LW}}(\bar \tau) = e^{2\pi i\bar \tau(\bar c+1)/24}\frac{1}{\eta(\bar \tau)}(1-e^{2\pi i \bar \tau})=\chi_{\bar c,0}^{\textrm{\tiny LW}}(-\bar \tau)=\chi_{-\bar c,0}^{\textrm{\tiny HW}}(\bar \tau)\,.
}{eq:appB7}

If we have a flipped CFT$_2$ where one chiral sector is built from highest-weight representations and the other chiral sector from lowest-weight representations, then vacuum dominance in the $S$-dual channel implies the partition function for this CFT is well-approximated by
\eq{
\tilde Z^{\textrm{\tiny CFT}}(\tau,\bar\tau) \approx \chi_{c,0}^{\textrm{\tiny HW}}\Big(-\frac{1}{\tau}\Big)  \bar\chi_{\bar c,0}^{\textrm{\tiny LW}}\Big(-\frac{1}{\bar \tau}\Big) = \frac{\big(1-e^{-2\pi i/\tau}\big)\big(1-e^{-2\pi i/\bar\tau}\big)}{\eta(-1/\tau)\,\eta(-1/\bar\tau)}\,e^{2\pi i\big(\frac{c-1}{24\tau}-\frac{\bar c+1}{24\bar\tau}\big)}\,.
}{eq:appB8}
The expression for the partition function \eqref{eq:appB8} is well-defined if both $\tau$ and $\bar\tau$ lie in the complex upper half-plane. This means they cannot be complex conjugates of each other.


\renewcommand{\emph}{\textit}
\renewcommand{\em}{\it}

\providecommand{\href}[2]{#2}\begingroup\raggedright\endgroup



\begin{thebibliography}{100}

\addcontentsline{toc}{section}{References}

\bibitem{Levy1965}
J.-M. L{\'e}vy-Leblond, {\it Une nouvelle limite non-relativiste du groupe de
  {P}oincar{\'e}},  {\em Annales de l'I.H.P. Physique th{\'e}orique} {\bf 3}
  (1965), no.~1 1--12.

\bibitem{Gupta1966}
N.~D. SenGupta, {\it On an analogue of the galilei group},  {\em Il Nuovo
  Cimento A Series 10} {\bf 44} (1966), no.~2 512--517.

\bibitem{Bagchi:2010zz}
A.~Bagchi, {\it {Correspondence between Asymptotically Flat Spacetimes and
  Nonrelativistic Conformal Field Theories}},  {\em Phys.Rev.Lett.} {\bf 105}
  (2010) 171601.

\bibitem{Barnich:2010eb}
G.~Barnich and C.~Troessaert, {\it {Aspects of the BMS/CFT correspondence}},
  {\em JHEP} {\bf 1005} (2010) 062, [\href{http://arxiv.org/abs/1001.1541}{{\tt
  1001.1541}}].

\bibitem{Barnich:2012aw}
G.~Barnich, A.~Gomberoff, and H.~A. Gonzalez, {\it {The Flat limit of three
  dimensional asymptotically anti-de Sitter spacetimes}},  {\em Phys.Rev.} {\bf
  D86} (2012) 024020, [\href{http://arxiv.org/abs/1204.3288}{{\tt 1204.3288}}].

\bibitem{Barnich:2012rz}
G.~Barnich, A.~Gomberoff, and H.~A. Gonzalez, {\it {BMS$_3$ invariant two
  dimensional field theories as flat limit of Liouville}},  {\em Phys. Rev.}
  {\bf D87:124032,} (2013) [\href{http://arxiv.org/abs/1210.0731}{{\tt
  1210.0731}}].

\bibitem{Bagchi:2012yk}
A.~Bagchi, S.~Detournay, and D.~Grumiller, {\it {Flat-Space Chiral Gravity}},
  {\em Phys.Rev.Lett.} {\bf 109} (2012) 151301,
  [\href{http://arxiv.org/abs/1208.1658}{{\tt 1208.1658}}].

\bibitem{Bagchi:2012xr}
A.~Bagchi, S.~Detournay, R.~Fareghbal, and J.~Simon, {\it {Holography of 3d
  Flat Cosmological Horizons}},  {\em Phys. Rev. Lett.} {\bf 110} (2013)
  141302, [\href{http://arxiv.org/abs/1208.4372}{{\tt 1208.4372}}].

\bibitem{Barnich:2012xq}
G.~Barnich, {\it {Entropy of three-dimensional asymptotically flat cosmological
  solutions}},  {\em JHEP} {\bf 1210} (2012) 095,
  [\href{http://arxiv.org/abs/1208.4371}{{\tt 1208.4371}}].

\bibitem{Bagchi:2013lma}
A.~Bagchi, S.~Detournay, D.~Grumiller, and J.~Simon, {\it {Cosmic Evolution
  from Phase Transition of Three-Dimensional Flat Space}},  {\em
  Phys.Rev.Lett.} {\bf 111} (2013) 181301,
  [\href{http://arxiv.org/abs/1305.2919}{{\tt 1305.2919}}].

\bibitem{Duval:2014uva}
C.~Duval, G.~W. Gibbons, and P.~A. Horvathy, {\it {Conformal Carroll groups and
  BMS symmetry}},  {\em Class. Quant. Grav.} {\bf 31} (2014) 092001,
  [\href{http://arxiv.org/abs/1402.5894}{{\tt 1402.5894}}].

\bibitem{Bagchi:2014iea}
A.~Bagchi, R.~Basu, D.~Grumiller, and M.~Riegler, {\it {Entanglement entropy in
  Galilean conformal field theories and flat holography}},  {\em
  Phys.Rev.Lett.} {\bf 114} (2015), no.~11 111602,
  [\href{http://arxiv.org/abs/1410.4089}{{\tt 1410.4089}}].

\bibitem{Bagchi:2015wna}
A.~Bagchi, D.~Grumiller, and W.~Merbis, {\it {Stress tensor correlators in
  three-dimensional gravity}},  {\em Phys. Rev. D} {\bf 93} (2016), no.~6
  061502, [\href{http://arxiv.org/abs/1507.05620}{{\tt 1507.05620}}].

\bibitem{Hartong:2015usd}
J.~Hartong, {\it {Holographic Reconstruction of 3D Flat Space-Time}},  {\em
  JHEP} {\bf 10} (2016) 104, [\href{http://arxiv.org/abs/1511.01387}{{\tt
  1511.01387}}].

\bibitem{Bagchi:2016bcd}
A.~Bagchi, R.~Basu, A.~Kakkar, and A.~Mehra, {\it {Flat Holography: Aspects of
  the dual field theory}},  {\em JHEP} {\bf 12} (2016) 147,
  [\href{http://arxiv.org/abs/1609.06203}{{\tt 1609.06203}}].

\bibitem{Ciambelli:2018wre}
L.~Ciambelli, C.~Marteau, A.~C. Petkou, P.~M. Petropoulos, and K.~Siampos, {\it
  {Flat holography and Carrollian fluids}},  {\em JHEP} {\bf 07} (2018) 165,
  [\href{http://arxiv.org/abs/1802.06809}{{\tt 1802.06809}}].

\bibitem{Donnay:2022aba}
L.~Donnay, A.~Fiorucci, Y.~Herfray, and R.~Ruzziconi, {\it {Carrollian
  Perspective on Celestial Holography}},  {\em Phys. Rev. Lett.} {\bf 129}
  (2022), no.~7 071602, [\href{http://arxiv.org/abs/2202.04702}{{\tt
  2202.04702}}].

\bibitem{Bagchi:2022emh}
A.~Bagchi, S.~Banerjee, R.~Basu, and S.~Dutta, {\it {Scattering Amplitudes:
  Celestial and Carrollian}},  {\em Phys. Rev. Lett.} {\bf 128} (2022), no.~24
  241601, [\href{http://arxiv.org/abs/2202.08438}{{\tt 2202.08438}}].

\bibitem{Donnay:2022wvx}
L.~Donnay, A.~Fiorucci, Y.~Herfray, and R.~Ruzziconi, {\it {Bridging Carrollian
  and celestial holography}},  {\em Phys. Rev. D} {\bf 107} (2023), no.~12
  126027, [\href{http://arxiv.org/abs/2212.12553}{{\tt 2212.12553}}].

\bibitem{Bagchi:2023fbj}
A.~Bagchi, P.~Dhivakar, and S.~Dutta, {\it {AdS Witten diagrams to Carrollian
  correlators}},  {\em JHEP} {\bf 04} (2023) 135,
  [\href{http://arxiv.org/abs/2303.07388}{{\tt 2303.07388}}].

\bibitem{Saha:2023hsl}
A.~Saha, {\it {Carrollian approach to 1 + 3D flat holography}},  {\em JHEP}
  {\bf 06} (2023) 051, [\href{http://arxiv.org/abs/2304.02696}{{\tt
  2304.02696}}].

\bibitem{Salzer:2023jqv}
J.~Salzer, {\it {An embedding space approach to Carrollian CFT correlators for
  flat space holography}},  {\em JHEP} {\bf 10} (2023) 084,
  [\href{http://arxiv.org/abs/2304.08292}{{\tt 2304.08292}}].

\bibitem{Saha:2023abr}
A.~Saha, {\it {w$_{1+\infty}$ and Carrollian holography}},  {\em JHEP} {\bf 05}
  (2024) 145, [\href{http://arxiv.org/abs/2308.03673}{{\tt 2308.03673}}].

\bibitem{Mason:2023mti}
L.~Mason, R.~Ruzziconi, and A.~Yelleshpur~Srikant, {\it {Carrollian amplitudes
  and celestial symmetries}},  {\em JHEP} {\bf 05} (2024) 012,
  [\href{http://arxiv.org/abs/2312.10138}{{\tt 2312.10138}}].

\bibitem{Chen:2023naw}
B.~Chen and Z.~Hu, {\it {Bulk reconstruction in flat holography}},  {\em JHEP}
  {\bf 03} (2024) 064, [\href{http://arxiv.org/abs/2312.13574}{{\tt
  2312.13574}}].

\bibitem{Nguyen:2023vfz}
K.~Nguyen and P.~West, {\it {Carrollian Conformal Fields and Flat Holography}},
   {\em Universe} {\bf 9} (2023), no.~9 385,
  [\href{http://arxiv.org/abs/2305.02884}{{\tt 2305.02884}}].

\bibitem{Alday:2024yyj}
L.~F. Alday, M.~Nocchi, R.~Ruzziconi, and A.~Yelleshpur~Srikant, {\it
  {Carrollian Amplitudes from Holographic Correlators}},
  \href{http://arxiv.org/abs/2406.19343}{{\tt 2406.19343}}.

\bibitem{Kraus:2024gso}
P.~Kraus and R.~M. Myers, {\it {Carrollian Partition Functions and the Flat
  Limit of AdS}},  \href{http://arxiv.org/abs/2407.13668}{{\tt 2407.13668}}.

\bibitem{Bagchi:2024efs}
A.~Bagchi, A.~Lipstein, M.~Mandlik, and A.~Mehra, {\it {3d Carrollian
  Chern-Simons theory \& 2d Yang-Mills}},  {\em JHEP} {\bf 11} (2024) 006,
  [\href{http://arxiv.org/abs/2407.13574}{{\tt 2407.13574}}].

\bibitem{Bagchi:2024gnn}
A.~Bagchi, P.~Dhivakar, and S.~Dutta, {\it {3D Stress Tensor for Gravity in 4D
  Flat Spacetime}},  \href{http://arxiv.org/abs/2408.05494}{{\tt 2408.05494}}.

\bibitem{Ruzziconi:2024kzo}
R.~Ruzziconi and A.~Saha, {\it {Holographic Carrollian Currents for Massless
  Scattering}},  \href{http://arxiv.org/abs/2411.04902}{{\tt 2411.04902}}.

\bibitem{Chakrabortty:2024bvm}
S.~Chakrabortty, S.~Hegde, and A.~Maurya, {\it {Differential Representation for
  Carrollian Correlators}},  \href{http://arxiv.org/abs/2411.09641}{{\tt
  2411.09641}}.

\bibitem{Fiorucci:2025twa}
A.~Fiorucci, S.~Pekar, P.~Marios~Petropoulos, and M.~Vilatte, {\it
  {Carrollian-holographic Derivation of BMS Flux-balance Laws}},
  \href{http://arxiv.org/abs/2505.00077}{{\tt 2505.00077}}.

\bibitem{Hawking:2016msc}
S.~W. Hawking, M.~J. Perry, and A.~Strominger, {\it {Soft Hair on Black
  Holes}},  {\em Phys. Rev. Lett.} {\bf 116} (2016), no.~23 231301,
  [\href{http://arxiv.org/abs/1601.00921}{{\tt 1601.00921}}].

\bibitem{Donnay:2015abr}
L.~Donnay, G.~Giribet, H.~A. Gonzalez, and M.~Pino, {\it {Supertranslations and
  Superrotations at the Black Hole Horizon}},  {\em Phys. Rev. Lett.} {\bf 116}
  (2016), no.~9 091101, [\href{http://arxiv.org/abs/1511.08687}{{\tt
  1511.08687}}].

\bibitem{Afshar:2016wfy}
H.~Afshar, S.~Detournay, D.~Grumiller, W.~Merbis, A.~Perez, D.~Tempo, and
  R.~Troncoso, {\it {Soft Heisenberg hair on black holes in three dimensions}},
   {\em Phys. Rev.} {\bf D93} (2016), no.~10 101503,
  [\href{http://arxiv.org/abs/1603.04824}{{\tt 1603.04824}}].

\bibitem{Grumiller:2019fmp}
D.~Grumiller, A.~P\'erez, M.~M. Sheikh-Jabbari, R.~Troncoso, and C.~Zwikel,
  {\it {Spacetime structure near generic horizons and soft hair}},  {\em Phys.
  Rev. Lett.} {\bf 124} (2020), no.~4 041601,
  [\href{http://arxiv.org/abs/1908.09833}{{\tt 1908.09833}}].

\bibitem{Adami:2023fbm}
H.~Adami, A.~Parvizi, M.~M. Sheikh-Jabbari, V.~Taghiloo, and H.~Yavartanoo,
  {\it {Hydro \& thermo dynamics at causal boundaries, examples in 3d
  gravity}},  {\em JHEP} {\bf 07} (2023) 038,
  [\href{http://arxiv.org/abs/2305.01009}{{\tt 2305.01009}}].

\bibitem{Freidel:2024emv}
L.~Freidel and P.~Jai-akson, {\it {Geometry of Carrollian Stretched Horizons}},
   \href{http://arxiv.org/abs/2406.06709}{{\tt 2406.06709}}.

\bibitem{Penna:2018gfx}
R.~F. Penna, {\it {Near-horizon Carroll symmetry and black hole Love numbers}},
   \href{http://arxiv.org/abs/1812.05643}{{\tt 1812.05643}}.

\bibitem{Donnay:2019jiz}
L.~Donnay and C.~Marteau, {\it {Carrollian Physics at the Black Hole Horizon}},
   {\em Class. Quant. Grav.} {\bf 36} (2019), no.~16 165002,
  [\href{http://arxiv.org/abs/1903.09654}{{\tt 1903.09654}}].

\bibitem{Marsot:2022qkx}
L.~Marsot, P.-M. Zhang, and P.~Horvathy, {\it {Anyonic spin-Hall effect on the
  black hole horizon}},  {\em Phys. Rev. D} {\bf 106} (2022), no.~12 L121503,
  [\href{http://arxiv.org/abs/2207.06302}{{\tt 2207.06302}}].

\bibitem{Gray:2022svz}
F.~Gray, D.~Kubiznak, T.~R. Perche, and J.~Redondo-Yuste, {\it {Carrollian
  motion in magnetized black hole horizons}},  {\em Phys. Rev. D} {\bf 107}
  (2023), no.~6 064009, [\href{http://arxiv.org/abs/2211.13695}{{\tt
  2211.13695}}].

\bibitem{Redondo-Yuste:2022czg}
J.~Redondo-Yuste and L.~Lehner, {\it {Non-linear black hole dynamics and
  Carrollian fluids}},  {\em JHEP} {\bf 02} (2023) 240,
  [\href{http://arxiv.org/abs/2212.06175}{{\tt 2212.06175}}].

\bibitem{Freidel:2022bai}
L.~Freidel and P.~Jai-akson, {\it {Carrollian hydrodynamics from symmetries}},
  {\em Class. Quant. Grav.} {\bf 40} (2023), no.~5 055009,
  [\href{http://arxiv.org/abs/2209.03328}{{\tt 2209.03328}}].

\bibitem{Freidel:2022vjq}
L.~Freidel and P.~Jai-akson, {\it {Carrollian hydrodynamics and symplectic
  structure on stretched horizons}},  {\em JHEP} {\bf 05} (2024) 135,
  [\href{http://arxiv.org/abs/2211.06415}{{\tt 2211.06415}}].

\bibitem{Bicak:2023rsz}
J.~Bi\v{c}\'ak, D.~Kubiz\v{n}\'ak, and T.~R. Perche, {\it {Migrating Carrollian
  particles on magnetized black hole horizons}},  {\em Phys. Rev. D} {\bf 107}
  (2023), no.~10 104014, [\href{http://arxiv.org/abs/2302.11639}{{\tt
  2302.11639}}].

\bibitem{deBoer:2021jej}
J.~de~Boer, J.~Hartong, N.~A. Obers, W.~Sybesma, and S.~Vandoren, {\it {Carroll
  Symmetry, Dark Energy and Inflation}},  {\em Front. in Phys.} {\bf 10} (2022)
  810405, [\href{http://arxiv.org/abs/2110.02319}{{\tt 2110.02319}}].

\bibitem{deBoer:2023fnj}
J.~de~Boer, J.~Hartong, N.~A. Obers, W.~Sybesma, and S.~Vandoren, {\it {Carroll
  stories}},  {\em JHEP} {\bf 09} (2023) 148,
  [\href{http://arxiv.org/abs/2307.06827}{{\tt 2307.06827}}].

\bibitem{Oling:2024vmq}
G.~Oling and J.~F. Pedraza, {\it {Mixmasters in Wonderland: Chaotic dynamics
  from Carroll limits of gravity}},
  \href{http://arxiv.org/abs/2409.05836}{{\tt 2409.05836}}.

\bibitem{Kasikci:2023tvs}
O.~Kasikci, M.~Ozkan, and Y.~Pang, {\it {Carrollian origin of spacetime
  subsystem symmetry}},  {\em Phys. Rev. D} {\bf 108} (2023), no.~4 045020,
  [\href{http://arxiv.org/abs/2304.11331}{{\tt 2304.11331}}].

\bibitem{Figueroa-OFarrill:2023vbj}
J.~Figueroa-O'Farrill, A.~P\'erez, and S.~Prohazka, {\it {Carroll/fracton
  particles and their correspondence}},  {\em JHEP} {\bf 06} (2023) 207,
  [\href{http://arxiv.org/abs/2305.06730}{{\tt 2305.06730}}].

\bibitem{Figueroa-OFarrill:2023qty}
J.~Figueroa-O'Farrill, A.~P\'erez, and S.~Prohazka, {\it {Quantum
  Carroll/fracton particles}},  {\em JHEP} {\bf 10} (2023) 041,
  [\href{http://arxiv.org/abs/2307.05674}{{\tt 2307.05674}}].

\bibitem{Bagchi:2022eui}
A.~Bagchi, A.~Banerjee, R.~Basu, M.~Islam, and S.~Mondal, {\it {Magic fermions:
  Carroll and flat bands}},  {\em JHEP} {\bf 03} (2023) 227,
  [\href{http://arxiv.org/abs/2211.11640}{{\tt 2211.11640}}].

\bibitem{Petkou:2022bmz}
A.~C. Petkou, P.~M. Petropoulos, D.~R. Betancour, and K.~Siampos, {\it
  {Relativistic fluids, hydrodynamic frames and their Galilean versus
  Carrollian avatars}},  {\em JHEP} {\bf 09} (2022) 162,
  [\href{http://arxiv.org/abs/2205.09142}{{\tt 2205.09142}}].

\bibitem{Bagchi:2023ysc}
A.~Bagchi, K.~S. Kolekar, and A.~Shukla, {\it {Carrollian Origins of Bjorken
  Flow}},  {\em Phys. Rev. Lett.} {\bf 130} (2023), no.~24 241601,
  [\href{http://arxiv.org/abs/2302.03053}{{\tt 2302.03053}}].

\bibitem{Armas:2023dcz}
J.~Armas and E.~Have, {\it {Carrollian Fluids and Spontaneous Breaking of Boost
  Symmetry}},  {\em Phys. Rev. Lett.} {\bf 132} (2024), no.~16 161606,
  [\href{http://arxiv.org/abs/2308.10594}{{\tt 2308.10594}}].

\bibitem{Bagchi:2023rwd}
A.~Bagchi, K.~S. Kolekar, T.~Mandal, and A.~Shukla, {\it {Heavy-ion collisions,
  Gubser flow, and Carroll hydrodynamics}},  {\em Phys. Rev. D} {\bf 109}
  (2024), no.~5 056004, [\href{http://arxiv.org/abs/2310.03167}{{\tt
  2310.03167}}].

\bibitem{Bagchi:2024ikw}
A.~Bagchi, A.~Banerjee, S.~Mondal, and S.~Sarkar, {\it {Carroll in Shallow
  Water}},  \href{http://arxiv.org/abs/2411.04190}{{\tt 2411.04190}}.

\bibitem{Bagchi:2013bga}
A.~Bagchi, {\it {Tensionless Strings and Galilean Conformal Algebra}},  {\em
  JHEP} {\bf 1305} (2013) 141, [\href{http://arxiv.org/abs/1303.0291}{{\tt
  1303.0291}}].

\bibitem{Bagchi:2015nca}
A.~Bagchi, S.~Chakrabortty, and P.~Parekh, {\it {Tensionless Strings from
  Worldsheet Symmetries}},  {\em JHEP} {\bf 01} (2016) 158,
  [\href{http://arxiv.org/abs/1507.04361}{{\tt 1507.04361}}].

\bibitem{Bagchi:2021ban}
A.~Bagchi, A.~Banerjee, S.~Chakrabortty, and R.~Chatterjee, {\it {A Rindler
  road to Carrollian worldsheets}},  {\em JHEP} {\bf 04} (2022) 082,
  [\href{http://arxiv.org/abs/2111.01172}{{\tt 2111.01172}}].

\bibitem{Bagchi:2022iqb}
A.~Bagchi, D.~Grumiller, S.~Sheikh-Jabbari, and M.~M. Sheikh-Jabbari, {\it
  {Horizon strings as 3D black hole microstates}},  {\em SciPost Phys.} {\bf
  15} (2023), no.~5 210, [\href{http://arxiv.org/abs/2210.10794}{{\tt
  2210.10794}}].

\bibitem{Cardona:2016ytk}
B.~Cardona, J.~Gomis, and J.~M. Pons, {\it {Dynamics of Carroll Strings}},
  {\em JHEP} {\bf 07} (2016) 050, [\href{http://arxiv.org/abs/1605.05483}{{\tt
  1605.05483}}].

\bibitem{Bagchi:2023cfp}
A.~Bagchi, A.~Banerjee, J.~Hartong, E.~Have, K.~S. Kolekar, and M.~Mandlik,
  {\it {Strings near black holes are Carrollian}},  {\em Phys. Rev. D} {\bf
  110} (2024), no.~8 086009, [\href{http://arxiv.org/abs/2312.14240}{{\tt
  2312.14240}}].

\bibitem{Hartong:2015xda}
J.~Hartong, {\it {Gauging the Carroll Algebra and Ultra-Relativistic Gravity}},
   {\em JHEP} {\bf 08} (2015) 069, [\href{http://arxiv.org/abs/1505.05011}{{\tt
  1505.05011}}].

\bibitem{Bergshoeff:2017btm}
E.~Bergshoeff, J.~Gomis, B.~Rollier, J.~Rosseel, and T.~ter Veldhuis, {\it
  {Carroll versus Galilei Gravity}},  {\em JHEP} {\bf 03} (2017) 165,
  [\href{http://arxiv.org/abs/1701.06156}{{\tt 1701.06156}}].

\bibitem{Ciambelli:2018ojf}
L.~Ciambelli and C.~Marteau, {\it {Carrollian conservation laws and Ricci-flat
  gravity}},  {\em Class. Quant. Grav.} {\bf 36} (2019), no.~8 085004,
  [\href{http://arxiv.org/abs/1810.11037}{{\tt 1810.11037}}].

\bibitem{Gomis:2019nih}
J.~Gomis, A.~Kleinschmidt, J.~Palmkvist, and P.~Salgado-Rebolledo, {\it
  {Newton-Hooke/Carrollian expansions of (A)dS and Chern-Simons gravity}},
  {\em JHEP} {\bf 02} (2020) 009, [\href{http://arxiv.org/abs/1912.07564}{{\tt
  1912.07564}}].

\bibitem{Ciambelli:2019lap}
L.~Ciambelli, R.~G. Leigh, C.~Marteau, and P.~M. Petropoulos, {\it {Carroll
  Structures, Null Geometry and Conformal Isometries}},  {\em Phys. Rev. D}
  {\bf 100} (2019), no.~4 046010, [\href{http://arxiv.org/abs/1905.02221}{{\tt
  1905.02221}}].

\bibitem{Grumiller:2020elf}
D.~Grumiller, J.~Hartong, S.~Prohazka, and J.~Salzer, {\it {Limits of JT
  gravity}},  {\em JHEP} {\bf 02} (2021) 134,
  [\href{http://arxiv.org/abs/2011.13870}{{\tt 2011.13870}}].

\bibitem{Gomis:2020wxp}
J.~Gomis, D.~Hidalgo, and P.~Salgado-Rebolledo, {\it {Non-relativistic and
  Carrollian limits of Jackiw-Teitelboim gravity}},  {\em JHEP} {\bf 05} (2021)
  162, [\href{http://arxiv.org/abs/2011.15053}{{\tt 2011.15053}}].

\bibitem{Hansen:2021fxi}
D.~Hansen, N.~A. Obers, G.~Oling, and B.~T. S\o{}gaard, {\it {Carroll Expansion
  of General Relativity}},  {\em SciPost Phys.} {\bf 13} (2022), no.~3 055,
  [\href{http://arxiv.org/abs/2112.12684}{{\tt 2112.12684}}].

\bibitem{Campoleoni:2022ebj}
A.~Campoleoni, M.~Henneaux, S.~Pekar, A.~P\'erez, and P.~Salgado-Rebolledo,
  {\it {Magnetic Carrollian gravity from the Carroll algebra}},  {\em JHEP}
  {\bf 09} (2022) 127, [\href{http://arxiv.org/abs/2207.14167}{{\tt
  2207.14167}}].

\bibitem{Ecker:2023uwm}
F.~Ecker, D.~Grumiller, J.~Hartong, A.~P\'erez, S.~Prohazka, and R.~Troncoso,
  {\it {Carroll black holes}},  {\em SciPost Phys.} {\bf 15} (2023), no.~6 245,
  [\href{http://arxiv.org/abs/2308.10947}{{\tt 2308.10947}}].

\bibitem{Grumiller:2024dql}
D.~Grumiller, L.~Montecchio, and M.~S. Nejati, {\it {Carroll dilaton
  supergravity in two dimensions}},  {\em JHEP} {\bf 12} (2024) 005,
  [\href{http://arxiv.org/abs/2409.17781}{{\tt 2409.17781}}].

\bibitem{Bagchi:2019xfx}
A.~Bagchi, A.~Mehra, and P.~Nandi, {\it {Field Theories with Conformal
  Carrollian Symmetry}},  {\em JHEP} {\bf 05} (2019) 108,
  [\href{http://arxiv.org/abs/1901.10147}{{\tt 1901.10147}}].

\bibitem{Chen:2021xkw}
B.~Chen, R.~Liu, and Y.-f. Zheng, {\it {On higher-dimensional Carrollian and
  Galilean conformal field theories}},  {\em SciPost Phys.} {\bf 14} (2023),
  no.~5 088, [\href{http://arxiv.org/abs/2112.10514}{{\tt 2112.10514}}].

\bibitem{Henneaux:2021yzg}
M.~Henneaux and P.~Salgado-Rebolledo, {\it {Carroll contractions of
  Lorentz-invariant theories}},  {\em JHEP} {\bf 11} (2021) 180,
  [\href{http://arxiv.org/abs/2109.06708}{{\tt 2109.06708}}].

\bibitem{Baiguera:2022lsw}
S.~Baiguera, G.~Oling, W.~Sybesma, and B.~T. S\o{}gaard, {\it {Conformal
  Carroll scalars with boosts}},  {\em SciPost Phys.} {\bf 14} (2023), no.~4
  086, [\href{http://arxiv.org/abs/2207.03468}{{\tt 2207.03468}}].

\bibitem{Ecker:2024czx}
F.~Ecker, D.~Grumiller, M.~Henneaux, and P.~Salgado-Rebolledo, {\it {Carroll
  swiftons}},  {\em Phys. Rev. D} {\bf 110} (2024), no.~4 L041901,
  [\href{http://arxiv.org/abs/2403.00544}{{\tt 2403.00544}}].

\bibitem{Bagchi:2024qsb}
A.~Bagchi, P.~Chakraborty, S.~Chakrabortty, S.~Fredenhagen, D.~Grumiller, and
  P.~Pandit, {\it {Boundary Carrollian Conformal Field Theories and Open Null
  Strings}},  {\em Phys. Rev. Lett.} {\bf 134} (2025), no.~7 071604,
  [\href{http://arxiv.org/abs/2409.01094}{{\tt 2409.01094}}].

\bibitem{Bagchi:2025vri}
A.~Bagchi, A.~Banerjee, P.~Dhivakar, S.~Mondal, and A.~Shukla, ``{The
  Carrollian Kaleidoscope}.'' invited review article for EPJC, 6, 2025.

\bibitem{Bagchi:2013qva}
A.~Bagchi and R.~Basu, {\it {3D Flat Holography: Entropy and Logarithmic
  Corrections}},  {\em JHEP} {\bf 1403} (2014) 020,
  [\href{http://arxiv.org/abs/1312.5748}{{\tt 1312.5748}}].

\bibitem{Bagchi:2019unf}
A.~Bagchi, A.~Saha, and Zodinmawia, {\it {BMS Characters and Modular
  Invariance}},  {\em JHEP} {\bf 07} (2019) 138,
  [\href{http://arxiv.org/abs/1902.07066}{{\tt 1902.07066}}].

\bibitem{Ghosh:2019rcj}
A.~Ghosh, H.~Maxfield, and G.~J. Turiaci, {\it {A universal Schwarzian sector
  in two-dimensional conformal field theories}},  {\em JHEP} {\bf 05} (2020)
  104, [\href{http://arxiv.org/abs/1912.07654}{{\tt 1912.07654}}].

\bibitem{Aggarwal:2022xfd}
A.~Aggarwal, A.~Castro, S.~Detournay, and B.~M\"uhlmann, {\it {Near-extremal
  limits of warped CFTs}},  {\em SciPost Phys.} {\bf 15} (2023), no.~2 056,
  [\href{http://arxiv.org/abs/2211.03770}{{\tt 2211.03770}}].

\bibitem{Aggarwal:2023peg}
A.~Aggarwal, A.~Castro, S.~Detournay, and B.~M\"uhlmann, {\it {Near-extremal
  limits of warped black holes}},  {\em SciPost Phys.} {\bf 15} (2023), no.~3
  083, [\href{http://arxiv.org/abs/2304.10102}{{\tt 2304.10102}}].

\bibitem{Cardy:1986ie}
J.~L. Cardy, {\it Operator content of two-dimensional conformally invariant
  theories},  {\em Nucl. Phys.} {\bf B270} (1986) 186--204.

\bibitem{Detournay:2012pc}
S.~Detournay, T.~Hartman, and D.~M. Hofman, {\it {Warped Conformal Field
  Theory}},  {\em Phys.Rev.} {\bf D86} (2012) 124018,
  [\href{http://arxiv.org/abs/1210.0539}{{\tt 1210.0539}}].

\bibitem{Zamolodchikov:1978xm}
A.~B. Zamolodchikov and A.~B. Zamolodchikov, {\it {Factorized s Matrices in
  Two-Dimensions as the Exact Solutions of Certain Relativistic Quantum Field
  Models}},  {\em Annals Phys.} {\bf 120} (1979) 253--291.

\bibitem{Yang:1968rm}
C.-N. Yang and C.~P. Yang, {\it {Thermodynamics of one-dimensional system of
  bosons with repulsive delta function interaction}},  {\em J. Math. Phys.}
  {\bf 10} (1969) 1115--1122.

\bibitem{Coleman:1974bu}
S.~R. Coleman, {\it {Quantum sine-Gordon equation as the massive Thirring
  model}},  {\em Phys. Rev.} {\bf D11} (1975) 2088.

\bibitem{Luther:1975wr}
A.~Luther and I.~Peschel, {\it {Calculation of critical exponents in
  two-dimensions from quantum field theory in one-dimension}},  {\em Phys. Rev.
  B} {\bf 12} (1975) 3908--3917.

\bibitem{'tHooft:1974bx}
G.~'t~Hooft and M.~J.~G. Veltman, {\it One loop divergencies in the theory of
  gravitation},  {\em Annales Poincare Phys. Theor.} {\bf A20} (1974) 69--94.

\bibitem{Witten:1998qj}
E.~Witten, {\it {Anti-de Sitter space and holography}},  {\em Adv. Theor. Math.
  Phys.} {\bf 2} (1998) 253--291,
  [\href{http://arxiv.org/abs/hep-th/9802150}{{\tt hep-th/9802150}}].

\bibitem{Witten:1986bf}
E.~Witten, {\it {Elliptic Genera and Quantum Field Theory}},  {\em Commun.
  Math. Phys.} {\bf 109} (1987) 525.

\bibitem{Banados:1992gq}
M.~Ba\~nados, M.~Henneaux, C.~Teitelboim, and J.~Zanelli, {\it Geometry of the
  (2+1) black hole},  {\em Phys. Rev.} {\bf D48} (1993) 1506--1525,
  [\href{http://arxiv.org/abs/gr-qc/9302012}{{\tt gr-qc/9302012}}].

\bibitem{Carlip:1994gy}
S.~Carlip, {\it The statistical mechanics of the (2+1)-dimensional black hole},
   {\em Phys. Rev.} {\bf D51} (1995) 632--637,
  [\href{http://arxiv.org/abs/gr-qc/9409052}{{\tt gr-qc/9409052}}].

\bibitem{Strominger}
A.~Strominger, {\it Black hole entropy from near-horizon microstates},  {\em
  JHEP} {\bf 02} (1998) 009, [\href{http://arxiv.org/abs/hep-th/9712251}{{\tt
  hep-th/9712251}}].

\bibitem{Hartman:2014oaa}
T.~Hartman, C.~A. Keller, and B.~Stoica, {\it {Universal Spectrum of 2d
  Conformal Field Theory in the Large c Limit}},  {\em JHEP} {\bf 09} (2014)
  118, [\href{http://arxiv.org/abs/1405.5137}{{\tt 1405.5137}}].

\bibitem{Gonzalez:2011nz}
H.~A. Gonzalez, D.~Tempo, and R.~Troncoso, {\it {Field theories with
  anisotropic scaling in 2D, solitons and the microscopic entropy of
  asymptotically Lifshitz black holes}},  {\em JHEP} {\bf 11} (2011) 066,
  [\href{http://arxiv.org/abs/1107.3647}{{\tt 1107.3647}}].

\bibitem{Shaghoulian:2015dwa}
E.~Shaghoulian, {\it {A Cardy formula for holographic hyperscaling-violating
  theories}},  {\em JHEP} {\bf 11} (2015) 081,
  [\href{http://arxiv.org/abs/1504.02094}{{\tt 1504.02094}}].

\bibitem{Perez:2016vqo}
A.~P{\'e}rez, D.~Tempo, and R.~Troncoso, {\it {Boundary conditions for General
  Relativity on AdS$_{3}$ and the KdV hierarchy}},  {\em JHEP} {\bf 06} (2016)
  103, [\href{http://arxiv.org/abs/1605.04490}{{\tt 1605.04490}}].

\bibitem{Afshar:2016kjj}
H.~Afshar, D.~Grumiller, W.~Merbis, A.~Perez, D.~Tempo, and R.~Troncoso, {\it
  {Soft hairy horizons in three spacetime dimensions}},  {\em Phys. Rev. D}
  {\bf 95} (2017), no.~10 106005, [\href{http://arxiv.org/abs/1611.09783}{{\tt
  1611.09783}}].

\bibitem{Grumiller:2017jft}
D.~Grumiller, A.~Perez, D.~Tempo, and R.~Troncoso, {\it {Log corrections to
  entropy of three dimensional black holes with soft hair}},  {\em JHEP} {\bf
  08} (2017) 107, [\href{http://arxiv.org/abs/1705.10605}{{\tt 1705.10605}}].

\bibitem{Melnikov:2018fhb}
D.~Melnikov, F.~Novaes, A.~P\'erez, and R.~Troncoso, {\it {Lifshitz Scaling,
  Microstate Counting from Number Theory and Black Hole Entropy}},  {\em JHEP}
  {\bf 06} (2019) 054, [\href{http://arxiv.org/abs/1808.04034}{{\tt
  1808.04034}}].

\bibitem{Chen:2019hbj}
B.~Chen, P.-X. Hao, and Z.-F. Yu, {\it {2d Galilean Field Theories with
  Anisotropic Scaling}},  {\em Phys. Rev. D} {\bf 101} (2020), no.~6 066029,
  [\href{http://arxiv.org/abs/1906.03102}{{\tt 1906.03102}}].

\bibitem{Chaturvedi:2020jyy}
P.~Chaturvedi, I.~Papadimitriou, W.~Song, and B.~Yu, {\it {AdS$_{3}$ gravity
  and the complex SYK models}},  {\em JHEP} {\bf 05} (2021) 142,
  [\href{http://arxiv.org/abs/2011.10001}{{\tt 2011.10001}}].

\bibitem{Afshar:2019tvp}
H.~R. Afshar, {\it {Warped Schwarzian theory}},  {\em JHEP} {\bf 02} (2020)
  126, [\href{http://arxiv.org/abs/1908.08089}{{\tt 1908.08089}}].

\bibitem{Bergshoeff:2022eog}
E.~Bergshoeff, J.~Figueroa-O'Farrill, and J.~Gomis, {\it {A non-lorentzian
  primer}},  {\em SciPost Phys. Lect. Notes} {\bf 69} (2023) 1,
  [\href{http://arxiv.org/abs/2206.12177}{{\tt 2206.12177}}].

\bibitem{Barnich:2006av}
G.~Barnich and G.~Comp{\`e}re, {\it {Classical central extension for asymptotic
  symmetries at null infinity in three spacetime dimensions}},  {\em
  Class.Quant.Grav.} {\bf 24} (2007) F15--F23,
  [\href{http://arxiv.org/abs/gr-qc/0610130}{{\tt gr-qc/0610130}}].

\bibitem{Brown:1986nw}
J.~D. Brown and M.~Henneaux, {\it {Central Charges in the Canonical Realization
  of Asymptotic Symmetries: An Example from Three-Dimensional Gravity}},  {\em
  Commun. Math. Phys.} {\bf 104} (1986) 207--226.

\bibitem{Bagchi:2020rwb}
A.~Bagchi, P.~Nandi, A.~Saha, and Zodinmawia, {\it {BMS Modular Diaries: Torus
  one-point function}},  {\em JHEP} {\bf 11} (2020) 065,
  [\href{http://arxiv.org/abs/2007.11713}{{\tt 2007.11713}}].

\bibitem{Bagchi:2023uqm}
A.~Bagchi, S.~Mondal, S.~Pal, and M.~Riegler, {\it {BMS modular covariance and
  structure constants}},  {\em JHEP} {\bf 11} (2023) 087,
  [\href{http://arxiv.org/abs/2307.00043}{{\tt 2307.00043}}].

\bibitem{Campoleoni:2016vsh}
A.~Campoleoni, H.~A. Gonzalez, B.~Oblak, and M.~Riegler, {\it {BMS Modules in
  Three Dimensions}},  {\em Int. J. Mod. Phys.} {\bf A31} (2016), no.~12
  1650068, [\href{http://arxiv.org/abs/1603.03812}{{\tt 1603.03812}}].

\bibitem{Bagchi:2020fpr}
A.~Bagchi, A.~Banerjee, S.~Chakrabortty, S.~Dutta, and P.~Parekh, {\it {A tale
  of three \textemdash{} tensionless strings and vacuum structure}},  {\em
  JHEP} {\bf 04} (2020) 061, [\href{http://arxiv.org/abs/2001.00354}{{\tt
  2001.00354}}].

\bibitem{Oblak:2015sea}
B.~Oblak, {\it {Characters of the BMS Group in Three Dimensions}},  {\em
  Commun. Math. Phys.} {\bf 340} (2015), no.~1 413--432,
  [\href{http://arxiv.org/abs/1502.03108}{{\tt 1502.03108}}].

\bibitem{Dedekind}
T.~Apostol, {\em The Dedekind eta function}, vol.~41 of {\em Graduate Texts in
  Mathematics}, pp.~47--73.
\newblock Springer, New York, 1990.

\bibitem{Carlip:1994gc}
S.~Carlip and C.~Teitelboim, {\it {Aspects of black hole quantum mechanics and
  thermodynamics in (2+1)-dimensions}},  {\em Phys. Rev. D} {\bf 51} (1995)
  622--631, [\href{http://arxiv.org/abs/gr-qc/9405070}{{\tt gr-qc/9405070}}].

\bibitem{Carlip:2000nv}
S.~Carlip, {\it {Logarithmic corrections to black hole entropy from the Cardy
  formula}},  {\em Class. Quant. Grav.} {\bf 17} (2000) 4175--4186,
  [\href{http://arxiv.org/abs/gr-qc/0005017}{{\tt gr-qc/0005017}}].

\bibitem{Sen:2012dw}
A.~Sen, {\it {Logarithmic Corrections to Schwarzschild and Other Non-extremal
  Black Hole Entropy in Different Dimensions}},  {\em JHEP} {\bf 1304} (2013)
  156, [\href{http://arxiv.org/abs/1205.0971}{{\tt 1205.0971}}].

\bibitem{Stanford:2017thb}
D.~Stanford and E.~Witten, {\it {Fermionic Localization of the Schwarzian
  Theory}},  {\em JHEP} {\bf 10} (2017) 008,
  [\href{http://arxiv.org/abs/1703.04612}{{\tt 1703.04612}}].

\bibitem{Almheiri:2016fws}
A.~Almheiri and B.~Kang, {\it {Conformal Symmetry Breaking and Thermodynamics
  of Near-Extremal Black Holes}},  {\em JHEP} {\bf 10} (2016) 052,
  [\href{http://arxiv.org/abs/1606.04108}{{\tt 1606.04108}}].

\bibitem{Preskill:1991tb}
J.~Preskill, P.~Schwarz, A.~D. Shapere, S.~Trivedi, and F.~Wilczek, {\it
  Limitations on the statistical description of black holes},  {\em Mod. Phys.
  Lett.} {\bf A6} (1991) 2353--2362.

\bibitem{Maldacena:1998uz}
J.~M. Maldacena, J.~Michelson, and A.~Strominger, {\it {Anti-de Sitter
  fragmentation}},  {\em JHEP} {\bf 02} (1999) 011,
  [\href{http://arxiv.org/abs/hep-th/9812073}{{\tt hep-th/9812073}}].

\bibitem{Page:2000dk}
D.~N. Page, ``{Thermodynamics of near extreme black holes}.'' preprint
  ALBERTA-THY-13-00, 2000.

\bibitem{Aggarwal:2025hkb}
A.~Aggarwal and J.~Sim{\'o}n, {\it {Modular matrices of 2d Carrollian and
  warped CFTs}},  {\em JHEP} {\bf 06} (2025) 181,
  [\href{http://arxiv.org/abs/2501.12450}{{\tt 2501.12450}}].

\bibitem{Cornalba:2002fi}
L.~Cornalba and M.~S. Costa, {\it {A New cosmological scenario in string
  theory}},  {\em Phys.Rev.} {\bf D66} (2002) 066001,
  [\href{http://arxiv.org/abs/hep-th/0203031}{{\tt hep-th/0203031}}].

\bibitem{Cornalba:2005je}
L.~Cornalba and M.~S. Costa, {\it {Unitarity in the presence of closed timelike
  curves}},  {\em Phys. Rev. D} {\bf 74} (2006) 064024,
  [\href{http://arxiv.org/abs/hep-th/0506104}{{\tt hep-th/0506104}}].

\bibitem{diFrancesco}
P.~Di~Francesco, P.~Mathieu, and D.~Senechal, {\em Conformal Field Theory}.
\newblock Springer, 1997.

\bibitem{Cotler:2018zff}
J.~Cotler and K.~Jensen, {\it {A theory of reparameterizations for AdS$_3$
  gravity}},  {\em JHEP} {\bf 02} (2019) 079,
  [\href{http://arxiv.org/abs/1808.03263}{{\tt 1808.03263}}].

\bibitem{Hao:2021urq}
P.-x. Hao, W.~Song, X.~Xie, and Y.~Zhong, {\it {BMS-invariant free scalar
  model}},  {\em Phys. Rev. D} {\bf 105} (2022), no.~12 125005,
  [\href{http://arxiv.org/abs/2111.04701}{{\tt 2111.04701}}].

\bibitem{Yu:2022bcp}
Z.-f. Yu and B.~Chen, {\it {Free field realization of the BMS Ising model}},
  {\em JHEP} {\bf 08} (2023) 116, [\href{http://arxiv.org/abs/2211.06926}{{\tt
  2211.06926}}].

\bibitem{Hao:2022xhq}
P.-X. Hao, W.~Song, Z.~Xiao, and X.~Xie, {\it {BMS-invariant free fermion
  models}},  {\em Phys. Rev. D} {\bf 109} (2024), no.~2 025002,
  [\href{http://arxiv.org/abs/2211.06927}{{\tt 2211.06927}}].

\bibitem{Banerjee:2022ocj}
A.~Banerjee, S.~Dutta, and S.~Mondal, {\it {Carroll fermions in two
  dimensions}},  {\em Phys. Rev. D} {\bf 107} (2023), no.~12 125020,
  [\href{http://arxiv.org/abs/2211.11639}{{\tt 2211.11639}}].

\bibitem{Bergshoeff:2023vfd}
E.~A. Bergshoeff, A.~Campoleoni, A.~Fontanella, L.~Mele, and J.~Rosseel, {\it
  {Carroll fermions}},  {\em SciPost Phys.} {\bf 16} (2024), no.~6 153,
  [\href{http://arxiv.org/abs/2312.00745}{{\tt 2312.00745}}].

\bibitem{Merbis:2019wgk}
W.~Merbis and M.~Riegler, {\it {Geometric actions and flat space holography}},
  {\em JHEP} {\bf 02} (2020) 125, [\href{http://arxiv.org/abs/1912.08207}{{\tt
  1912.08207}}].

\bibitem{Cotler:2024xhb}
J.~Cotler, K.~Jensen, S.~Prohazka, A.~Raz, M.~Riegler, and J.~Salzer, {\it
  {Quantizing Carrollian field theories}},  {\em JHEP} {\bf 10} (2024) 049,
  [\href{http://arxiv.org/abs/2407.11971}{{\tt 2407.11971}}].

\bibitem{Prohazka:pc}
S.~Prohazka, ``2025.'' personal communication.

\bibitem{Cotler:2024cia}
J.~Cotler, K.~Jensen, S.~Prohazka, M.~Riegler, and J.~Salzer, {\it {Soft
  gravitons in three dimensions}},  \href{http://arxiv.org/abs/2411.13633}{{\tt
  2411.13633}}.

\bibitem{Poulias:2025eck}
G.~Poulias and S.~Vandoren, {\it {On Carroll partition functions and flat space
  holography}},  {\em JHEP} {\bf 06} (2025) 232,
  [\href{http://arxiv.org/abs/2503.20615}{{\tt 2503.20615}}].

\bibitem{Barnich:2015mui}
G.~Barnich, H.~A. Gonzalez, A.~Maloney, and B.~Oblak, {\it {One-loop partition
  function of three-dimensional flat gravity}},  {\em JHEP} {\bf 1504} (2015)
  178, [\href{http://arxiv.org/abs/1502.06185}{{\tt 1502.06185}}].

\bibitem{Simon:2024dwm}
J.~Sim\'on and B.~Yu, {\it {BMS$_{3}$ fermionic localization}},  {\em JHEP}
  {\bf 04} (2025) 137, [\href{http://arxiv.org/abs/2412.05038}{{\tt
  2412.05038}}].

\bibitem{Hao:2025btl}
P.-X. Hao, K.~Shinmyo, Y.-k. Suzuki, S.~Takahashi, and T.~Takayanagi, {\it
  {Bulk Reconstruction of Scalar Excitations in Flat$_3$/CCFT$_2$ and the Flat
  Limit from (A)dS$_3$/CFT$_2$}},  \href{http://arxiv.org/abs/2505.20084}{{\tt
  2505.20084}}.

\bibitem{Riegler:2014bia}
M.~Riegler, {\it {Flat space limit of higher-spin Cardy formula}},  {\em
  Phys.Rev.} {\bf D91} (2015), no.~2 024044,
  [\href{http://arxiv.org/abs/1408.6931}{{\tt 1408.6931}}].

\bibitem{Fareghbal:2014oba}
R.~Fareghbal and A.~Naseh, {\it {Rindler/Contracted-CFT Correspondence}},  {\em
  JHEP} {\bf 1406} (2014) 134, [\href{http://arxiv.org/abs/1404.3937}{{\tt
  1404.3937}}].

\bibitem{Duval2014a}
C.~Duval, G.~W. Gibbons, and P.~A. Horvathy, {\it {Conformal Carroll groups and
  BMS symmetry}},  {\em Class. Quant. Grav.} {\bf 31} (2014) 092001,
  [\href{http://arxiv.org/abs/1402.5894}{{\tt 1402.5894}}].

\bibitem{Despontin:2025dog}
E.~Despontin, S.~Detournay, S.~Dutta, and D.~Fontaine, {\it {Anisotropic
  conformal Carroll field theories and their gravity duals}},
  \href{http://arxiv.org/abs/2505.23755}{{\tt 2505.23755}}.

\bibitem{FarahmandParsa:2018ojt}
A.~Farahmand~Parsa, H.~R. Safari, and M.~M. Sheikh-Jabbari, {\it {On Rigidity
  of 3d Asymptotic Symmetry Algebras}},  {\em JHEP} {\bf 03} (2019) 143,
  [\href{http://arxiv.org/abs/1809.08209}{{\tt 1809.08209}}].

\end{thebibliography}
\end{document}